\documentclass[times,twocolumn,final]{elsarticle}
\usepackage{cag}
\usepackage[table]{xcolor}

\usepackage{framed, multirow, bigstrut}
\usepackage{array}
\usepackage{pgfplots}
\usepackage{import}

\pgfplotsset{compat=newest}
\usetikzlibrary{positioning}
\usepgfplotslibrary{groupplots}
\makeatletter
\pgfplotsset{
    groupplot xlabel/.initial={},
    every groupplot x label/.style={
        at={($({\pgfplots@group@name\space c1r\pgfplots@group@rows.west}|-{\pgfplots@group@name\space c1r\pgfplots@group@rows.outer south})!0.5!({\pgfplots@group@name\space c\pgfplots@group@columns r\pgfplots@group@rows.east}|-{\pgfplots@group@name\space c\pgfplots@group@columns r\pgfplots@group@rows.outer south})$)},
        anchor=north,
        font=\footnotesize
    },
    groupplot ylabel/.initial={},
    every groupplot y label/.style={
            rotate=90,
        at={($({\pgfplots@group@name\space c1r1.north}-|{\pgfplots@group@name\space c1r1.outer
west})!0.5!({\pgfplots@group@name\space c1r\pgfplots@group@rows.south}-|{\pgfplots@group@name\space c1r\pgfplots@group@rows.outer west})$)},
        anchor=south
    },
    execute at end groupplot/.code={%
      \node [/pgfplots/every groupplot x label]
{\pgfkeysvalueof{/pgfplots/groupplot xlabel}};  
      \node [/pgfplots/every groupplot y label] 
{\pgfkeysvalueof{/pgfplots/groupplot ylabel}};  
    }
}

\def\endpgfplots@environment@groupplot{%
    \endpgfplots@environment@opt%
    \pgfkeys{/pgfplots/execute at end groupplot}%
    \endgroup%
}
\makeatother

\newcommand{\colorBoxGraphics}[5]
{
    \begin{tikzpicture}
        \node[anchor=south west,inner sep=0] (#1) at (0,0) {\includegraphics[width=#4]{#2}};
        \begin{scope}[x={(#1.south east)},y={(#1.north west)}]
            \draw[#3,#5] (0.0,0.0) rectangle (1,1);
        \end{scope}
    \end{tikzpicture}
}

\usepackage{amssymb}
\usepackage{latexsym}
\usepackage{ulem}

\usepackage{url}
\definecolor{newcolor}{rgb}{0.8,0.349,0.1}

\definecolor{myBlue}{rgb}{0.3,0.3,0.7}
\definecolor{myRed}{rgb}{0.81,0.33,0.53}
\definecolor{myGreen}{rgb}{0.23,0.70,0.29}
\definecolor{myPurple}{rgb}{0.56,0.11,0.70}
\definecolor{myTeal}{rgb}{0.05,0.52,0.52}
\definecolor{myOrange}{rgb}{0.87,0.52,0.18}
\definecolor{myGray}{rgb}{0.6,0.6,0.6}
\definecolor{myTableWhite}{rgb}{0.9,0.9,0.9}
\definecolor{myTableGray}{rgb}{0.8,0.8,0.8}

\newcommand{\cellW}{\cellcolor{myTableWhite}}
\newcommand{\cellG}{\cellcolor{myTableGray}}

\usepackage[pdfauthor=author]{hyperref}
\usepackage[export]{adjustbox}
\usepackage[switch,pagewise]{lineno} %Required by command \linenumbers below

\usepackage{algorithm}
\usepackage{algpseudocode}

\usepackage{amsmath}
\usepackage{mathtools}

\usepackage{subfig}

\usepackage{booktabs}

\usepackage{tabularx}

\usepackage{tikz}
\usepackage[outline]{contour}
\contourlength{1pt}

\newcommand*{\mline}[1]{%
\begingroup
   \renewcommand*{\arraystretch}{1.0}%
   \begin{tabular}[c]{@{}>{\raggedleft\arraybackslash}p{1.2cm}@{}}#1\end{tabular}%
  \endgroup
}

\DeclareGraphicsExtensions{.pdf, .jpg, .jpeg}

\journal{Computers \& Graphics}

\begin{document}

\setlength{\extrarowheight}{0.1mm}

\verso{Preprint Submitted for review}

\begin{frontmatter}

\title{
    Path Guiding for Wavefront Path Tracing: A Memory Efficient Approach for GPU Path Tracers
    %\tnoteref{tnote1}
}%
%\tnotetext[tnote1]{Only capitalize first word and proper nouns in the title.}

\author[1]{Bora \snm{Yalçıner}\corref{cor1}}
\cortext[cor1]{Corresponding author: 
  Tel.: +90-312-210-5545;  
  }
%\emailauthor{yalciner.bora@metu.edu.tr}{Bora Yalçıner}
\ead{yalciner.bora@metu.edu.tr}
    
\author[1]
{Ahmet Oğuz \snm{Akyüz}}
\ead{akyuz@ceng.metu.edu.tr}
%\fntext[fn1]{Footnote 1.}  

\address[1]{Middle East Technical University, Computer Engineering Department, Ankara, Turkey}

%\received{1 February 2017}
\received{\today}
%%%% Do not use the below for submitted manuscripts
%\finalform{28 March 2017}
%\accepted{2 April 2017}
%\availableonline{15 May 2017}
%\communicated{S. Sarkar}

\begin{abstract}
%%%
We propose a path-guiding algorithm to be incorporated into the wavefront style of path tracers (WFPTs). As WFPTs are primarily implemented on graphics processing units (GPUs), the proposed method aims to leverage the capabilities of the GPUs and reduce the hierarchical data structure and memory usage typically required for such techniques. To achieve this, our algorithm only stores the radiant exitance on a single global sparse voxel octree (SVO) data structure. Probability density functions required to guide the rays are generated \textit{on-the-fly} using this data structure. The proposed approach reduces the scene-related persistent memory requirements compared to other path-guiding techniques while producing similar or better results depending on scene characteristics. To our knowledge, our algorithm is the first one that incorporates path guiding into a WFPT.
%%%%
\end{abstract}

\begin{keyword}
%\MSC 68U07\sep %41A10\sep 65D05\sep 65D17
%% Keywords
\KWD Graphics Processors\sep Monte Carlo Rendering\sep Path Tracing\sep Path Guiding
\end{keyword}

\end{frontmatter}

%\linenumbers

\section{Introduction}
\label{sec:introduction}

Path tracing family of techniques became one of the standard methods for generating photo-realistic imagery~\cite{Kajiya:1986:REQ}. The primary motivation to use these methods is their implementation simplicity and generated image quality. Furthermore, recent advances in graphics hardware enable interactive implementations of such techniques. These techniques tackle the complex recursive light-transport integral by applying numeric Monte Carlo integration, a process known as sampling.

Many sampling schemes are proposed throughout the literature that either sample sub-sections of the integral (e.g., next-event estimation) or reflectance portion of the integral~\cite{Heitz:2014:VNDF}. Such sampling schemes are comparatively simpler because their data is readily available in the initial scene definition. Other parts of the integrand mostly depend on the layout of the elements described in the scene. Extracting a probability field of light distribution over the scene is a critical component of a robust photo-realistic image estimator. This problem is tackled by a family of algorithms which are collectively known as \textit{path guiding} algorithms~\cite{Jensen:1995:ImportancePhoton, Vorba:GMM:2014, Muller:PPG:2017}.

Most path-guiding methods utilize a hierarchical discretization of the light field or a combination of analytically defined functions that fit this light field. The generation of this probability field relies on the light transport simulation itself; thus, path-guiding methods progressively learn this field from path tracing either during runtime or in a preprocessing step. This progressive nature of path guiding necessitates the usage of highly adaptive data structures, which inherently do not suit the GPU architecture well. Adaptive discretization, which relies on adaptive memory management, is not a GPU-friendly operation. Another problem is that such a probability field has a large memory requirement due to its being high-dimensional.

To this end, we propose a wavefront path guiding algorithm that is GPU-friendly and designed to fully utilize the GPU's capabilities. Our main contributions are thus (1) on-the-fly generation of the radiant exitance field, which resides on an SVO data structure; (2) hardware-accelerated approximate cone tracing for an efficient query of the radiant exitance, (3) GPU-friendly parallel product path guiding scheme that utilizes warp-level intrinsics, and (4) a heuristic that judiciously combines generated samples for improved final image quality.

\section{Previous Work}
\label{sec:prev}

The rendering equation is defined by the following integral~\cite{Kajiya:1986:REQ}:
\begin{equation}
L_o(x, \omega_o) = L_e(x, \omega_o) + \int\displaylimits_{\Omega} f_s(x, \omega_i, \omega_o) L_i(x, \omega_i)\cos\theta_i d\omega_i,
\label{eq:re}
\end{equation}
where the outgoing radiance \(L_o(x, \omega_o)\) may contribute to another location \(x_k\), thus, becoming \(L_i(x_k, -\omega_i)\). As such, the above equation can be recursively expanded, resulting in a series of chained integrals with a theoretically infinite recursion depth. Because of that, it is not analytically integrable and is usually evaluated by using the following Monte Carlo estimator, where the emitted radiance $L_e(.)$ is typically omitted:
\begin{equation}
\hat{L}_o(x, \omega_o)  = \frac{1}{N} \sum^{N}_{i = 1} \frac{f_s(x, \omega_i, \omega_o) L_i(x, \omega_i)\cos\theta_i}{p(\omega_i | x, \omega_o)}.
\label{eq:re:mc}
\end{equation}

The variance of such an estimator, visually noticed as noise, is directly related to the similarity between PDF $p$ and the integrand. As finding a single optimal PDF is usually not practical, multiple PDFs are typically combined using multiple importance sampling (MIS)~\cite{Veach:1995:MIS}.

The BSDF portion of the integrand, $f_s(.)$, is traditionally used for importance sampling due to its fully or partially analytic nature. Additionally, parameters required for evaluating it do not recursively depend on other surfaces; thus, $f_s(.)$ can be directly sampled in an efficient manner. However, fitting a density function for the incoming radiance portion of the integrand, $L_i(.)$, is more cumbersome. Finding a plausible density function for the incoming radiance field falls into the domain of path-guiding algorithms.

\begin{figure*}[ht]
\centering

\includegraphics[width=0.9\textwidth]{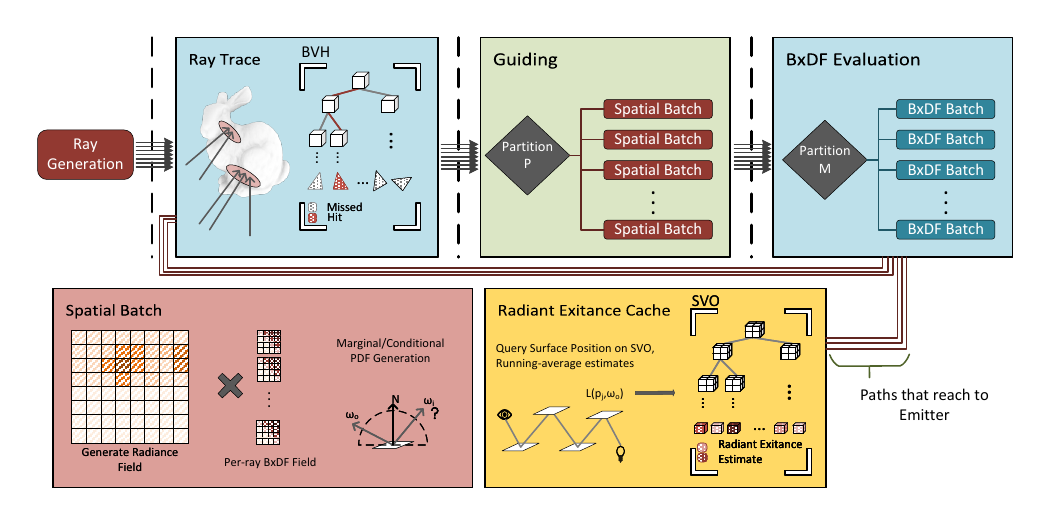}

\caption{The top-down view of the entire path-guiding algorithm. Blue rectangles of the image show the wavefront path tracing operations. Other colored parts are the additional steps required for guiding the rays. \(Partition_p\) and \(Partition_m\) sections represent partitioning the rays by position and material, respectively. Device code is executed for each spatial batch. Each batch generates an incoming radiance field, which is incorporated into the sampling scheme. Paths that reach an emitter contribute to an approximation of the radiant exitance, which is cached on an SVO.}
\label{fig:overview}
\end{figure*}

\subsection{Path Guiding}
\label{sec:prev:pathGuide}

Scene radiance field is a high-dimensional data requiring three dimensions for spatial information and two for spherical direction information. Since the directional distribution of the incoming radiance field is not available from the outset, most path-guiding methods either pre-generate it~\cite{Vorba:GMM:2014, Muller:NIS:2019, Huo:AICF:2020, Bako:DIS:2019} or progressively learn it using already computed path chains~\cite{Muller:PPG:2017, Vevoda:BOR:2018}.

 Path guiding was pioneered by the works of Lafortune et al.~\cite{Lafortune:1995:5D} and Jensen~\cite{Jensen:1995:ImportancePhoton}. These approaches utilize histogram-based techniques, which suffer from memory scalability issues when dealing with high-resolution data. Schüßler et al. proposed a 9D Gaussian mixture model (GMM) incorporating the incident and the previous point's information on the path chain~\cite{Schubler:2022:5DGAUSS}. Combinations of analytically defined functions for approximating the radiance field are proposed as well ~\cite{Vorba:GMM:2014, Dodik:2021:SDDM}. Product sampling, which is the sampling of not only the incoming radiance but its product with BSDF, is also incorporated into these approaches~\cite{Herholz:PIS:2016}.

Ruppert et al. utilize a kd-tree for spatial subdivision and a series of von Mises–Fisher distributions (vMF) for the directional portion of the field~\cite{Ruppert:2020:ParalaxPG}. This method tackles the variance seams that occur when a spatial portion of the data structure is at a low-resolution state by adjusting the vMFs to the sampler path's point of view. Müller et al. proposed the practical path guiding method~\cite{Muller:PPG:2017}. This algorithm utilizes sparse data structures in the global radiance field's spatial and directional portions. Product extension is also proposed for this method~\cite{Diolatzis:2020:PPPG}.

Learning-based methods are also proposed for path guiding. Among these, neural network techniques are either pre-trained via generic scenes~\cite{Bako:DIS:2019, Huo:AICF:2020, Zhu:2021:PhotonDrivenPG} or trained on a per-scene basis~\cite{Muller:NIS:2019}. Reinforcement learning-based techniques are also used by resembling this problem into Q-learning~\cite{Sutton:2018:RL,Dahm:LLT:2017, Kim:2021:SARSAQLearn}. However, these approaches store the directional portion of the radiance field densely, which is not scalable in terms of memory. For the spatial portion, point samples~\cite{Dahm:LLT:2017} or dense arrays~\cite{Kim:2021:SARSAQLearn} are utilized. The Bayesian regression model is also applied to efficiently sample light sources for improved next-event estimation sampling~\cite{Vevoda:BOR:2018}. 

There are real-time and GPU-focused path-guiding approaches as well. Derevyannykh proposed a screen space parametric mixture model for path guiding~\cite{Derevyannykh:2022:PGMixRT}. Due to the screen space nature of the method, only the first bounce is guided. Dittebrandt et al. propose a real-time path-guiding scheme that utilizes compressed quad-trees and a visibility cache for light sampling~\cite{Dittebrandt:2020:Quake}.

Despite several techniques being available for both CPU- and GPU-based path guiding, to our knowledge, none of these algorithms are tailored toward a WFPT style path tracer, which has unique design requirements, as discussed next.

\subsection{Wavefront Path Tracing (WFPT)}
\label{sec:proposed:wavefrontPT}

WFPT is the state-of-the-art design of the graphics device path tracing algorithm~\cite{Laine:2013:MEGAHARM}. Although API-backed, host-style execution methods exist in the device~\cite{Zheng:2022:Luisa}, efficient warp execution mandates some form of partitioning internally.

A straightforward method for parallelizing the path tracing algorithm on a host (i.e., CPU) system is to assign each recursive random walk sequence to a single logical core. However, such parallelization is ill-suited for GPUs because each warp executes in a lock-step fashion. Due to the random nature of the walks, neighboring threads may execute different evaluation routines (e.g., different material and shading computations), which forces the warps to serialize the execution.

The wavefront method segregates the ray casting and material (BSDF) evaluation/sampling routines, allowing recursion to be evaluated in a lock-step fashion. Specifically, for each recursion depth, threads perform the ray-casting operation to determine the incident location. Once the evaluating locations are found, rays are partitioned with respect to the material evaluation parameters. By partitioning the rays in this way, the computational routines are common among the threads, which leads to improved performance and efficiency in execution.

The main issue with this approach is the high memory requirement. This approach requires storing walk states after each operation, which becomes increasingly burdensome as the number of parallel walks increases. In practice, graphics devices often require thousands of walks to be executed to saturate the device, leading to a significant memory burden. 

 Our primary motivation to describe this proposed method comes from this central issue. The accompanying path-guiding methods for path tracing on graphics devices should not further hinder the available memory or, at the very least, minimize its impact.

\section{Proposed Method}
\label{sec:proposed}

\subsection{Wavefront Path Guiding}
\label{sec:proposed:wavefrontPG}

Wavefront path guiding (WFPG) introduces additional steps to the WFPT algorithm. Before partitioning with respect to the BSDF, rays are partitioned by position. Then, the partitioned rays \textit{collaboratively} generate an estimated radiance field, which is utilized for path guiding. After guiding is conducted, rays continue the WFPT steps as usual. The main overview of this algorithm is given in Algorithm~\ref{alg:overview} and visualized in Figure~\ref{fig:overview}. In the following, we describe each part of the algorithm in detail.

\begin{algorithm}[!ht]

\caption{Wavefront path guiding.}\label{alg:overview}

\begin{algorithmic}

\State \textbf{Input-Output}
\State $R_1 = \{r_1, r_2 \dots \}$ \Comment{Set of initial rays}
\State $B = \{(p_1, R_{p,1}), (p_2, R_{p,2}) \dots \}$ \Comment{Position bins}
\State $M = \{(m_1, R_{m,1}), (m_2, R_{m,2}) \dots \}$ \Comment{Material bins}
\State\textbf{Start}
\State \textbf{Initially} Generate rays from the camera and populate $R_1$
\For{$i = 1$ to MaxDepth}

    \State $B_i=\{(p_1, R_{p_1}), (p_2, R_{p_2})\dots\} \leftarrow$ \textsc{Partition-S}$(R_i)$ 
    \State $N_i=\{(n_1, R_{n_1}), (n_2, R_{n_2})\dots\} \leftarrow$ \textsc{Partition-M}$(R_i)$ 
    % Font chars do not have the same spacing
    % Changed to narrow char N
    %\State $M=\{(m_1, R_{m_1}), (p_2, R_{m_2})\dots\} \leftarrow$ \textsc{Partition-M}($R_i$)

    % Guiding
    \ForAll{$(p_j, R_{p_j}) \in B_i$}
        \State $R_{i+1}^j \leftarrow$ \textsc{GuideRays}$((p_j, R_{p_j}))$
    \EndFor

    % Material Evaluation
    \ForAll{$(n_j, R_{n_j}) \in N_i$}
        \ForAll{$r_k \in R_{n_j}$}
            \State \textsc{EvaluateBSDF}$(n_j, r_k)$
        \EndFor    
    \EndFor

    \State $R_{i+1} = \{R_{i+1}^1, R_{i+1}^2 \dots\}$ \Comment{Next set of rays}
\EndFor

\ForAll{paths that reach an emitter}
    \State \textsc{UpdateExitance(SVO)}
\EndFor

\end{algorithmic}
\end{algorithm}

\subsection{Radiant Exitance Caching using Sparse Voxel Octree}

 Unlike other methods that store an incoming radiance field over a spatially discretized volume, we approximate radiant exitance. The main reason for this approach is to reduce memory usage, as radiant exitance is a directionless quantity. This quantity is extracted during Monte Carlo integration and cached in a sparse voxel octree (SVO)~\cite{Laine:2010:Octree:NVIDIA, Crassin:2009:GIGAVOX}.

To explain the caching scheme, we resort to the recursively expanded version of the rendering equation, in which path chains are explicitly written~\cite{PBRBOOK:2016}:
\begin{equation}
    L(p_1 \to p_0) = \sum_{k=1}^{\infty} P(\bar{p}_{k}), 
    \label{eq:gi1}
\end{equation}
where $\bar{p}_k = p_0, p_1, ... p_k$ represents all points along the path of a ray, with $p_0$ on the image plane. The radiance that reaches $p_0$ from such a path is then:
\begin{equation}
\begin{split}
    P(\bar{p}_{k}) & = \underbrace{\idotsint_{\Omega}}_{k-1} L_e(p_k \to p_{k-1}) \\ 
    & \times \left(\prod_{j = 1}^{k-1}f_s(p_{j+1} \to p_j \to p_{j-1}) G(p_{j+1}\to p_j) \right) \\
    & \times dA(p_2) \cdots dA(p_k).
\end{split}
\label{eq:pathSpace}
\end{equation}
where \(f_s\) is the BSDF and \(G\) is the geometry term. The radiance from the path vertex $p_k$ toward $p_{k-1}$ can be extracted from the total throughput as follows:
\begin{equation}
    L(p_k \to p_{k-1}) = \frac{T(\bar{p}_{n})}{T(\bar{p}_{k})} L_e(p_n \to p_{n-1}). 
\label{eq:pathSpaceLocalRad}
\end{equation}
Thus, for every path, the position \(p_k\) and the throughput $T(\bar{p}_{k})$ are stored for every depth on the path. When an emitter is found, its radiance is backpropagated at every depth, and its local radiance estimate is found. The position $p_k$ is used to query the SVO to find the leaf voxel, and finally, these local radiance estimates are accumulated to approximate the radiant exitance for that leaf. This operation corresponds to \textsc{UpdateExitance(SVO}) routine in Algorithm \ref{alg:overview}.

The SVO is generated using Crassin et al.'s approach~\cite{Crassin:2012:HWVOXEL}. The scene is conservatively voxelized in 3D space, and voxels are generated. After the voxelization step, the tree hierarchy is generated using the method described by Karras et al. using Morton code sorting~\cite{Karras:2012:BVH}. 

Since the SVO has a limited voxel resolution, scenes with thin objects would not be adequately represented due to the SVO's volumetric subdivision. To alleviate this, we approximate the surface orientation via normals. Thus, each node of the SVO stores two surface normals, and the radiant exitance corresponding to each normal direction is separately stored. The radiant exitance that is stored on the leaf nodes of the SVO is propagated toward the inner nodes of the tree structure. In our experiments, we found a simple bottom-up averaging scheme to be sufficient. This information is required to query incoming radiance using cone tracing. 

\begin{figure}
    \centering
    \includegraphics[width=0.85\columnwidth]{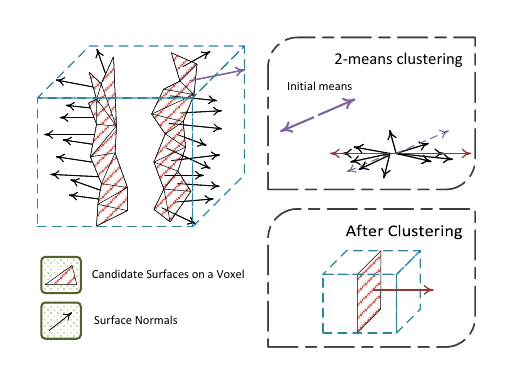}
    \caption{The approximate normals of the voxel are calculated by conducting a k-means clustering with $k=2$. This prevents opposite normals from canceling each other out.}
    \label{fig:voxelMismatch}
\end{figure}

To estimate normals, the surface fragment normals are obtained during the voxelization process and are subjected to a simple k-means clustering procedure with $k=2$. This process is demonstrated in Figure~\ref{fig:voxelMismatch}. The initial means for the algorithm is a random vector selected from the surface elements inside the voxel and its opposite. Through iterative refinement, these vectors are updated to represent the two dominant directions of the surface elements better. At the end, the first cluster's representative vector \(\Vec{N}\) and its opposite \(-\Vec{N}\) are selected as the representative directions for that voxel. The motivation behind this approach is to allow a voxel to become an omnidirectional source. If the clustering results were directly used, certain fragments whose normals point away from the dominant directions would not make a contribution. The clustering approach also prevents normals from canceling each other out, which could occur if a simple average was used.

\begin{figure}[!ht]
\captionsetup[subfigure]{labelformat=empty}
\centering

\subfloat[]{
\begin{tikzpicture}
    \node[anchor=south west,inner sep=0] (image) at (0,0){\includegraphics[width=0.73\columnwidth]{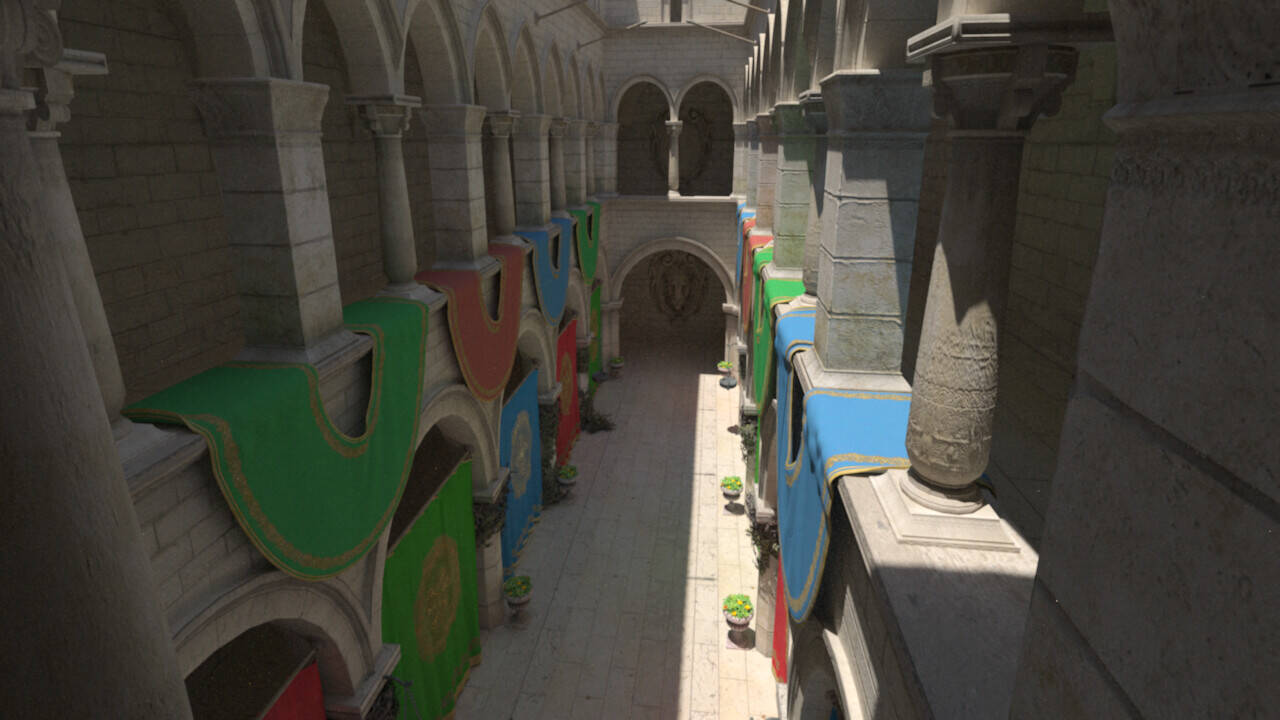}};
    \begin{scope}[x={(image.south east)},y={(image.north west)}]
       \fill[red] (0.11, 0.65) ellipse (0.006 and 0.0106);
       \fill[green] (0.50, 0.59) ellipse (0.006 and 0.0106);
        
    \end{scope}
\end{tikzpicture}
}

\hfill%
\bigskip\\[-7.5ex]
%\hspace*{\fill}%
\hfill%
\subfloat[]{\includegraphics[width=.22\columnwidth,  cframe=green 1.25pt]{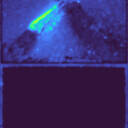}} \hspace{2mm}%
\subfloat[]{\includegraphics[width=.22\columnwidth, cframe=green 1.25pt]{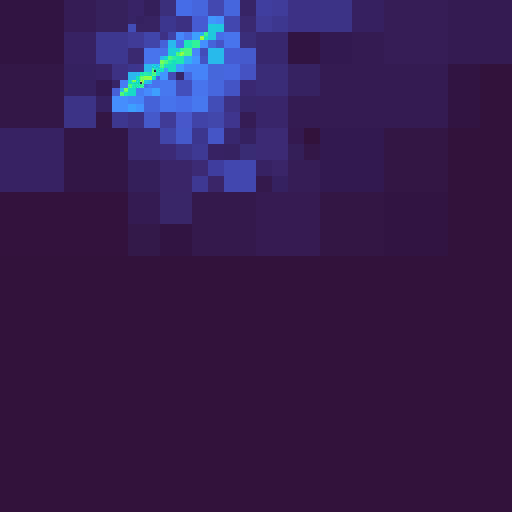}} \hspace{2mm}%
\subfloat[]{\includegraphics[width=.22\columnwidth, cframe=green 1.25pt]{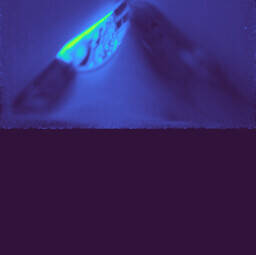}}
\hfill%
\bigskip\\[-7.5ex]
\hfill%
\subfloat[WFPG]{\includegraphics[width=.22\columnwidth, cframe=red 1.25pt]{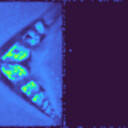}} \hspace{2mm}%
\subfloat[PPG]{\includegraphics[width=.22\columnwidth, cframe=red 1.25pt]{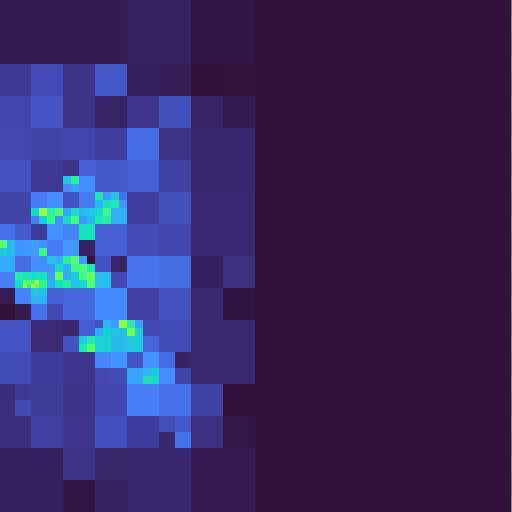}} \hspace{2mm}%
\subfloat[Reference]{\includegraphics[width=.22\columnwidth, cframe=red 1.25pt]{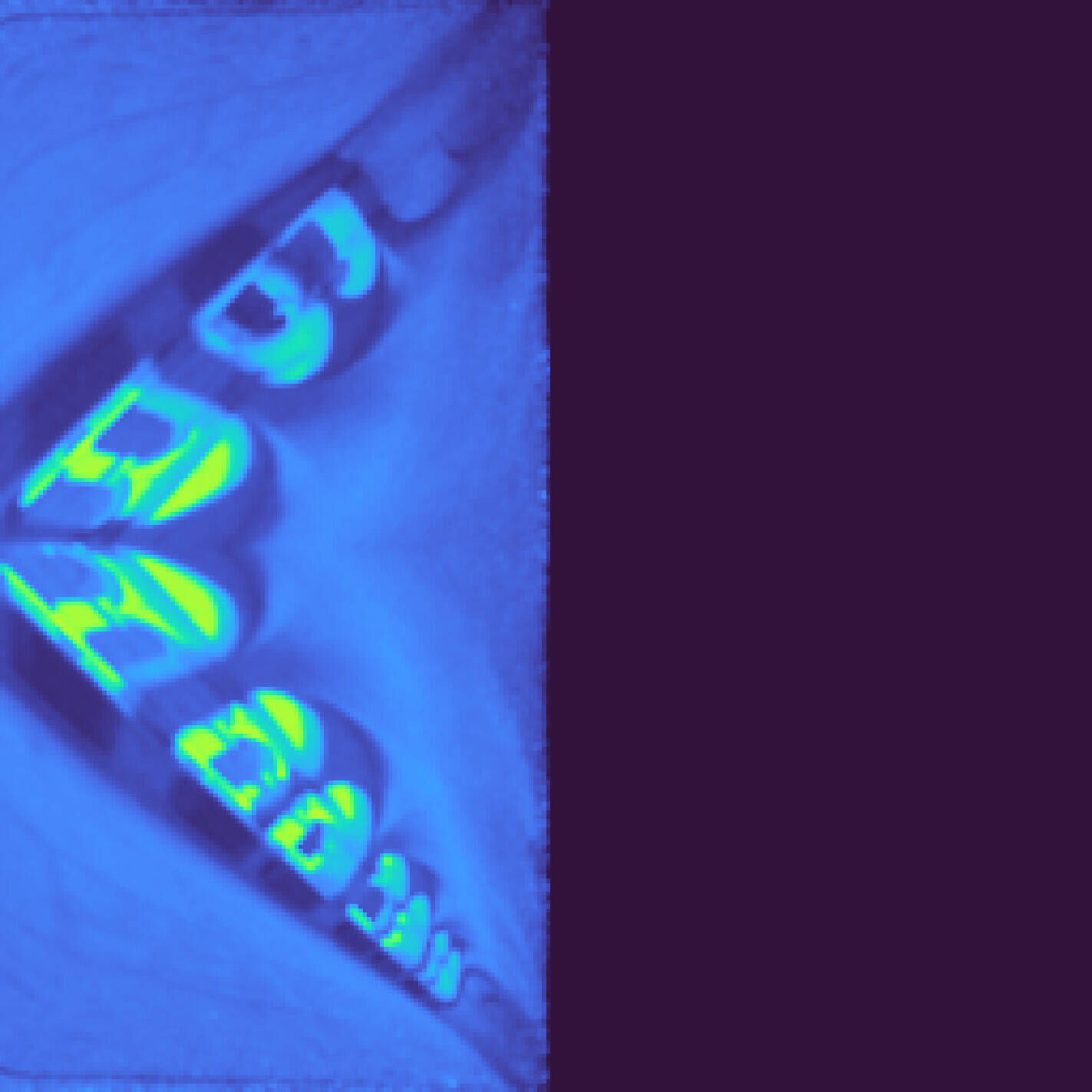}}
\hfill%
\bigskip\\

\caption{Normalized radiance fields of our method (WFPG) and Müller et al.'s method (PPG). In this instance, our method produces a radiance field with a resolution of \(128^2\). Both methods are trained using an equal number of samples (2048 per pixel). The reference radiance field is generated via path tracing and has a resolution of \(256^2\) (\(2^{16}\) samples per pixel).}
\label{fig:pgRef}
\end{figure}

\subsection{On-the-fly Generation of Local Radiance Field}
\label{sec:proposed:wavefrontPG:onthefly}

To generate the incoming radiance over a surface, we utilize a modified cone tracing approach to reduce aliasing artifacts that would be caused by sampling the environment using infinitesimally thin rays~\cite{amanatides1984ray,Crassin:2011:SVO}.  

Given a location on the scene, $p_k$, the omnidirectional incoming radiance field \(L(p_k, \omega_i)\) is stratified into equal solid angle patches, $\omega$. The SVO is queried for each patch by tracing a cone towards that direction to find the incident hit position, which can be found using different approaches. The volumetric estimation proposed by Crassin et al.  \cite{Crassin:2011:SVO} is efficient but is prone to light leaks. Empty space skipping cone tracing~\cite{Laine:2010:Octree:NVIDIA} can also be used, but we found it to be slower compared to a third alternative. In this alternative, we use the underlying device's hardware-accelerated ray tracing capabilities to find the intersection point. After the hit point is found, the radiant exitance stored in the SVO is queried using the cone aperture, hit position, and distance.

As the cone with an aperture of $\omega$ travels into the scene, the area $A$ of the disk at the base of the cone increases. This (projected) area can be computed by using the following formula:
\begin{equation}
    A = r^2 \omega,
    \label{eq:coneArea}
\end{equation}
where $r$ is the distance between the apex and the cone base. At this distance, the area of the disk would be equal to $\pi R^2$, with $R$ being the disk's radius. As we already know the leaf voxel that contains the intersection point, we traverse up the SVO to find the voxel whose area is closest to the disk's area (the square of its side length approximates the voxel's cross-section area). The radiant exitance in this node is then sampled by multiplying the corresponding voxel normal with the cone's principal direction.  

To show the effectiveness of the proposed approach, Figure~\ref{fig:pgRef} compares our method's generated radiance field and that of the practical path-guiding technique~\cite{Muller:PPG:2017} together with the ground-truth reference as seen from two different viewpoints. It can be seen from the figures that our scheme better approximates the actual incoming radiance field.

\begin{algorithm}[!t]

\caption{\textsc{Partition-S} Routine. Partition the paths that have hit \(p_i\) to series of bins \(b_j\) using an SVO with the depth \(d\).}\label{alg:partition}

\begin{algorithmic}

\State \textbf{Input}
\State $R = \{{r_{1}, r_{2} \dots}\} $ \Comment{Rays that are going to be partitioned}
\State $SVO = \{(n_1)^1, (n_1, \dots)^2 \dots (n_1, \dots)^d\}$  \Comment{$(n_1)^1$ is the root}
\State \textbf{Output}
\State $B = \{(p_1, R_{p_1}), (p_2, R_{p_2}) \dots\} $ \Comment{Pair of positions and ray sets}
\State\textbf{Buffer}
\State $I = \{{b_{1}, b_{2} \dots}\} $ \Comment{Bin id for each ray} 
\State\textbf{Start}
\State Clear $B$, $I$
\ForAll{$r_{i} \in R$}
    \State $p_i\leftarrow$ \textsc{RayPosition}$(r_i)$
    \State $n^d_i \leftarrow$ \textsc{DescendLeaf}$(p_{i})$ \Comment{Find the leaf node}
    \State \textsc{AtomicAdd}($n^d_i$, 1)
    \State $b_i \leftarrow $ \textsc{NodeId$(n^d_i)$}
\EndFor

\ForAll{$l \in $ SVO  (in bottom-up fashion, up to $l_{min}$)}
\ForAll{$ n^l \in (n \dots)^l$ in SVO level $l$}
    \State $C = \{c_1, c_2, ... c_8\}$ \Comment{Node children's path count}
    \State $ T \leftarrow c_1 + \dots + c_8 $
    \If{$T \geq c_{ray}$ \textbf{or} $l = l_{min}$}
        \State \textsc{MarkNode$(n^l_i)$} \Comment{This node has sufficient rays}
    \EndIf
\EndFor
\EndFor

\ForAll{$b_{i} \in I$} \Comment{Find the node for each bin}
    \State $n^d_i \leftarrow$ \textsc{ToNode}$(b_{i})$
    \State $n^l_i \leftarrow$ \textsc{AscendAndFindMarked}$(n^d_i)$
    \State $b_i \leftarrow $ \textsc{NodeId$(n^l_i)$}
\EndFor

\State $ B \leftarrow $ \textsc{Partition$(I, R)$} 
\end{algorithmic}
\end{algorithm}

\subsection{Positional Binning using SVO}

Executing the aforementioned radiance field generation scheme would be too costly if evaluated at every point. To amortize this cost, we employ a binning scheme that generates a single radiance field for nearby points. The primary assumption of this approach is that similar regions of the scene would receive similar radiance. 

\begin{figure}
    \centering
    \begin{tabular}{cc}
        \(c_{ray}=128, l_{min}=3\) & 
        \(c_{ray}=256, l_{min}=3\) 
        \\
        \includegraphics[width=0.33\columnwidth]{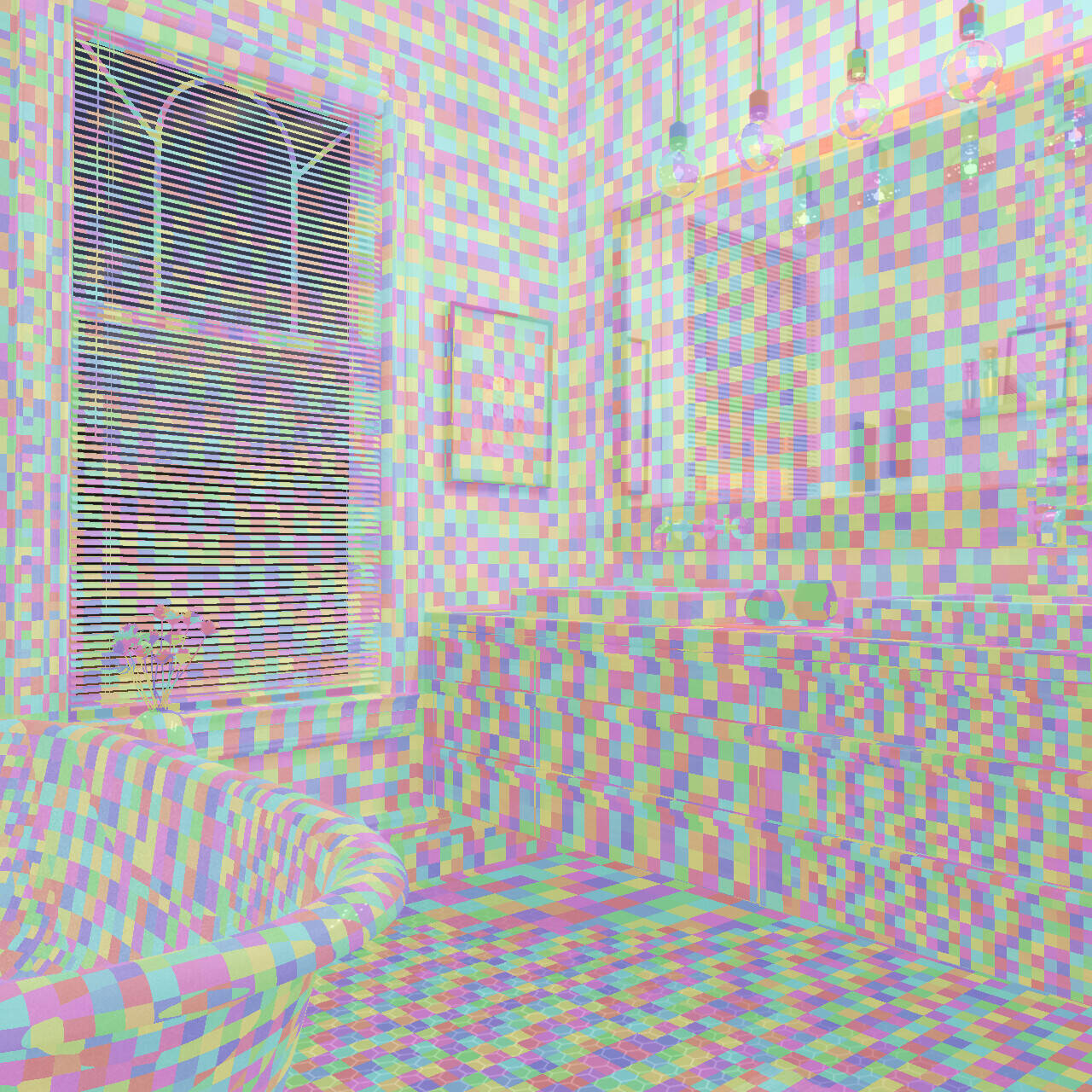}
        &
        \includegraphics[width=0.33\columnwidth]{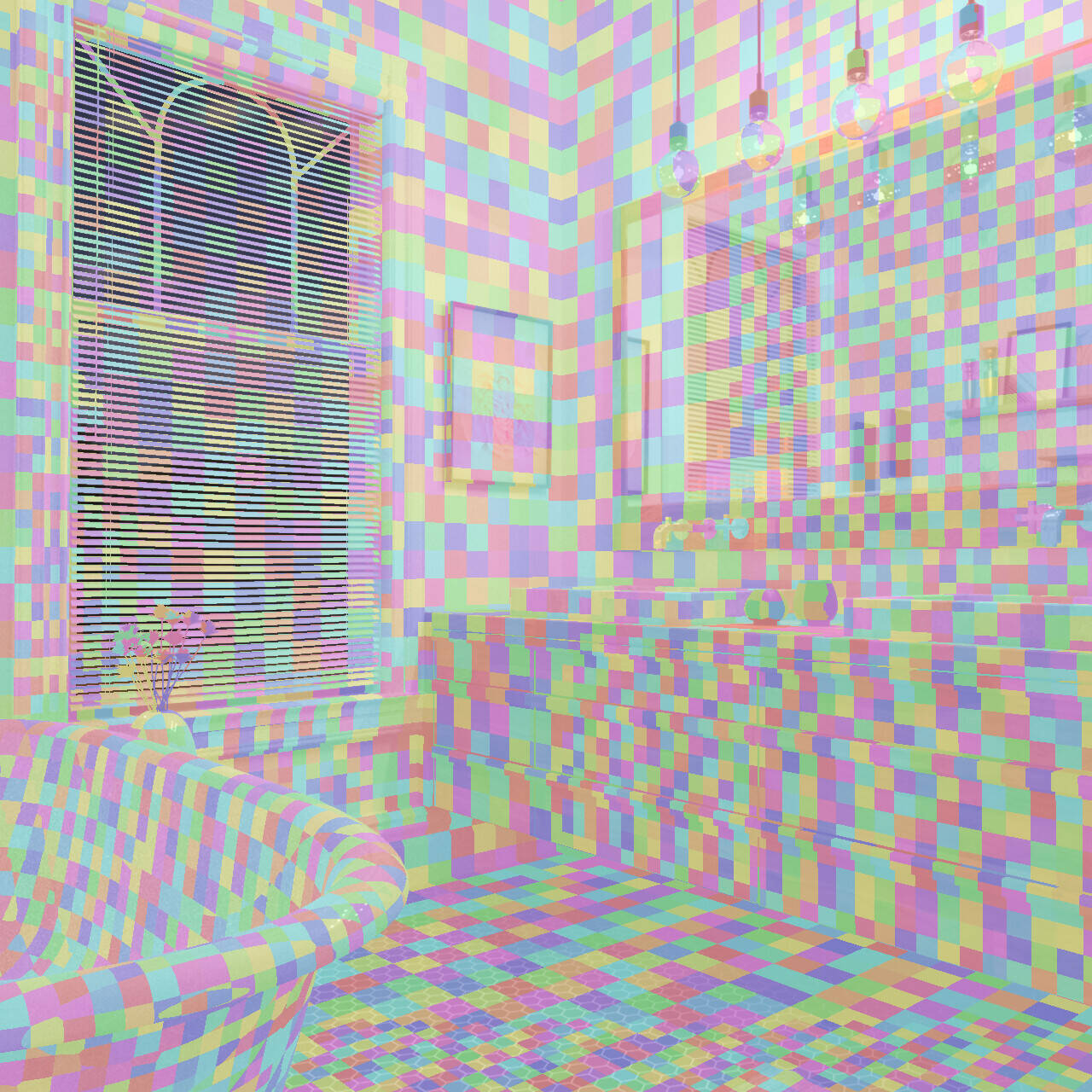} \\
        
        \(c_{ray}=512, l_{min}=3\) & 
        \(c_{ray}=1024, l_{min}=3\) \\

        \includegraphics[width=0.34\columnwidth]{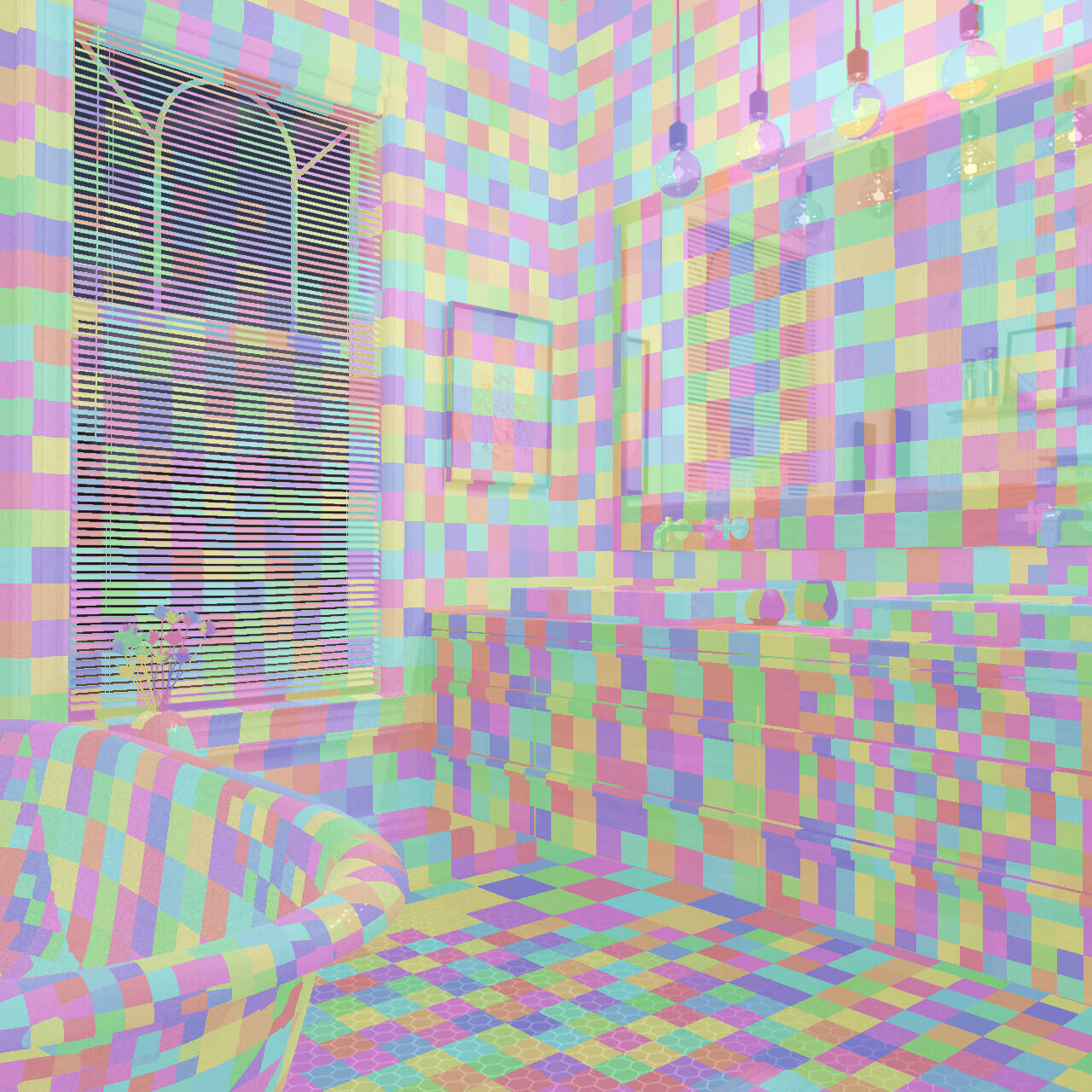}
        &
        \includegraphics[width=0.34\columnwidth]{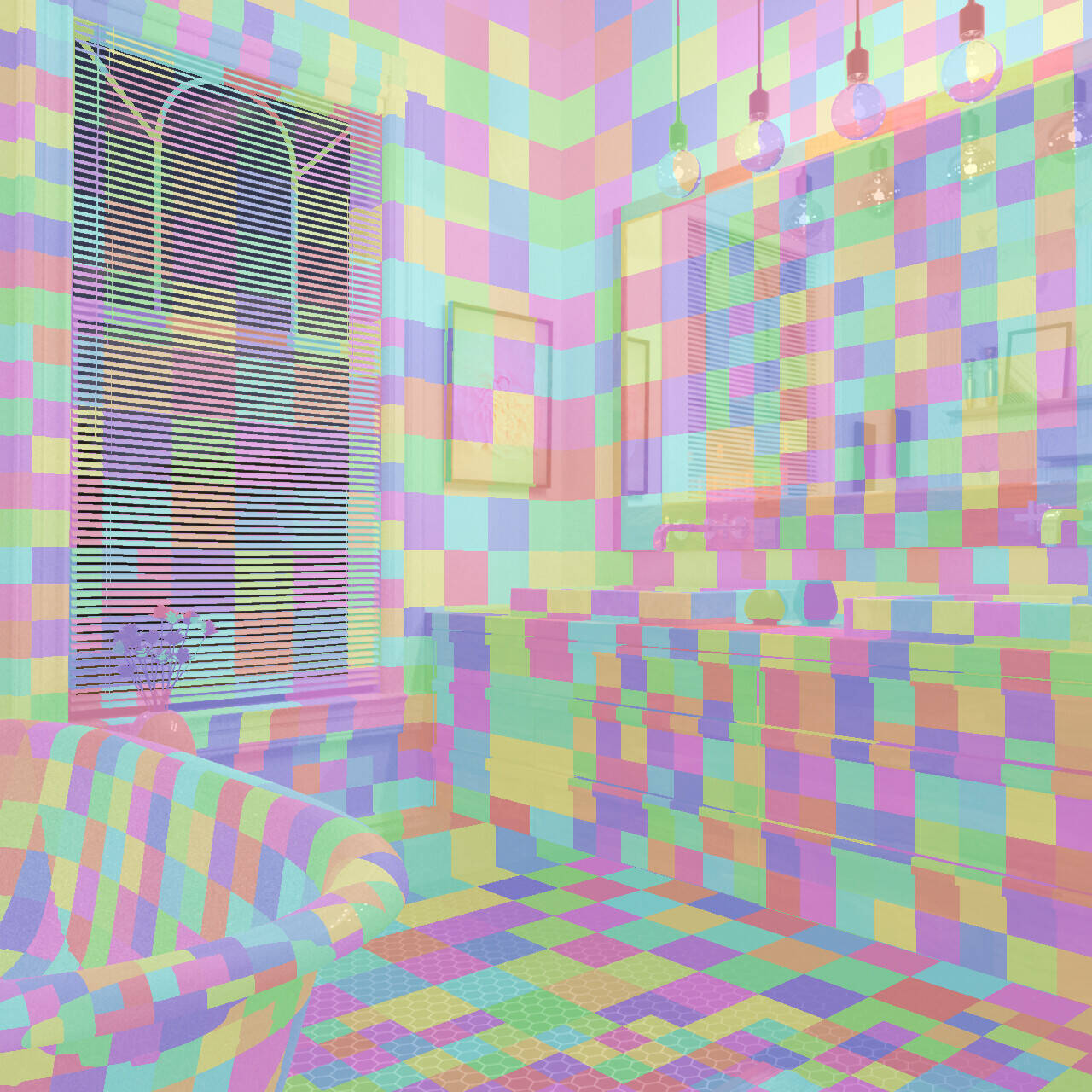} \\
    \end{tabular}
    \caption{False color representation for the binning process for the initial rays coming from the camera. For each colored region, a single local radiance field is generated. Note that the region size increases with the ray count threshold parameter, $c_{\textrm{ray}}$.}
    \label{fig:binningAbl}
\end{figure}

Our positional binning scheme is described in Algorithm~\ref{alg:partition}. Since we already have the scene's SVO hierarchy, we utilize that for the partitioning scheme. Each path atomically increments a value on the leaves of the SVO. Then, these values are accumulated for each level of the SVO in a bottom-up fashion. Two user-defined parameters, referred to as \(l_{min}\)  and  \(c_{\textrm{ray}}\), are employed to control the partitioning process. The \(l_{min}\) parameter sets the minimum limit for the tree level up to which binning can be performed. The \(c_{ray}\) parameter, on the other hand, determines the threshold for the number of rays considered sufficient for each bin. A sample output for this process is shown in Figure~\ref{fig:binningAbl}.

Once binning is complete, the radiance field is generated for each bin in a GPU-oriented manner. That is, a single device block is utilized for each partition, and the threads on that block \textit{simultaneously} generate the radiance field. The resolution of this radiance field is another parameter of our method. In our experiments, we used a maximum resolution of \(128 \times 128\) (corresponds to 64KiB of memory) due to shared memory limitations. This local radiance field is stored in the shared memory available for each block -- in other words, no persistent GPU memory is used.

\subsection{Path Guiding}

The local radiance field generated after the binning process can now be used for path guiding. To this end, we first compute the probability and cumulative density functions (PDF) and (CDF) for sampling according to these distributions. For the sampling process, the radiance field is treated as a piecewise constant 2D function. Traditional inverse sampling methods can be used in parallel~\cite{Shirley:2019:RTGEMS:ZOO}. Each row of the 2D field is assigned to a single warp, which applies an inclusive scan (prefix sum) using warp-level intrinsics to generate the  CDF for each row. The marginal portion of the CDF is calculated similarly. For non-product path guiding (see below), the results of this approach can be used directly. The pseudocode for this phase of our approach is shared in Algorithm~\ref{alg:sample}.

\begin{algorithm}[!t]

\caption{\textsc{GuideRays} routine. Given a bin with partitioned rays, generate incident radiance field, generate PDF and CDF, and sample either using path guiding or BSDF via MIS.}\label{alg:sample}

\begin{algorithmic}
    \State \textbf{Input}
    \State $(p_j, R_{p_j})$ \Comment{Partitioned position and rays}
    %\State $R\{r_{1}, r_{2} \dots r_{c_i}\}$ \Comment{rays of the bin}
    \State \textbf{Output}
    \State ${R^j}_{i+1}$ \Comment{Guided rays}
    \State \textbf{Buffer}
    \State $L(p_i, \omega_i)$ \Comment{Incoming Radiance Field on shared memory}
    \State $PDF(\omega_i), CDF(\omega_i)$ \Comment{PDF and CDF on shared memory}
    \State \textbf{Start}
    \State $p_o \leftarrow$ \textsc{SelectOrigin}$(R_{p_j})$
    \ForAll{$\omega_i \in \Omega$}
        \State $L(p_o, \omega_i) \leftarrow $ \textsc{ConeTrace}$(SVO, p_o, \omega_i)$
    \EndFor
    \State $CDF(\omega_i), PDF(\omega_i)  \leftarrow$ \textsc{GeneratePDF-CDF$(L)$}
    \ForAll{$r_k \in R_{p_j}$}
        \State $M \leftarrow$ \textsc{AcquireMaterial}$(r_k)$
        \State $r_{k_{next}} \leftarrow$ \textsc{MIS}$({PDF(w_i)}, {CDF(w_i)}, M)$
    \EndFor
    \State ${R^j}_{i+1} = \{r_{1_{next}} \dots\}$  
\end{algorithmic}
\end{algorithm}

\subsection{Product Path Guiding}
\label{sec:proposed:wavefrontPG:product}

In product path guiding, we multiply the BSDF at each point with its corresponding radiance field. The main problem that needs to be solved is the efficient computation of this product. For this purpose, we utilize the approach of Estevez et al.~\cite{Estevez:2018:PRODENVMAP}, which was initially proposed for environment mapping. This method utilizes a two-layer hierarchical system. The lower level is the original radiance field. The higher level is the subsampled representation of this field into the resolution of the BSDF field. Based on the constraints of the underlying GPU, we found using an $8\times8$ resolution appropriate for the higher level. The process then involves element-wise multiplication of the BSDF and low-resolution radiance fields. This approach uses a different parallelization scheme compared to the previous one. In this scheme, each warp (group of threads) handles a single point instead of each thread. Using warp-level intrinsics, each warp collaboratively generates a multiplied upper layer and samples from it.

The result is then used for the first stage of sampling, which can be done as explained in the previous section. We then find the corresponding block in the lower level (i.e., higher resolution) radiance field and perform a second stage of sampling according to the distribution in this block. For example, if the higher and lower levels are $8\times8$ and $128\times128$ respectively, the second sampling samples from a $16\times16$ field.

\subsection{Sample Combination Heuristic}
\label{sec:proposed:sample}

As our approach progressively learns about the radiant exitance distribution, initial samples may not benefit sufficiently from path guiding. With each primary ray sample, the distribution will be better learned, and the benefits will improve. However, after a certain number of samples, the field may saturate and only undergo incremental changes. This section describes several heuristics that experimentally combine different sampling schemes given the described behavior. Given a set of \(N\) full image samples and weights \(S = \{ (S_1, W_1), (S_2, W_2), \dots, (S_N, W_N)\)\}, the resulting radiance-field of the generated image \(I\) can be computed with the given heuristics function \(h(i)\) as follows:
\begin{equation}
    I = \frac{\sum\limits_{i=1}^N W_i S_i h(i)}
             {\sum\limits_{i=1}^n W_i h(i)}.
    \label{eq:sampleCombination}
\end{equation}

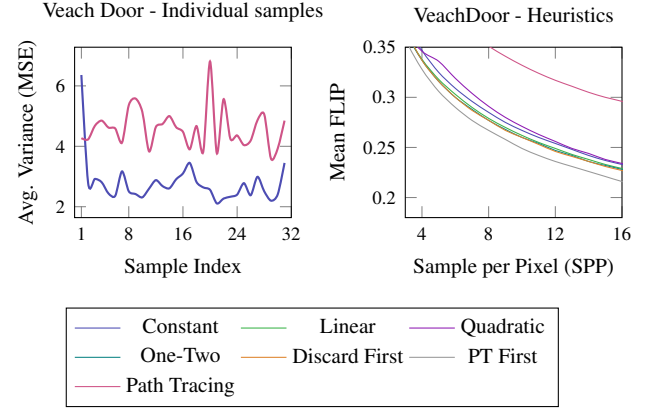
\begin{figure}
\centering
\begin{tikzpicture}
\begin{groupplot}
[
    group style=
    {
        group size= 2 by 1,
        horizontal sep=1.5cm
    },
    %groupplot xlabel=Sample Index (Each sample is one sample per pixel),
    title style={font=\footnotesize},
    label style={font=\footnotesize},
    legend style=
    {
        at={(14.6,-0.5)},
        anchor=north east,
        font=\footnotesize,
        legend columns=3
    },
    width=\columnwidth
]
    
\nextgroupplot
[    
    title=Veach Door - Individual samples,
    %xlabel=Sample per Pixel (SPP),
    ylabel=Avg. Variance (MSE),
    xmin=0, xmax=32,
    xtick={1,8,16,24, 32},
    width=.5\columnwidth,
    xlabel=Sample Index,
    %legend style={at={(0.5,-0.45)},anchor=south,font=\footnotesize},
    %label style={font=\footnotesize},
    tick label style={font=\scriptsize} ,
    legend to name=mainLegend
]
    
    \addplot[smooth,myBlue,thick] table[x=Sample, y=WFPG]{figures/graphs/door-singlesample-data.txt}; \addlegendentry{Constant}

    \addplot[smooth,myRed,thick] table[x=Sample, y=PT]{figures/graphs/door-singlesample-data.txt}; \addlegendentry{Path Tracing}
    
 \nextgroupplot
[    
    title=VeachDoor - Heuristics,
    xlabel=Sample per Pixel (SPP),
    ylabel=Mean FLIP,
    xmin=3, xmax=16,
    ymin=0.18, ymax=0.35,
    xtick={2,4,8,...,16},
    width=.5\columnwidth,
    tick label style={font=\scriptsize},
    legend to name=mainLegendHeu
]
    
    % WFPG GRAPHS
    \addplot[smooth,myBlue,thin] table[x=Sample, y=WFPG]{figures/graphs/door-method-data.txt}; \addlegendentry{Constant}
    
    \addplot[smooth,myGreen,thin] table[x=Sample, y=WFPG_lin]{figures/graphs/door-method-data.txt}; \addlegendentry{Linear}
    
    \addplot[smooth,myPurple,thin] table[x=Sample, y=WFPG_quad]{figures/graphs/door-method-data.txt}; \addlegendentry{Quadratic}
    
    \addplot[smooth,myTeal,thin] table[x=Sample, y=WFPG_onetwo]{figures/graphs/door-method-data.txt}; \addlegendentry{One-Two}
    
    \addplot[smooth,myOrange,thin] table[x=Sample, y=WFPG_discardfirst]{figures/graphs/door-method-data.txt}; \addlegendentry{Discard First}
    
    \addplot[smooth,myGray,thin] table[x=Sample, y=WFPG_ptfirst]{figures/graphs/door-method-data.txt}; \addlegendentry{PT First}

    % PT GRAPH
    \addplot[smooth,myRed,thin] table[x=Sample, y=PT]{figures/graphs/door-method-data.txt}; \addlegendentry{Path Tracing}
    
\end{groupplot}
\end{tikzpicture}
\pgfplotslegendfromname{mainLegendHeu}
\caption{Single sample variance of the proposed and traditional path-tracking methods. Each sample on the graph is considered in isolation. This graph exposes our method's learning scheme. The general trend of the light distribution is immediately learned in a couple of samples. In this example, the benefits stabilize after about the $20^\textrm{th}$ sample. The first sample of path guiding has higher error than pure path tracing because we sample an omnidirectional field, whereas path tracing samples a hemispherical one.}
\label{fig:proposed:heuristics}
\end{figure}

Several heuristic functions are shown below: 

\begin{align}
%h(i) &= 1  \tag{Constant} \\
h(i) &= \begin{dcases*}
            i & \(i < 5\) \\
            5 & otherwise
        \end{dcases*} 
        \label{eq:linear}
        \tag{Linear} 
\\
h(i) &= \begin{dcases*}
            i^2 & \(i < 5\) \\
            25 & otherwise
        \end{dcases*} 
        \label{eq:quadratic}
        \tag{Quadratic}
\\
h(i) &= \begin{dcases*}
            1 &  \(i = 1 \) \\
            2 & otherwise
        \end{dcases*} \tag{One-Two}
        \label{eq:oneTwo}
\\
h(i) &= \begin{dcases*}
            0 &  \(i = 1 \) \\
            1 & otherwise
        \end{dcases*} \tag{Discard First}
        \label{eq:discardFirst}
\end{align}
In addition to these heuristics, we experimented with two more, namely ``Constant'' and ``PT First''. In the former, each sample has constant weight, and in the latter, the first sample directly comes from the first path-tracing sample without path guiding being applied. The remaining samples are generated with path guiding and are equally weighted.

The results for different combinations are shown in Figure~\ref{fig:proposed:heuristics}. Here, the left graph shows the mean squared error for each sample in isolation. The pink curve corresponds to pure path tracing and the purple curve to our approach. It can seen that for the first sample, our approach has higher error as we sample an omnidirectional field, whereas path tracing samples a hemispherical one. In our case, the following samples produce lower variance than path tracing as the light distribution is learned. The benefits stabilize after a certain point. On the right-hand side of the same figure, we show the results of different sample combination heuristics with respect to the HDR-FLIP metric~\cite{Andersson:2021:HDRFLIP}. It can be observed that among the proposed strategies, the best combination strategy is ``PT First'', which we use for the results produced in this paper.

\begin{figure}[!ht]
\centering

\includegraphics[width=0.75\columnwidth]{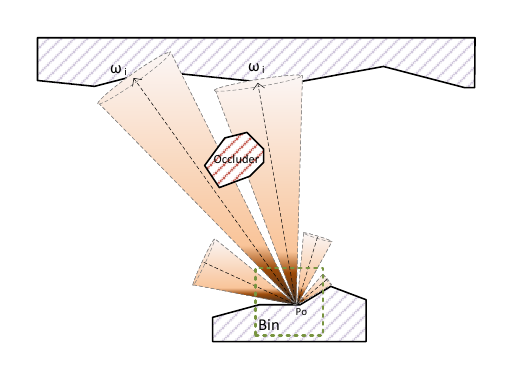}

\caption{Aliasing illustration, assuming radiance field is generated over the volume represented by the green dashed square. The radiance field is generated from a point \(p_o\). The contribution of a small occluder (shaded red) could not be captured due to the low-resolution radiance field. Cone rays miss the occluder, and the radiant exitance of the surface behind is queried.}
\label{fig:alias}
\end{figure}

\section{Implementation}
\label{sec:implementation}

We have implemented our algorithm using CUDA. For hardware-accelerated ray tracing, we use the OptiX Framework~\cite{Parker:2010:OptiX}. Our source code is publicly available in~\cite{Yalciner:2023:MRay}. In the following, we discuss several important implementation issues.

\textbf{OptiX \& Shared Memory:} Since OptiX does not expose inline ray-tracing capabilities, we could not utilize the shared memory and the device's hardware-accelerated ray-tracing capabilities in a single kernel execution. Therefore, we use a small persistent buffer to segregate the OptiX ray-tracing pipeline launches with the sampling kernel launches. The allocation amount depends on the number of processors on the device. In our experiments, 8 to 16 MiB of memory was enough to saturate mid to high-end GPU. This segregation is unnecessary for other APIs, such as DXR and Vulkan, since hardware-accelerated ray tracing inside a compute shader and access to the shared memory can be done simultaneously.

\textbf{Captured radiance field's origin and aliasing:} Selecting the reference point for the radiance field generation is not simple. Directly selecting the center point of the collaborating rays would create self-occlusions, or directly using the center of the partitioned region would make variance seams towards the edges of the partitioned area. Instead, we randomly select a candidate hit location and use it as a radiance field origin at each iteration.

Like in the real-time rendering paradigm, aliasing is an issue, even with the cone tracing approach (Figure \ref{fig:alias}). High-frequency illumination or occlusion could not be captured due to the relatively low-resolution radiance field. To reduce this issue, we jitter the sampling directions to capture these high-frequency regions, at least on some iterations.

\begin{figure}
    \centering
    \small
    \setlength{\tabcolsep}{3.5mm}
    \begin{tabular}{cccc}
    
        No Jitter &  Jitter + Gaussian \\        
        \includegraphics[width=0.33\columnwidth]{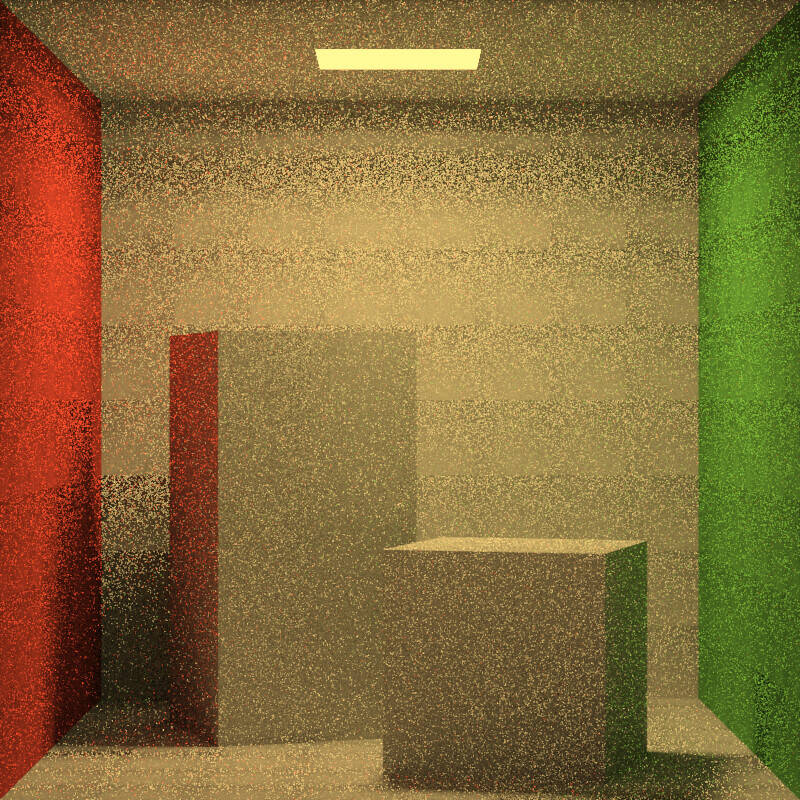}
        &
        \includegraphics[width=0.33\columnwidth]{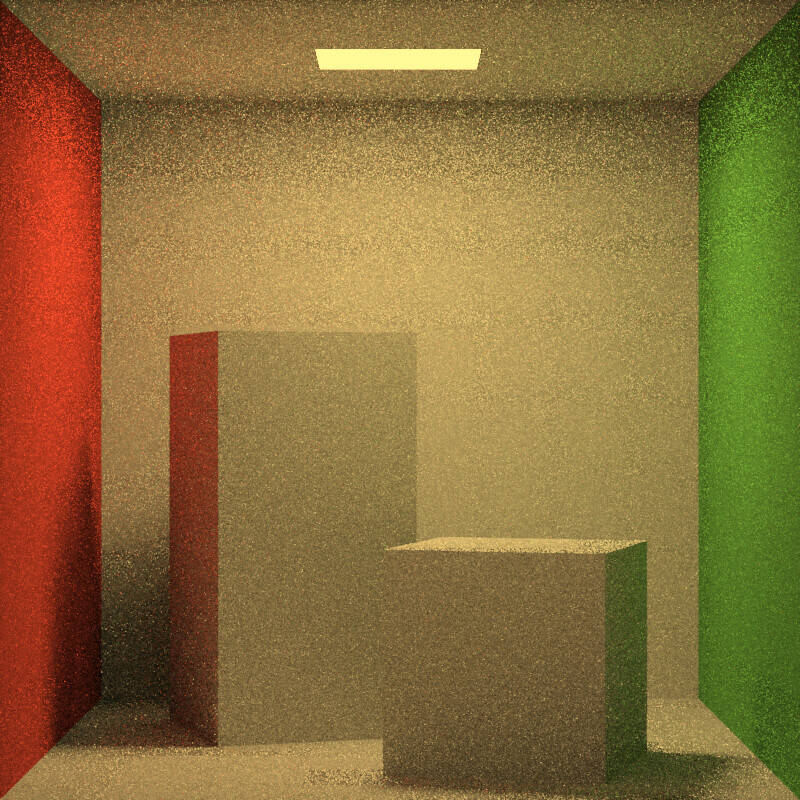}        
        \\        
    \end{tabular}
    \caption{Demonstration of jittering and filtering. NEE is turned off for demonstration purposes. Variance seams are visible without jittering. Positional and directional jitter and Gaussian blur minimize variance seams.}
    \label{fig:antiAlias}
\end{figure}

Both directional and spatial jittering minimize variance seams. However, the radiance field we generate corresponds precisely to the spatial location being rendered; any other rays trailing a slightly different part of the scene would need a slightly modified field to compensate for the difference in perspective. To make the generated radiance field plausible for all the rays in the bin, we filter the radiance field using Gaussian blur. The effects of these antialiasing approaches are illustrated in Figure~\ref{fig:antiAlias}.

\textbf{PDF Domains:} For an estimator to be unbiased, the sampling domain should encapsulate the sampled radiance field, meaning that its PDF should be non-zero where the radiance field is non-zero. Generated radiance fields are common for groups of rays, and they are generated from a singular location on the spatial domain; sampling for some rays would not satisfy this unbiasedness constraint. Blurring the generated radiance field alleviates this issue somewhat, but there can still be zero values on the radiance field due to self-occlusion. Because of that, we set a constant non-zero epsilon ($\epsilon = 10^{-2}$) value as the initial value of the radiance field.

\textbf{Multiple Importance Sampling (MIS):} Although we explain our algorithm as if the radiance field guiding scheme is the only sampler for the given path, in practice, traditional next-event estimation (NEE) and BSDF sampling schemes are combined using MIS. MIS requires the combined PDFs to be present for calculation. To acquire the BSDF sampler's PDF, BSDF data needs to be accessed, which may raise an issue of branching discrepancies between threads within a warp. However, due to the spatial binning scheme, nearby regions usually have the same BSDF, and branch divergence is minimal. This is not an issue with the product path guiding scheme since the entire warp is responsible for a single ray.

\textbf{Radiance field projection on to 2D Cartesian space.} To represent the radiance field on a 2D Cartesian grid, we utilize concentric octahedral mapping~\cite{Clarberg:2008:COOCTO}. This method is known to give better results on lower resolutions than other classical projection techniques, such as spherical projection.

\section{Results \& Validation}
\label{sec:results}

In this section, we evaluate our method under various test scenarios. For clarity, our method is named WFPG throughout this section. In all tests, next-event estimation is enabled. All of the GPU measurements are done using a 3070Ti Mobile GPU. Lastly, we generate radiance fields in a decaying manner, meaning that with each path depth, we decrease the size of the generated radiance field by two in each dimension. This is motivated by the fact that earlier bounces make a greater contribution to the final image due to higher throughput.

\subsection{Profiling}
\label{sec:results:profiling}

\begin{table}[!ht]
    \small
    %\centering
    \raggedright
    \setlength{\tabcolsep}{1.0mm} % Default value: 6pt
    
    \caption{Single sample per pixel timings (ms) of the wavefront path guiding stages. Each Depth\textsubscript{n}  is the total computation time of our algorithm with (bottom) and without (top) product path guiding for the $n^\textrm{th}$ bounce of a ray path. The miscellaneous portion includes partitioning and material evaluation routines. 
    % Except for the Bathroom scene, the total traversal depth is 6. For the bathroom scene, the traversal depth is 10. 
    For product path guiding, BSDF resolution is \(8 \times 8\). The bottom row shows the time of pure path tracing.}
    \label{tab:profiling}
    
    \begin{tabular}{c c c c c}
    \toprule
    % HEADERS    
    & \multicolumn{2}{c}{\(1920 \times 1080\)}
    & \multicolumn{2}{c}{\(1280 \times 1280\)}
    \\
    %& \rotatebox[origin=l]{45}{\textsc{Sponza}}
    & \textsc{Sponza}
    %& \rotatebox[origin=l]{45}{\textsc{VeachDoor}}
    & \textsc{VeachDoor}
    %& \rotatebox[origin=l]{45}{\textsc{Bathroom}}
    & \textsc{Bathroom}
    & \textsc{CornellBox}
    \\ \midrule %\cmidrule{2-5}
    % WFPG
    \textbf{WFPG} & \multicolumn{4}{c}{ \(l_{min} = 5\), \(c_{ray} = 512 \), SVO \(=256^3\)}
    \\ \cmidrule{2-5}
    % ======================================================%
    % Depth 1 - NP
    \multicolumn{1}{l}{
    \multirow{2}{*}{Depth\textsubscript{1} \footnotesize(\(128 \times 128\))}}
    & \cellW 48.29 & \cellW 35.76 & \cellW 33.78 & \cellW 74.46
    \\
    % Depth 1 - P
    & \cellG 199.67 & \cellG 180.66 & \cellG 143.37 & \cellG 134.88
    \\ \cmidrule{2-5}
    % ======================================================%
    % Depth 2 - NP
    \multicolumn{1}{l}{
    \multirow{2}{*}{Depth\textsubscript{2} \footnotesize(\(64 \times 64\))}}
    & \cellW 25.36 & \cellW 21.23 & \cellW 22.54 & \cellW 31.24
    \\
    % Depth 2 - P
    & \cellG 109.46 & \cellG 122.96 & \cellG 99.94 & \cellG 82.72
    \\ \cmidrule{2-5}
    % ======================================================%
    % Depth 3 - NP
    \multicolumn{1}{l}{
    \multirow{2}{*}{Depth\textsubscript{3} \footnotesize(\(32 \times 32\))}}
    & \cellW 16.68 & \cellW 16.84 & \cellW 14.20 & \cellW 12.39
    \\
    % Depth 3 - P
    & \cellG 76.67 & \cellG 107.43 & \cellG 79.90 & \cellG 56.16
    \\ \cmidrule{2-5}
    % ======================================================%
    % Depth 4 - NP
    \multicolumn{1}{l}{
    \multirow{2}{*}{Depth\textsubscript{4} \footnotesize(\(16 \times 16\))}}
    & \cellW 10.96 & \cellW 17.25 & \cellW 11.10 & \cellW 7.15
    \\
    % Depth 4 - P
    & \cellG 58.03 & \cellG 105.77 & \cellG 73.90 & \cellG 47.56
    \\ \cmidrule{2-5}
    % ======================================================%
    % Update Exitance
    \multicolumn{1}{l}{Update Exitance}
    & \cellW 4.21 & \cellW 4.07 & \cellW 6.5 & \cellW 4.37
    \\ \cmidrule{2-5}
    % ======================================================%
    % Misc
    \multicolumn{1}{l}{Miscellaneous}
    & \cellG 75.21 & \cellG 70.80 & \cellG 102.76 & \cellG 48.33
    %\\ \cmidrule{2-5}
    %\addlinespace 
    \\ \cmidrule{2-5}\morecmidrules\cmidrule{2-5}
    % ======================================================%
    % Total - NP
    \multicolumn{1}{l}{
    \multirow{2}{*}{Total}}
    & \cellW 180.71 & \cellW 165.95 & \cellW 190.88 & \cellW 177.84
    \\
    % Total - P
    & \cellG 523.25 & \cellG 591.69 & \cellG 506.37 & \cellG 374.02
    \\ \midrule
    \multicolumn{1}{l}{SVO Generation}
    & \cellW 45.45 & \cellW 17.81 & \cellW 35.54 & \cellW 44.39
    \\ \midrule \midrule
    % Path Tracing
    \textbf{PT} & 74.00 & 80.8 & 92.73 & 54.42
    \\ 
    \bottomrule
    
    \end{tabular}
\end{table}

In Table~\ref{tab:profiling}, we share the overall timing calculations for our algorithm. The WFPG portion shows the time spent at each depth of our algorithm. The lightly shaded rows (top) are for pure path guiding, and the darkly shaded ones (bottom) are for product path guiding. The time for SVO generation, which is done only once, is shown at the bottom. The PT row shows the time of pure path tracing without using our method. It can be seen that the overall cost of our path-guiding algorithm is approximately two times of path tracing. The cost of product path guiding is higher due to per-ray product field calculation, multiplication, and layered sampling.

Table \ref{tab:stats} shows several statistics of the WFPG on different scenes. These are the bin counts, average rays per bin, and the combined PDF and CDF memory sizes at each bounce. Dense methods such as Dahm and Keller~\cite{Dahm:LLT:2017} and Kim et al.~\cite{Kim:2021:SARSAQLearn} would be required to hold these dense structures in persistent memory, whereas in our case, this memory is transient in the sense that it is used as shared memory for each bounce and released for the next one.

\begin{table}[!ht]
    \small
    \setlength{\tabcolsep}{1.5mm}
    \caption{WFPG statistics for selected scenes. Parameters for our method are the same as in Table \ref{tab:profiling}. Statistics for only the first three path depths are provided. }
    \label{tab:stats}
    \centering
        \begin{tabular}{cc c c c}
        %\cline{3-10}
        \toprule
        && \textbf{Depth\textsubscript{1}}
        &  \textbf{Depth\textsubscript{2}}
        &  \textbf{Depth\textsubscript{3}}
        %& \textbf{Depth\textsubscript{4}}
        \\
        && \textbf{\(128 \times 128\)}
        & \textbf{\(64 \times 64\)}
        & \textbf{\(32 \times 32\)}
        \\
        \cmidrule{2-5}
        %===========================
        %       SPONZA
        %===========================
        \multirow{2}{*}{\textsc{Sponza}}
        & \cellW (1)
        & \cellW 2005 / 1034
        & \cellW 2713 / 570
        & \cellW 2604 / 444
        %& \cellG1 2444 / 257
        \\
        & \cellG (2)
        & \cellG 262.8
        & \cellG 88.9
        & \cellG 21.3
        %& \cellG2 5.0
        \\ \cmidrule{2-5}
        %===========================
        %       VEACH DOOR
        %===========================
        \multirow{2}{*}{\textsc{VeachDoor}}
        & \cellW (1)
        & \cellW 2218 / 935
        & \cellW 2637 / 707
        & \cellW 2859 / 602
        %& xxx / xxx
        \\
        & \cellG (2)
        & \cellG 290.7
        & \cellG 86.41
        & \cellG 23.42
        %& \multicolumn{2}{c}{xxx}
        \\ \cmidrule{2-5}
        %===========================
        %       BATHROOM
        %===========================
        \multirow{2}{*}{\textsc{Bathroom}}
        & \cellW (1)
        & \cellW 1867 / 829
        & \cellW 3158 / 444
        & \cellW 3022 / 418
        %& xxx & xxx
        \\ 
        & \cellG (2)
        & \cellG 244.7
        & \cellG 103.5
        & \cellG 24.8
        %& xxx
        \\ \cmidrule{2-5}
        %===========================
        %       CornellBox
        %===========================
        \multirow{2}{*}{\textsc{CornellBox}}
        & \cellW (1)
        & \cellW 4861 / 337
        & \cellW 7151 / 180
        & \cellW 7048 / 136
        %& xxx & xxx
        \\ 
        & \cellG (2)
        & \cellG 637.1
        & \cellG 234.3
        & \cellG 57.7
        %& xxx
        \\ \midrule
        \multicolumn{5}{l}
        {
            \begin{tabular}{cl}
                (1)& Bin count / avg. ray per bin  \\
                (2)& Generated PDF and CDF memory (MiB) \\ 
            \end{tabular}
        } \\
    \bottomrule
    \end{tabular}    
\end{table}

\subsection{Equal Sample/Time Comparison}
\label{sec:results:eqTimeEqSample}

\newcommand{\plotGen}[9]
{
    \nextgroupplot
    [    
        title=#5,
        xlabel=#7,
        %ylabel=Mean FLIP,
        xmin=#2, xmax=#4,
        ymin=#8, ymax=#9,
        xtick={#2,#3,\fpeval{2*(#3)},...,#4},
        width=.21\textwidth,
        tick label style={font=\scriptsize},
        legend to name=#6,
        yticklabel style={
                        /pgf/number format/precision=3,
                        /pgf/number format/fixed},
    ]

    \addplot[smooth,myRed,thick]
        table[x=#1, y=PT , col sep=semicolon]{#6};
    \addlegendentry{Path Tracing}

    \addplot[smooth,myGreen,thin]
        table[x=#1, y=WFPG-np-mis, col sep=semicolon]{#6};
    \addlegendentry{WFPG MIS}

    \addplot[smooth,myGray,thin]
        table[x=#1, y=WFPG-p-mis, col sep=semicolon]{#6};
    \addlegendentry{WFPG Product MIS}

}

\newcommand{\logPlotGen}[9]
{
    \nextgroupplot
    [    
        title=#5,
        xlabel=#7,
        %ylabel=Mean FLIP,
        ymode=log,
        xmin=#2, xmax=#4,
        ymin=#8, ymax=#9,
        xtick={#2,#3,\fpeval{2*(#3)},...,#4},
        width=.21\textwidth,
        tick label style={font=\scriptsize},
        legend to name=#6,
        yticklabel style={
                        /pgf/number format/precision=3,
                        /pgf/number format/fixed},
    ]

    \addplot[smooth,myRed,thick]
        table[x=#1, y=PT , col sep=semicolon]{#6};
    \addlegendentry{Path Tracing}

    \addplot[smooth,myGreen,thin]
        table[x=#1, y=WFPG-np-mis, col sep=semicolon]{#6};
    \addlegendentry{WFPG MIS}

    \addplot[smooth,myGray,thin]
        table[x=#1, y=WFPG-p-mis, col sep=semicolon]{#6};
    \addlegendentry{WFPG Product MIS}

}

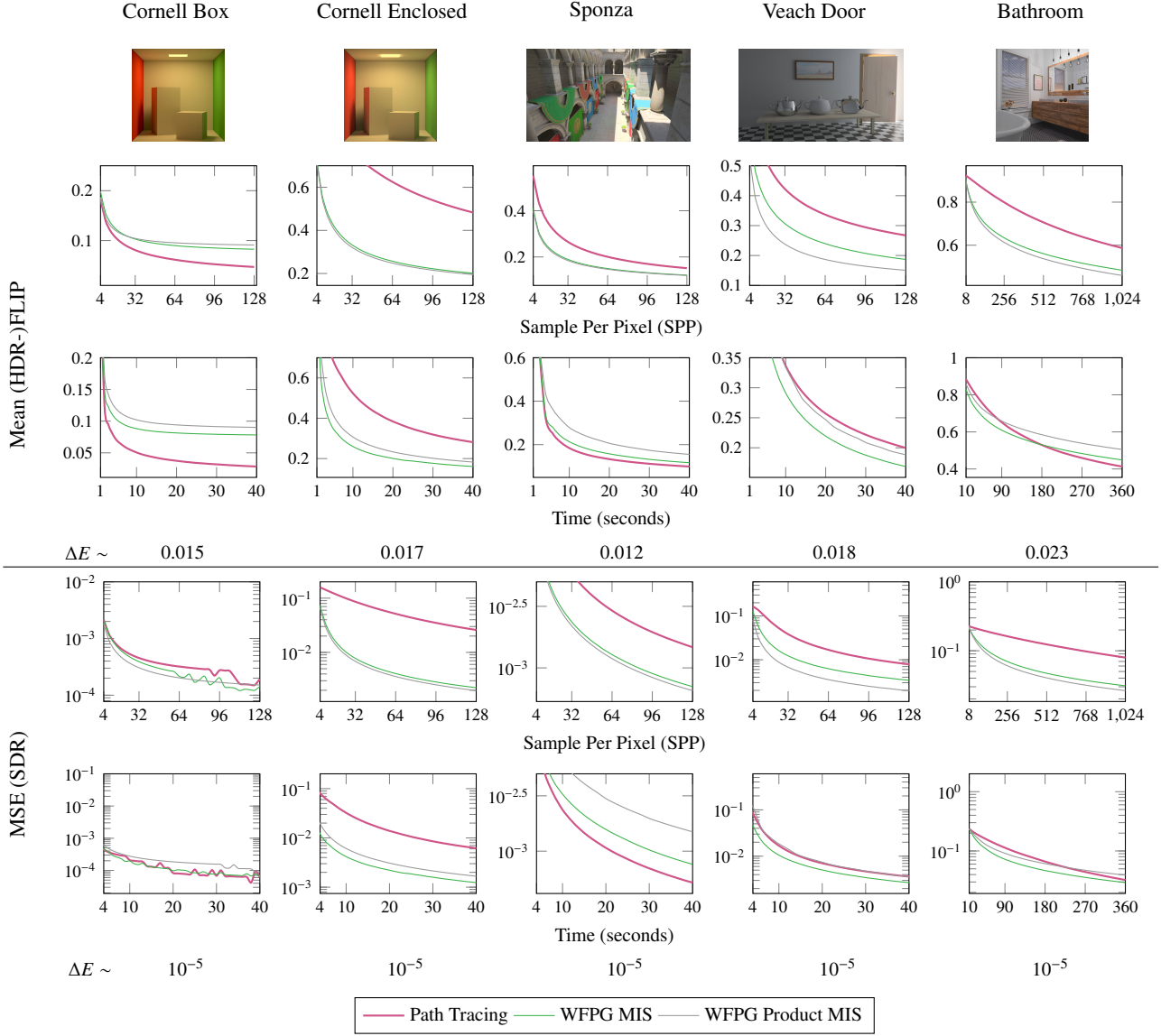
\begin{figure*}[!t]
    \renewcommand{\arraystretch}{0.9}
    \centering
    \small
    \setlength{\tabcolsep}{1.0mm}
    \begin{tabular}{cc}
    
    \multirow{1}{*}[12em]{\rotatebox[origin=l]{90}{Mean (HDR-)FLIP}} &
    
    \begin{tikzpicture}
        % Don't know the tikz coord system so manually setting
        % coords
        \node[anchor=center,inner sep=0] at (1.1,4.0) {Cornell Box};
        \node[anchor=south west,inner sep=0] (image) at (0.45,2.1){\includegraphics[height=.0731\textwidth]{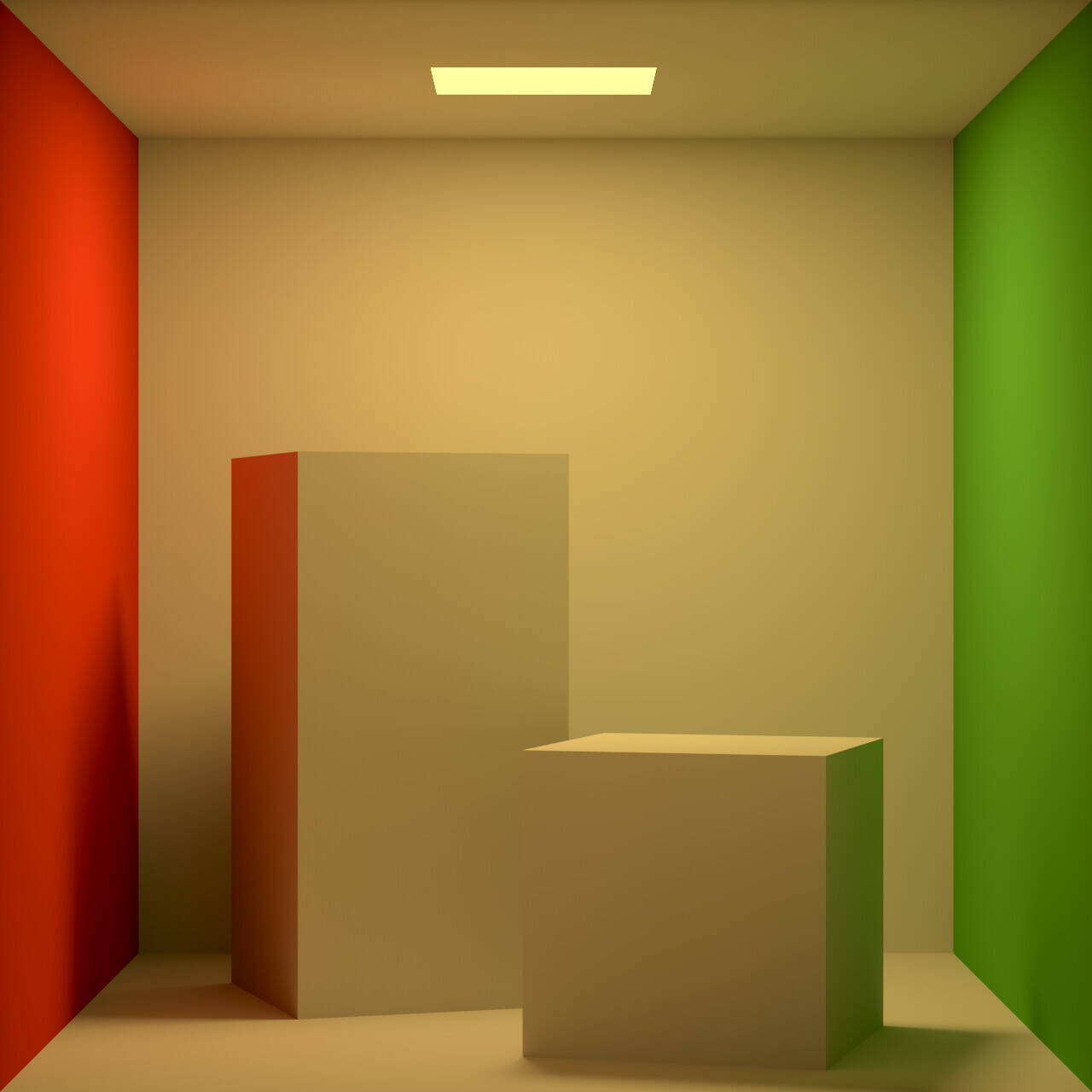}};

        \node[anchor=center,inner sep=0] at (4.25,4.0) {Cornell Enclosed};
        \node[anchor=south west,inner sep=0] (image) at (3.55,2.1){\includegraphics[height=.0731\textwidth]{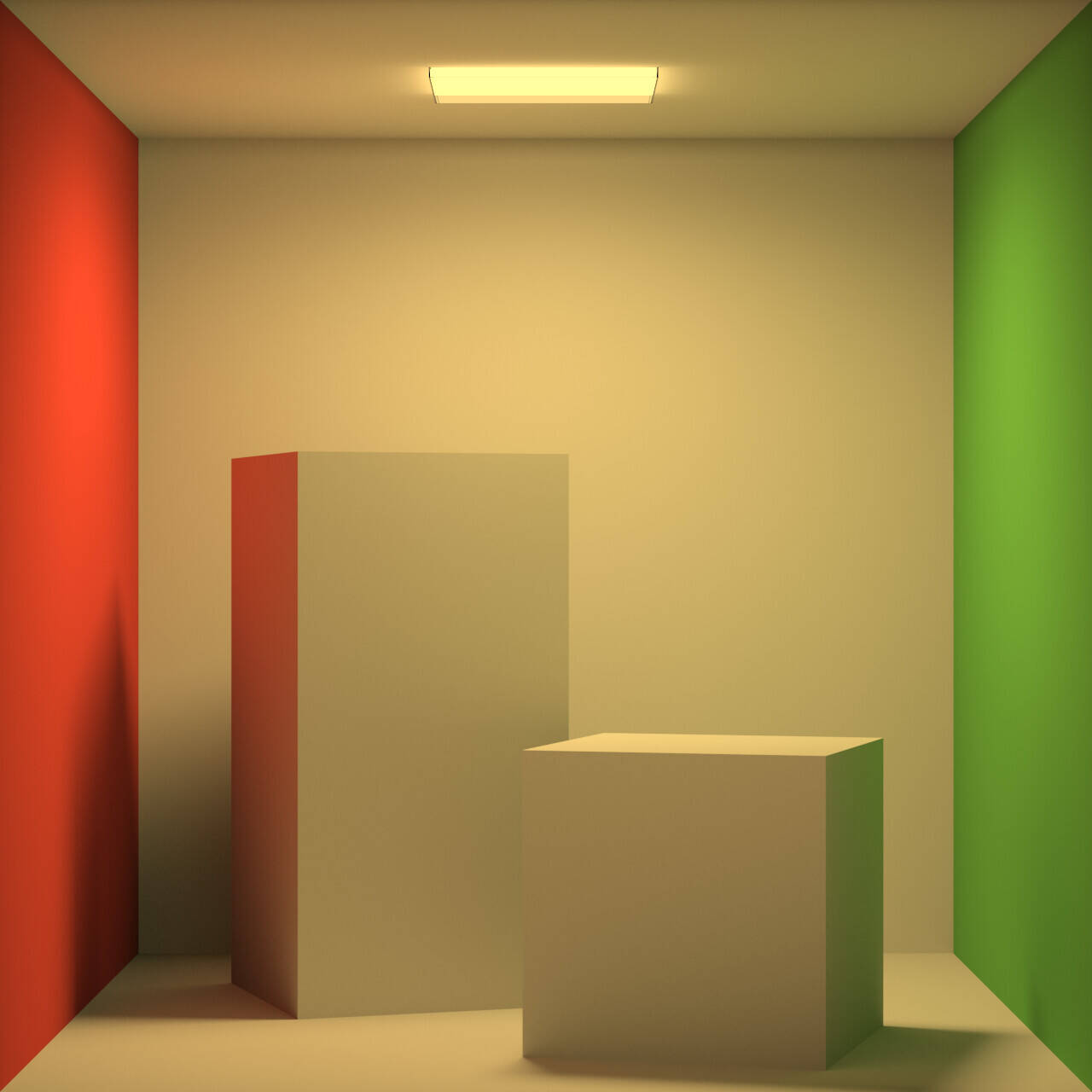}};

        \node[anchor=center,inner sep=0] at (7.3,4.0) {Sponza};
        \node[anchor=south west,inner sep=0] (image) at (6.2,2.1){\includegraphics[width=.13\textwidth]{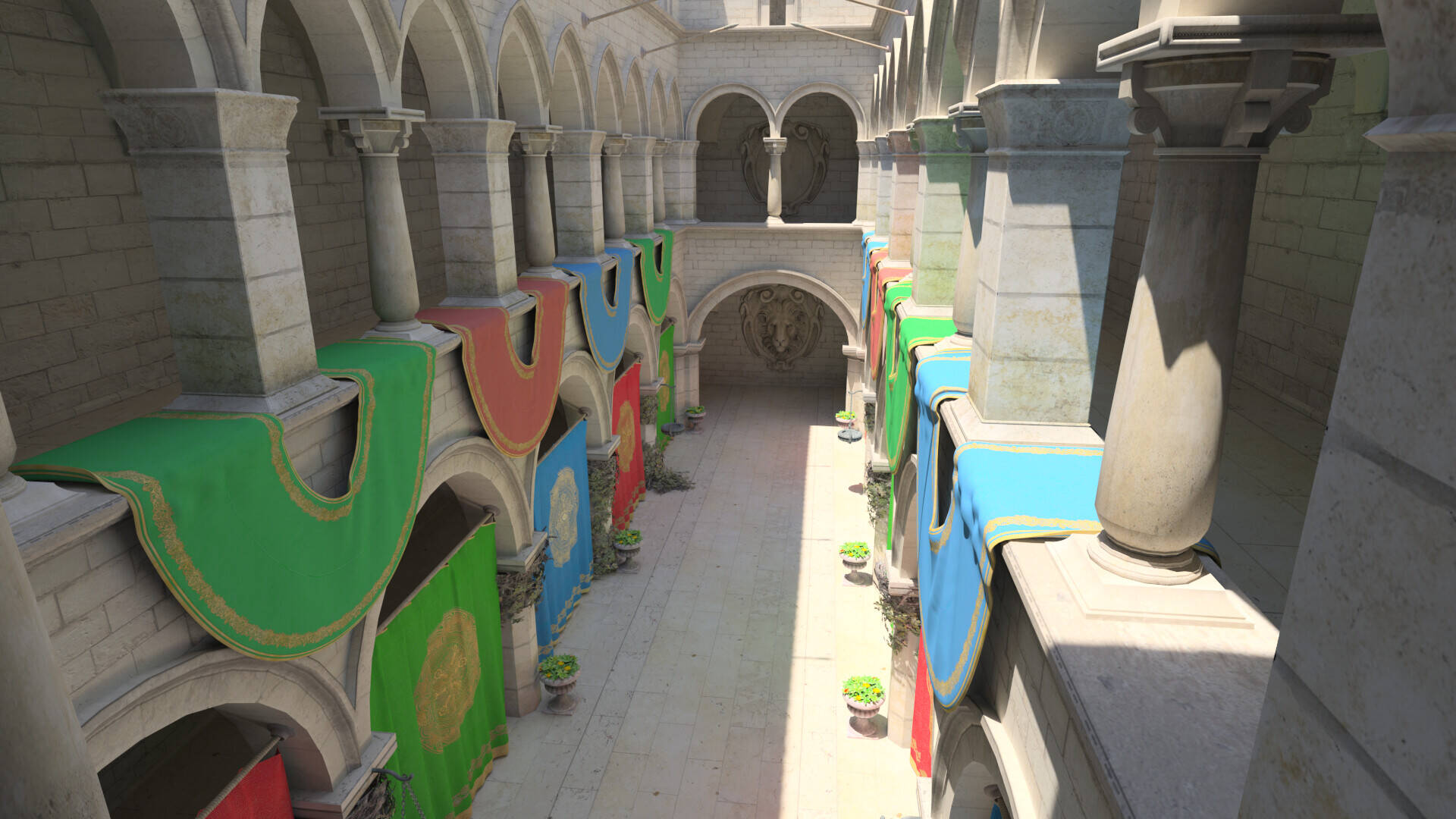}};

        \node[anchor=center,inner sep=0] at (10.4,4.0) {Veach Door};
        \node[anchor=south west,inner sep=0] (image) at (9.3,2.1){\includegraphics[width=.13\textwidth]{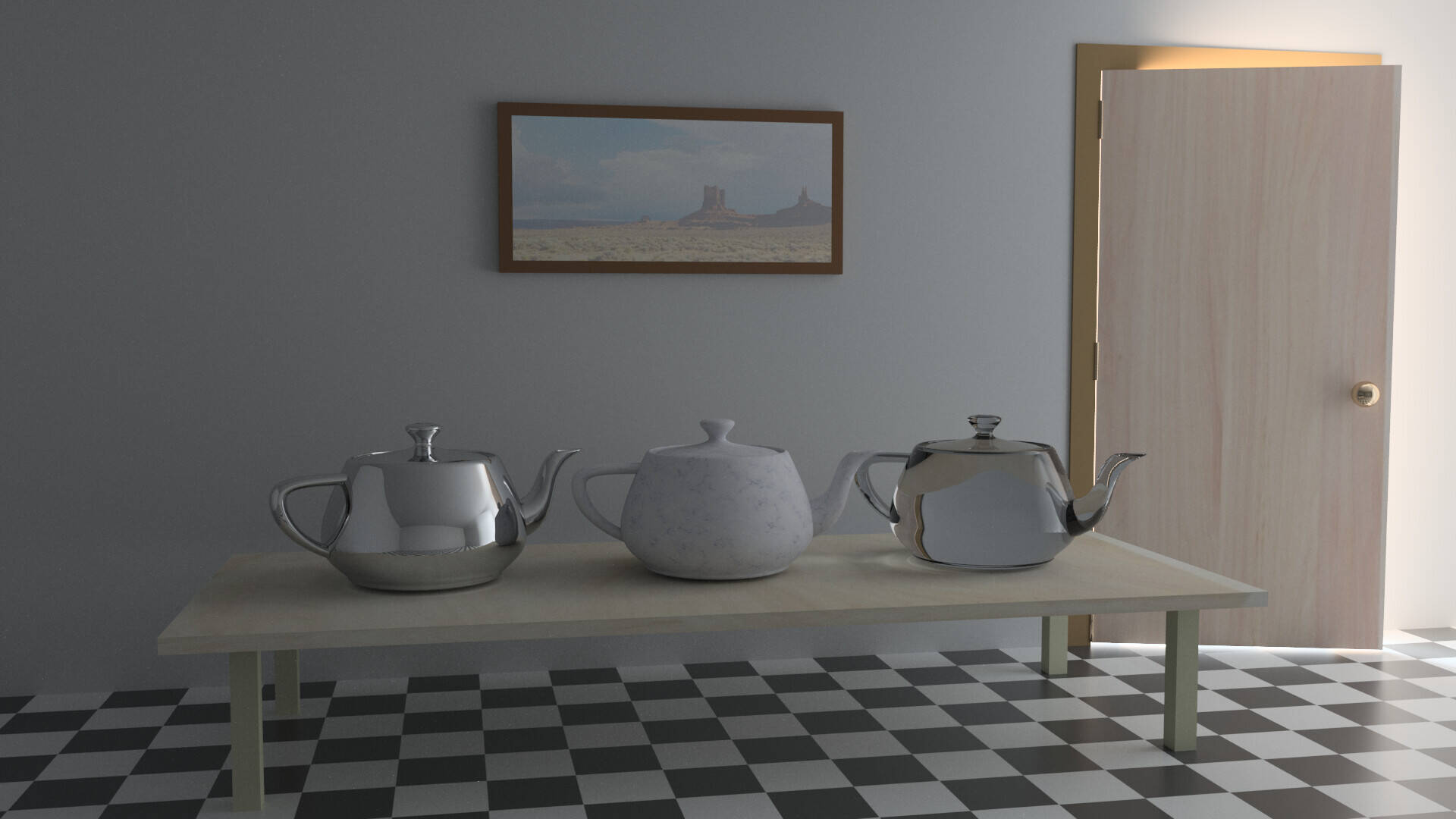}};

        \node[anchor=center,inner sep=0] at (13.7,4.0) {Bathroom};
        \node[anchor=south west,inner sep=0] (image) at (13.05,2.1){\includegraphics[height=.0731\textwidth]{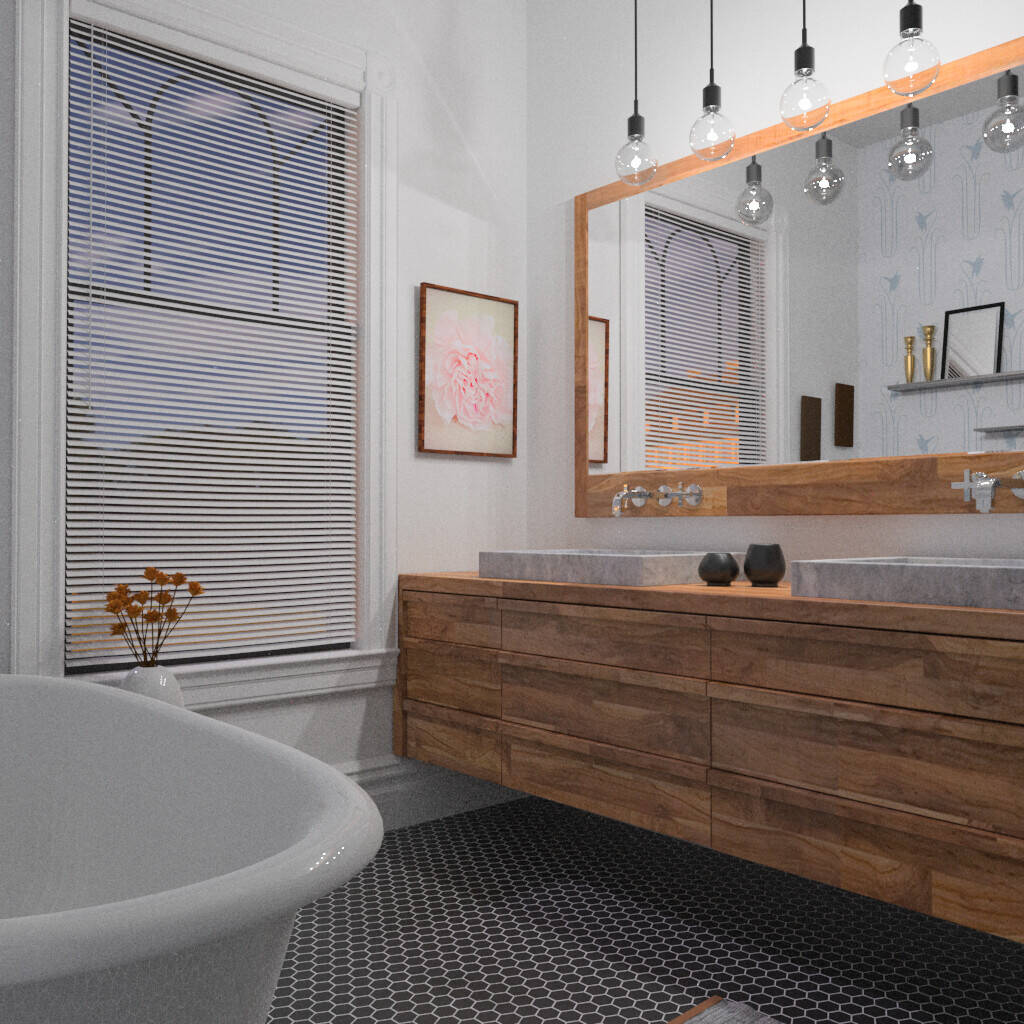}};

        % ========================================
        % FLIP
        % ========================================
        \begin{groupplot}
        [
            group style=
            {
                group size= 5 by 2,
                vertical sep=30pt,
                horizontal sep=25pt
            },
            title style={font=\footnotesize},
            label style={font=\footnotesize},
            legend style=
            {
                at={(4.1,-0.5)},
                anchor=north east,
                font=\footnotesize,
                legend columns=5
            },
            width=\textwidth
        ]

        % ========================================
        % SAMPLE GRAPHS
        % ========================================
        \plotGen{Sample}{4}{32}{130}{}
                {figures/maingraphs/cornellBox-sample-merged.csv}
                {}
                {0.01}{0.25}
    
        \plotGen{Sample}{4}{32}{128}{}
                {figures/maingraphs/cornellEnclosed-sample-merged.csv}
                {}
                {}{0.7}
    
        \plotGen{Sample}{4}{32}{130}{}
                {figures/maingraphs/sponza-sample-merged.csv}
                {Sample Per Pixel (SPP)}
                {}{}
    
        \plotGen{Sample}{4}{32}{128}{}
                {figures/maingraphs/veachDoor-sample-merged.csv}
                {}
                {0.1}{0.5}
    
        \plotGen{Sample}{8}{256}{1024}{}
                {figures/maingraphs/bathroom-sample-merged.csv}
                {}
                {}{}
    
        % ========================================
        % TIME GRAPHS
        % ========================================
        \plotGen{Time}{1}{10}{40}{}
                {figures/maingraphs/cornellBox-time-merged.csv}
                {}
                {}{0.2}
    
        \plotGen{Time}{1}{10}{40}{}
                {figures/maingraphs/cornellEnclosed-time-merged.csv}
                {}
                {}{0.7}
    
         \plotGen{Time}{1}{10}{40}{}
                 {figures/maingraphs/sponza-time-merged.csv}
                 {Time (seconds)}
                 {}{0.6}
    
        \plotGen{Time}{1}{10}{40}{}
                {figures/maingraphs/veachDoor-time-merged.csv}
                {}
                {}{0.35}
    
        \plotGen{Time}{10}{90}{360}{}
                {figures/maingraphs/bathroom-time-extended.csv}
                {}
                {}{1.0}
    
        \end{groupplot}

        %\node[anchor=south,inner sep=0] at (7.5,-4) {\footnotesize \(\Delta E \sim (0.015 - 0.022)\) at 150,000 spp};

        \node[anchor=south,inner sep=0] at (-0.2,-4) {\footnotesize \(\Delta E \sim\)};
        
        \node[anchor=south,inner sep=0] at (1.2,-4) {\footnotesize \(0.015\)};

        \node[anchor=south,inner sep=0] at (4.4,-4) {\footnotesize \(0.017\)};

        \node[anchor=south,inner sep=0] at (7.6,-4) {\footnotesize \(0.012\)};

        \node[anchor=south,inner sep=0] at (10.7,-4) {\footnotesize \(0.018\)};

        \node[anchor=south,inner sep=0] at (13.8,-4) {\footnotesize \(0.023\)};

        \end{tikzpicture}
    \\
    \hline

    \multirow{1}{*}[13.5em]{\rotatebox[origin=l]{90}{MSE (SDR)}} &

    % ========================================
    % MSE
    % ========================================
    \begin{tikzpicture}
        \begin{groupplot}
        [
            group style=
            {
                group name=mseGraph,
                group size= 5 by 2,
                vertical sep=30pt,
                horizontal sep=25pt
            },
            title style={font=\footnotesize},
            label style={font=\footnotesize},
            legend style=
            {
                at={(4.1,-0.5)},
                anchor=north east,
                font=\footnotesize,
                legend columns=5
            },
            width=\textwidth
        ]

        % ========================================
        % SAMPLE GRAPHS
        % ========================================
        %\plotGen{Sample}{4}{32}{128}{}
        \logPlotGen{Sample}{4}{32}{128}{}
                {figures/maingraphs/cornellBox-mse-sample-merged.csv}
                {}
                {}{0.01}

        %\plotGen{Sample}{1}{10}{40}{}
        \logPlotGen{Sample}{4}{32}{128}{}
                {figures/maingraphs/cornellEnclosed-mse-sample-merged.csv}
                {}
                {}{0.2}
    
         %\plotGen{Sample}{4}{10}{40}{}
         \logPlotGen{Sample}{4}{32}{128}{}
                 {figures/maingraphs/sponza-mse-sample-merged.csv}
                 {Sample Per Pixel (SPP)}
                 {}{0.005}
    
        %\plotGen{Sample}{4}{10}{40}{}
        \logPlotGen{Sample}{4}{32}{128}{}
                {figures/maingraphs/veachDoor-mse-sample-merged.csv}
                {}
                {}{0.6}

        %\plotGen{Time}{10}{90}{360}{}
        \logPlotGen{Sample}{8}{256}{1024}{}
                {figures/maingraphs/bathroom-mse-sample-merged.csv}
                {}
                {}{1}

        % ========================================
        % TIME GRAPHS
        % ========================================
        %\plotGen{Time}{1}{10}{40}{}
        \logPlotGen{Time}{4}{10}{40}{}
                {figures/maingraphs/cornellBox-mse-time-merged.csv}
                {}
                {}{0.1}

        %\plotGen{Time}{1}{10}{40}{}
        \logPlotGen{Time}{4}{10}{40}{}
                {figures/maingraphs/cornellEnclosed-mse-time-merged.csv}
                {}
                {}{0.2}
    
         %\plotGen{Time}{4}{10}{40}{}
         \logPlotGen{Time}{4}{10}{40}{}
                 {figures/maingraphs/sponza-mse-time-merged.csv}
                 {Time (seconds)}
                 {}{0.005}
    
        %\plotGen{Time}{4}{10}{40}{}
        \logPlotGen{Time}{4}{10}{40}{}
                {figures/maingraphs/veachDoor-mse-time-merged.csv}
                {}
                {}{0.6}

        %\plotGen{Time}{10}{90}{360}{}
        \logPlotGen{Time}{10}{90}{360}{}
                {figures/maingraphs/bathroom-mse-time-extended.csv}
                {}
                {}{2.0}
    
        \end{groupplot}

        \node[anchor=south,inner sep=0] at (-0.2,-4) {\footnotesize \(\Delta E \sim\)};
        
        \node[anchor=south,inner sep=0] at (1.2,-4) {\footnotesize \(10^{-5}\)};

        \node[anchor=south,inner sep=0] at (4.4,-4) {\footnotesize \(10^{-5}\)};

        \node[anchor=south,inner sep=0] at (7.6,-4) {\footnotesize \(10^{-5}\)};

        \node[anchor=south,inner sep=0] at (10.7,-4) {\footnotesize \(10^{-5}\)};

        \node[anchor=south,inner sep=0] at (13.8,-4) {\footnotesize \(10^{-5}\)};

        \node[below = 1.4cm of mseGraph c3r2.south]
        {
            \pgfplotslegendfromname{figures/maingraphs/sponza-mse-time-merged.csv}
        };
    
    \end{tikzpicture}
    
    \end{tabular}

    \caption{Equal sample (top row) and equal time (bottom row) comparisons between the proposed method and path tracing. The results of the proposed method are shown with product path guiding both activated and deactivated. WFPG Parameters for ``Sponza Lion'', ``Veach Door'', and ``Bathroom'' scenes are the same as in Table \ref{tab:profiling}. Due to scene simplicity, Cornell Box parameters are \(l_{min} = 4\), \(c_{ray} = 512 \), and SVO \(=256^3\). Overall, parameters are selected to generate around \(2000-3000\) bins per depth iteration consistently. \( \Delta E \) represents the mean difference of our methods from the ground-truth for each scene rendered with \(150000\) samples to demonstrate unbiasedness.}
    \label{fig:mainGraphs}
\end{figure*}

In the top row of Figure~\ref{fig:mainGraphs}, we compare our algorithm's path-guiding and product path-guiding versions with each other and against path tracing under an equal sample scenario for five test scenes. Two error metrics are used. For the top two rows, HDR-FLIP~\cite{Andersson:2021:HDRFLIP}, which is a perceptual metric, is used, and for the bottom two rows, the Mean Square Error (MSE), which is a numeric metric, is used. It is worth noting that we applied the MSE metric on tone-mapped~\cite{reinhard2023photographic} frames as otherwise critical but minor errors in dark pixels would be subjugated by relatively less important but higher magnitude errors in lighter pixels.

In these results, the ``Cornell Box'' scene is a reproduction of the original scene, while the ``Cornell Enclosed'' is the same scene except that a glass enclosure surrounds the light source. This effectively disables NEE, as no shadow ray can directly reach the light source. The comparison of these two scenes shows that our method performs noticeably better under indirect illumination. This is expected as when NEE connects a scene point to the light source, it dominates the shading of the point, minimizing the impact of path guiding. As for the other three scenes, both path guiding and product path guiding variants of our method perform better than path tracing. 

As our method requires extra computations that increase the run time, we evaluate its performance under an equal time setting. This is shown at the bottom row of Figure~\ref{fig:mainGraphs}. Except for the ``Cornell Box'', ``Sponza'', and ``Bathroom'' scenes for which direct lighting is dominant, our method outperforms the pure path tracing approach.

An interesting behavior can be seen in the ``Bathroom'' scene. Here, our path-guiding methods outperform regular path tracing under the equal sample scenario; however, in the equal-time setting, path tracing eventually outperforms both proposed methods. One explanation for this is that the dominant light sources (the filaments of the light bulbs) are very small; therefore, even if the rays are guided toward the voxels that contain the light sources, they may not reach the filaments. This also depends on the resolution of the voxels and the radiance field. Secondly, because the mirror is perfectly specular, path guiding using the radiant exitance information produces sub-optimal results. These combined constraints and the additional overhead of our guiding schemes result in this slight underperformance of our approach in the long run for this scene.

\begin{table}[!ht]
    \renewcommand{\arraystretch}{1.3}
    \setlength{\tabcolsep}{1.0mm}
    \small
    \centering
    \caption{Memory requirements of different methods. WFPG method parameters are \(l_{min} = 5\), \(c_{ray} = 512 \) and SVO \(=128^3\). All other methods use their default parameters. The iterative training SPP value of Ruppert et al.'s method is 4. For ``Sponza'' and ``Veach Door'' scenes, the resolution is $1920\times1080$. For the Bathroom scene, the resolution is $1280\times1280$.}
    \label{tab:memory}
    
    \begin{tabular}{c c c cc c c }
        \toprule
        &&& \multicolumn{3}{c}{\textbf{Scene-related Memory (MiB)}} 
        \\ 
        & Depth & Ours && Training 
        & Müller et al.
        & Ruppert et al.
        \\ \cmidrule{2-7}
        \multirow{3}{*}[-2mm]{\rotatebox[origin=l]{90}{\textsc{Sponza}}}
        & \multirow{3}{*}[-1mm]{4}
        & \cellW
        && 16 t
        & \cellW 19.0
        & \cellW 3.2
        \\
        && \cellW 3.96
        && 32 t
        & \cellG 26.0
        & \cellG 5.5
        \\
        && \cellW 
        && 64 t
        & \cellW 37.0
        & \cellW 20.1
        \\ \cmidrule{3-7}
        \multirow{3}{*}[-1mm]{\rotatebox[origin=l]{90}{\textsc{VeachDoor}}}
        & \multirow{3}{*}[-1mm]{6}        
        & \cellG 
        && 32 t
        & \cellG 40.3
        & \cellG 7.4
        \\
        && \cellG 1.26
        && 64 t
        & \cellW 56.3
        & \cellW 14.8        
        \\
        && \cellG 
        && 128 t
        & \cellG 77.0
        & \cellG 30.0        
        \\ \cmidrule{3-7}
        \multirow{3}{*}[-1mm]{\rotatebox[origin=l]{90}{\textsc{Bathroom}}}
        & \multirow{3}{*}[-1mm]{10}        
        & \cellW 
        && 128 t
        & \cellW 83.4
        & \cellW 15.4
        \\
        && \cellW 2.00
        && 256 t
        & \cellG 121.32
        & \cellG 31.1
        \\
        && \cellW
        && 512 t
        & \cellW 167.03
        & \cellW 63.9        
        %\\ \cmidrule{3-6}
        \\ 
        
        %\midrule
        %\multicolumn{6}{l}
        %{
        %    \begin{tabular}{cl}
        %        (1) & Number of training samples, if applicable.  
        %        \\
        %        *   & Potential massively parallel implementation.
        %    \end{tabular}
        %} \\
        \bottomrule
    \end{tabular}
\end{table}

\begin{figure*}%[!ht]
\centering
\setlength{\tabcolsep}{0.3mm} % Default value: 6pt
\footnotesize

\begin{tabular}{ccc}

    \begin{tabular}{c}
        \begin{tikzpicture}
        \node[anchor=south west,inner sep=0] (convRef) at (0,0){\includegraphics[width=0.2\textwidth]{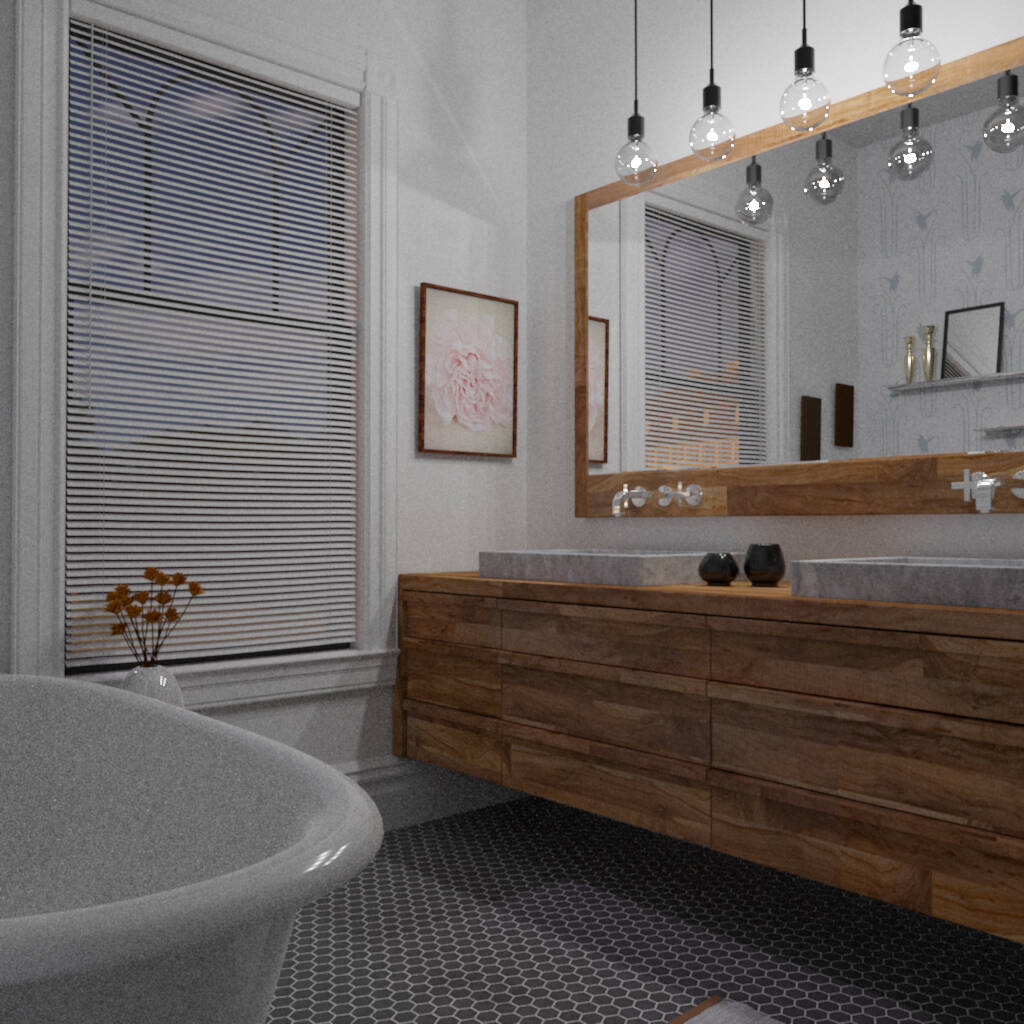}};
        \begin{scope}[x={(convRef.south east)},y={(convRef.north west)}]
            \fill[red] (0.45, 0.63) ellipse (0.0106 and 0.0106);
        \end{scope}
        \end{tikzpicture}  
        \\
        Camera Reference
    \end{tabular}    
    & 
    \begin{tabular}{ccccc}
        \multirow{1}{*}[2.5em]{\mline{{\footnotesize PPG}}} &
        \includegraphics[width=.07\textwidth]{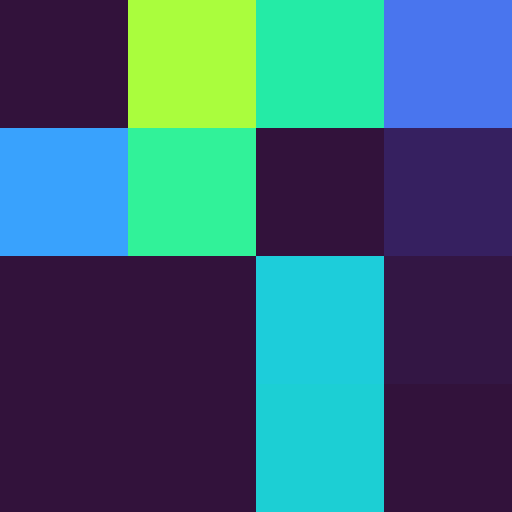} &
        \includegraphics[width=.07\textwidth]{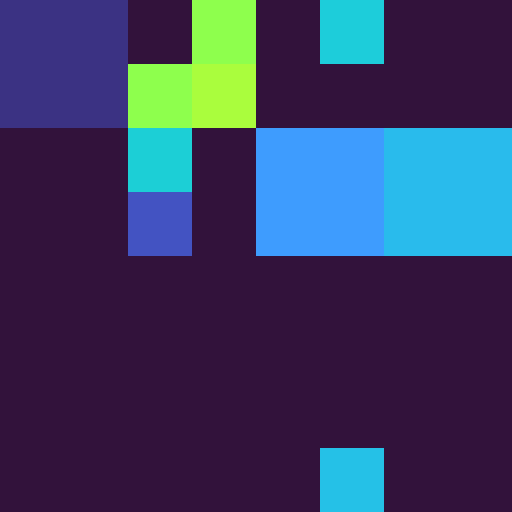} &
        \includegraphics[width=.07\textwidth]{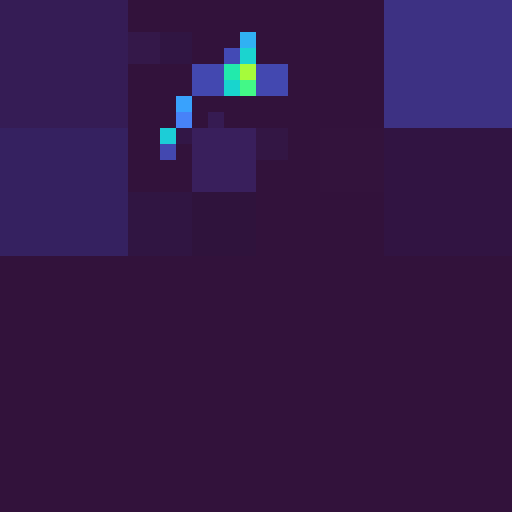} &
        \includegraphics[width=.07\textwidth]{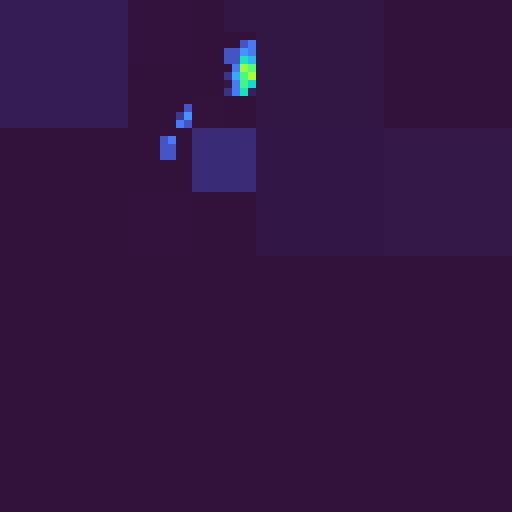}
        \\

        \multirow{1}{*}[3em]{\mline{{\footnotesize WFPG $64\times64$}}} &
        \includegraphics[width=.07\textwidth]{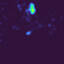} &
        \includegraphics[width=.07\textwidth]{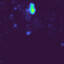} &
        \includegraphics[width=.07\textwidth]{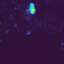} &
        \includegraphics[width=.07\textwidth]{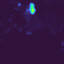}
        \\

        \multirow{1}{*}[3em]{\mline{{\footnotesize WFPG $128\times128$}}} &
        \includegraphics[width=.07\textwidth]{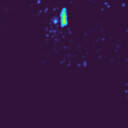} &
        \includegraphics[width=.07\textwidth]{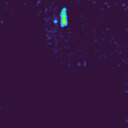} &
        \includegraphics[width=.07\textwidth]{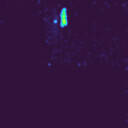} &
        \includegraphics[width=.07\textwidth]{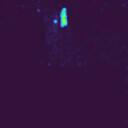}
        \\
        & 2 & 4 & 8 & 16
    \end{tabular}
    &
    \begin{tabular}{c}
         \includegraphics[width=0.2\textwidth]{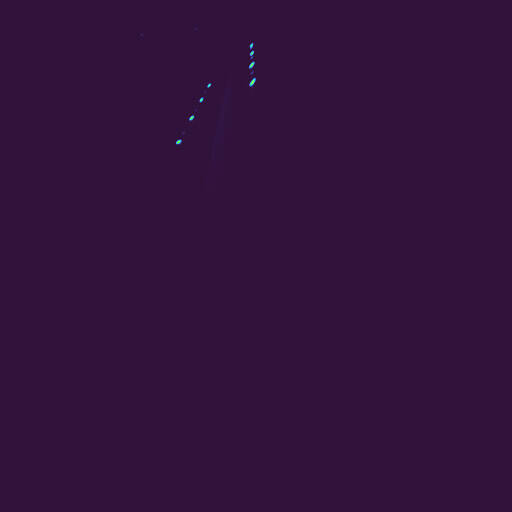} \\
         Reference PDF
    \end{tabular}

\end{tabular}

\caption{Convergence of Müller et al.'s method and our method. The reference PDF is generated using path tracing over the red region. For our method, \(64 \times 64\) and \(128 \times 128\) radiance fields are generated. Müller et al.'s method uses default parameters.}

\label{fig:training}
\end{figure*}

\subsection{Memory Utilization Comparison}
\label{sec:results:memory}

Table~\ref{tab:memory} shows the scene-related memory requirements of our method and other state-of-the-art path-guiding methods. In our case, the scene-related memory is the SVO memory. For Müller et al.~\cite{Muller:PPG:2017}, it is the memory consumed by sd-trees. Finally, for Ruppert et al.~\cite{Ruppert:2020:ParalaxPG}, it is the total memory of vMF coefficients and the kd-tree. As can be seen from the table, the scene-related memory cost of other methods increases together with the training time due to the refinement of the data structures. On the other hand, our method's memory requirement is not only smaller but also does not increase over time.

In addition to the scene-related memory cost, our method also requires path memory. In complexity notation, it can be represented as $\Theta(p\times d)$ where $p$ is the path count, and $d$ is the maximum traversal depth of the paths. Both of the other approaches are CPU-based path-guiding algorithms. However, we can argue that their potential wavefront-style implementation on the GPU would also require the same amount of path memory as ours.

\begin{figure}
    \renewcommand{\arraystretch}{1.0}
    \setlength{\tabcolsep}{0.4mm} % Default value: 6pt
    \centering
    \small
    \begin{tabular}{rccccc}
    & \(16^3\) & \(32^3\) & \(64^3\) & \(128^3\) \\    
    & 
    \includegraphics[width=.22\columnwidth]{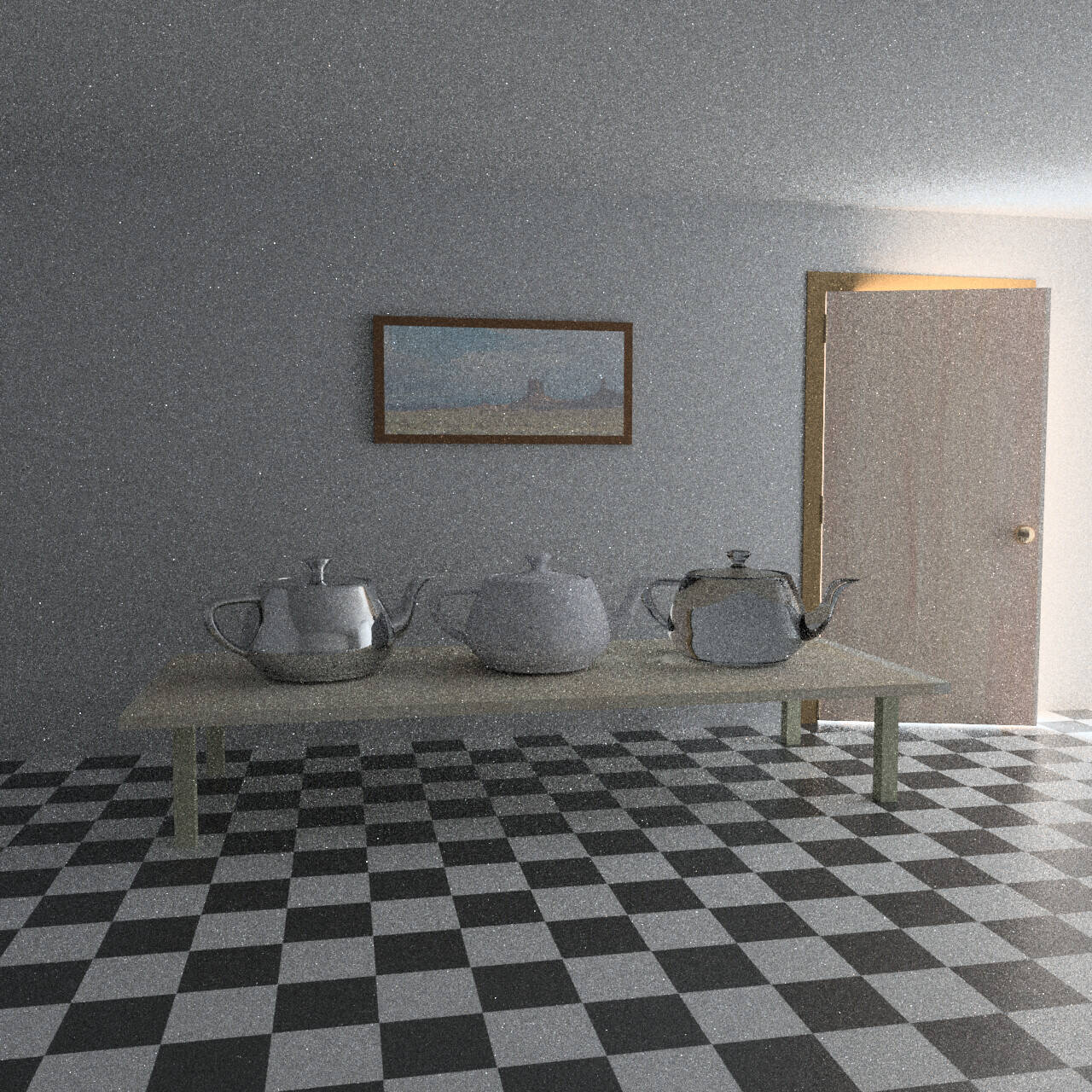} &
    \includegraphics[width=.22\columnwidth]{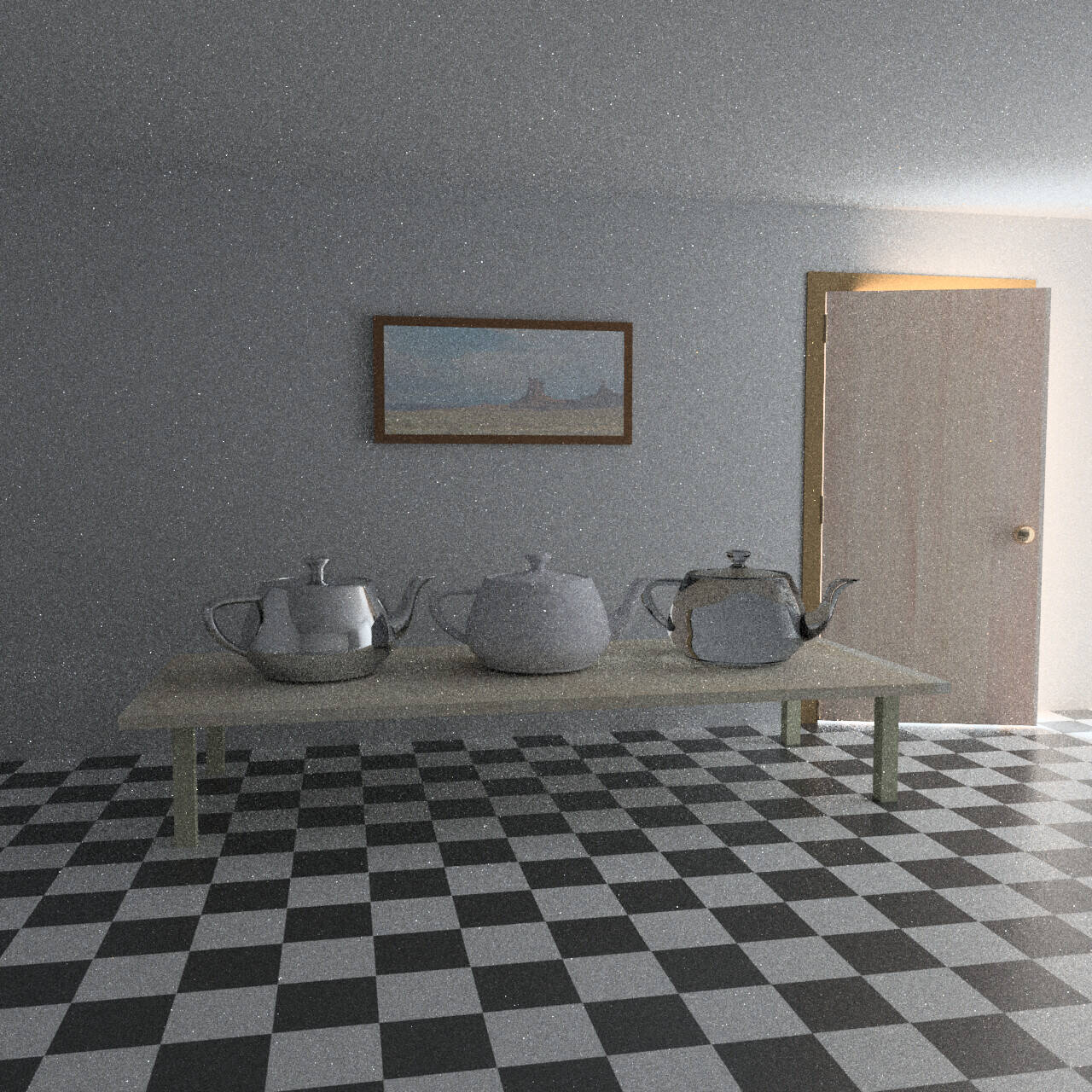} &
    \includegraphics[width=.22\columnwidth]{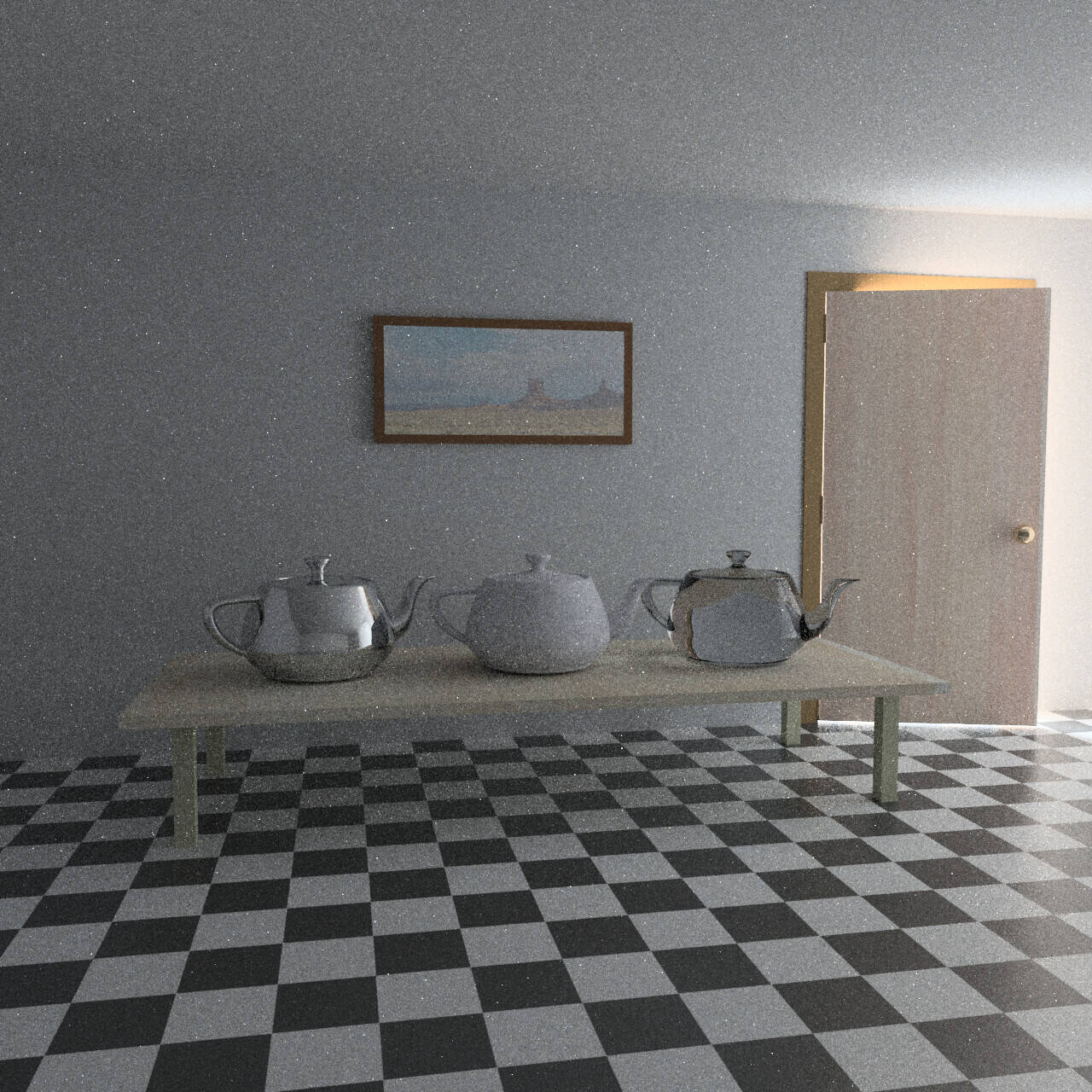} &
    \includegraphics[width=.22\columnwidth]{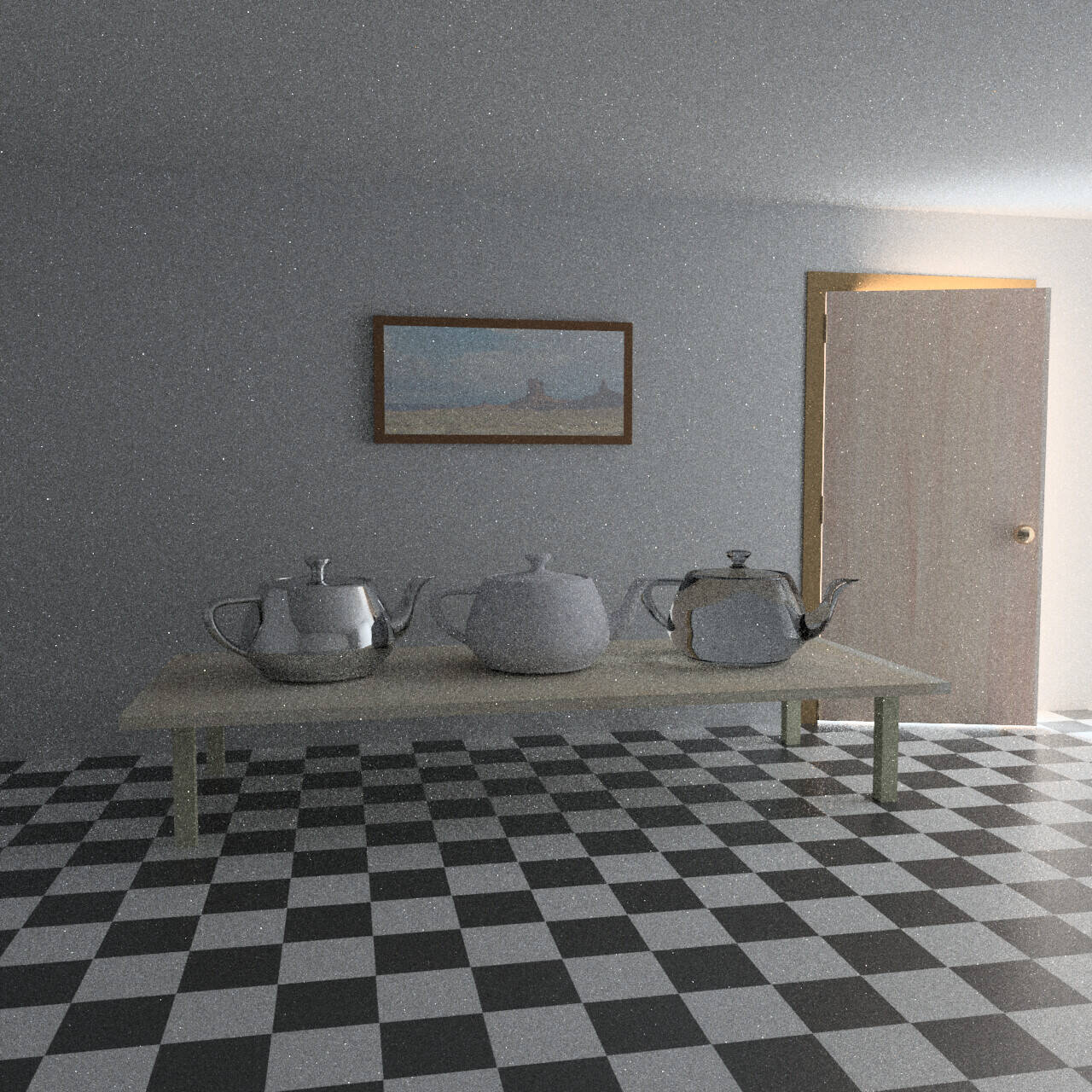} \\
    (1) & 0.179 & 0.154 & 0.144 & 0.144 \\
    (2) & \(3.19 \cdot 10^{-3}\) & \(2.23 \cdot 10^{-3}\)
        & \(1.91 \cdot 10^{-3}\) & \(1.92 \cdot 10^{-3}\) \\
    (3) & 0.02 & 0.07 & 0.31 & 1.26 \\
        
    \end{tabular}
    \caption{Performance change with respect to SVO resolution. The row values are as follows: (1) Mean FLIP. 
    (2) Mean Square Error. (3) SVO Memory MiB. }
    \label{fig:mem:ablation}
\end{figure}

Given that the SVO memory is the only path guiding related data structure that we hold in persistent memory, we analyze the effect of different SVO resolutions on render quality. This is shown in Figure~\ref{fig:mem:ablation} for resolutions from $16^3$ to $128^3$. As can be seen from this figure, the render quality increases up to a point but remains intact afterward. The memory usage of the SVO also increases proportionally. In our analysis, we found the $128^3$ SVO resolution to work well for all of our test scenes.

\subsection{Radiance Field Validation}

We compare the reference and generated radiance fields over specific regions in different scenes to validate the generated radiance field. Reference radiance fields are generated via path tracing. Two such comparisons can be seen in Figures~\ref{fig:pgRef} and \ref{fig:training}. The latter figure exposes the advantage of our method compared to Müller et al.'s method. Our method does not require adapting its directional data structure. The directional data is dense and more or less immediately captures the radiance field; further refinement reduces residual noise. 

\subsection{Product path guiding}
\label{sec:results:productPG}

In our experiments, low-resolution (\(8 \times 8\)) product path guiding mostly eliminates sidedness problems that occur when rays are partitioned around a thin reflective surface with dramatic radiance differences between their sides. Due to omnidirectional generation, rays may probabilistically select the other side of the thin object. The ``Veach Door'' scene is an excellent example, as the illumination in this scene comes from a bright light source in the back room. As such, regular path guiding may steer more rays toward this direction. With product path guiding, however, if this direction cannot illuminate a surface due to its normal facing away from it, rays will not be guided toward it. In Figure~\ref{fig:mainGraphs}, the scene that most benefits from product path guiding is ``Veach Door'' due to this characteristic of the scene. Despite this, in equal time comparison, the regular path guiding appears to be still better due to the extra computations involved in product path guiding.

\begin{figure}%[!ht]
    \setlength{\tabcolsep}{1.5mm}
    \small
    \centering
    \includegraphics[width=0.85\columnwidth]{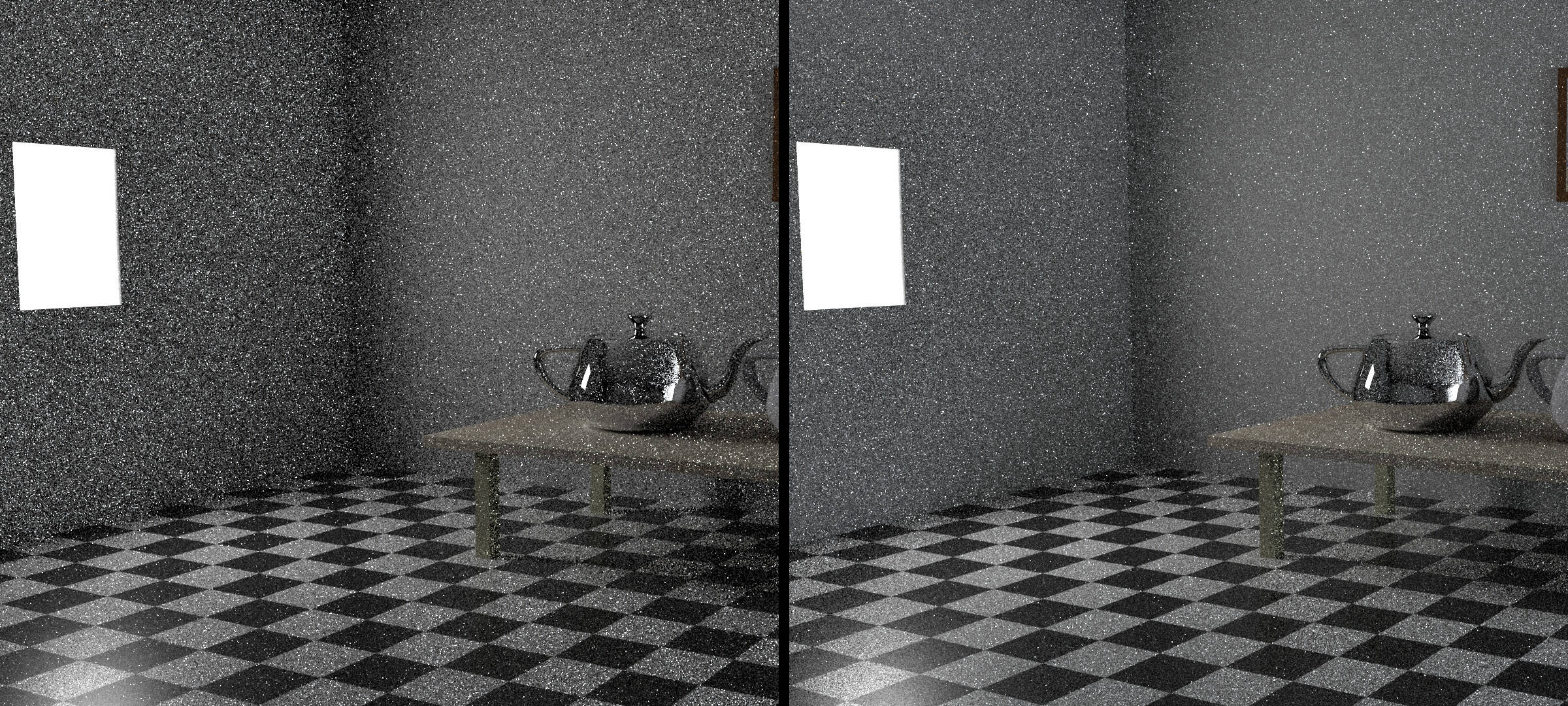}
    
    %\begin{tabularx}{0.75\columnwidth}{>{\raggedright}X>{\raggedleft}X}
    \begin{tabularx}{0.85\columnwidth}    
    {   
        >{\centering\arraybackslash}X 
        >{\centering\arraybackslash}X
    }
        (a) Non-product & (b) Product \\
    \end{tabularx}
    \\
    \begin{tabularx}{0.85\columnwidth}
    {   
        >{\raggedright\arraybackslash}X 
        r
        r
        %>{\raggedleft\arraybackslash}X
        %>{\raggedleft\arraybackslash}X
    }     
        \toprule
        % Header
        & \textbf{(a)} & \textbf{(b)} \\
        \midrule
        Runtime (60s) &  404 spp & 126 spp        \\
        Mean FLIP     & 0.467    & \textbf{0.310} \\
        MSE           & 0.027    & \textbf{0.009} \\

        \bottomrule
    \end{tabularx}

    \caption{An equal-time comparison between the non-product (a) and product (b) version of the proposed method showcasing a scenario where product path guiding outperforms regular path guiding in equal time.}
    \label{fig:productBetterCase}
\end{figure}

However, if we slightly modify this scene by placing the room in an omnidirectional environment map where the majority of the illumination comes from a window on the wall, product path guiding may outperform regular path guiding even under an equal-time setting. This is illustrated in Figure~\ref{fig:productBetterCase}. It can be seen that despite fewer rays being traced for product path guiding, each ray ``counts'' more, yielding a final image with reduced noise. Product path guiding can be helpful in these scenarios where thin walls separate an intense illumination between two regions.

\subsection{Comparison with the Literature}
\label{sec:results:stateoftheart}

We conducted an equal sample comparison between Müller et al.'s~\cite{Muller:PPG:2017}, Ruppert et al.'s~\cite{Ruppert:2020:ParalaxPG}, and our methods. We refrain from conducting equal-time comparisons due to underlying architectural differences. Moreover, to prevent renderer-based differences from altering the results, each technique is compared against the reference image of that renderer.

\begin{table*}[!t]
    \small
    \renewcommand{\arraystretch}{0.8}
    \setlength{\tabcolsep}{0.5mm} % Default value: 6pt
    \caption{The comparison between the two state-of-the-art and our path-guiding algorithms is shown. We used the default parameters of the literature methods except for the sample counts. We set our sample count (1536) to the sum of the training and rendering sample counts used for each method. Our parameters were \(l_{min} = 5\), \(c_{ray}=512\), and the SVO resolution equal to \(128^3\). The maximum path depth of the scenes were 10 for ``Bathroom'', 4 for ``Sponza'', and 6 for ``Veach Door''.}
    \label{tab:vmmComparison}
    \centering
    \begin{tabular}{ccc|ccc|cc}

    & \multicolumn{2}{c}{Reference} & PT & WFPG & WFPG Product & Ruppert et al. & Müller et al. \\
    &&& \multicolumn{3}{c|}{1536spp} & \multicolumn{2}{c}{512t + 1024spp} \\

    %=======================================%
    % Bathroom Scene
    %=======================================%
    \multirow{3}{*}[1em]{\rotatebox[origin=l]{90}{\textsc{\normalsize Bathroom}}} 
    &
    \multirow{1}{*}[3.3em]{\rotatebox[origin=l]{90}{\textsc{Ours}}}
    &
    \includegraphics[width=0.125\textwidth]{figures/comparisons/Bathroom/mray-bathroom-ref}
    &
    \includegraphics[width=0.125\textwidth]{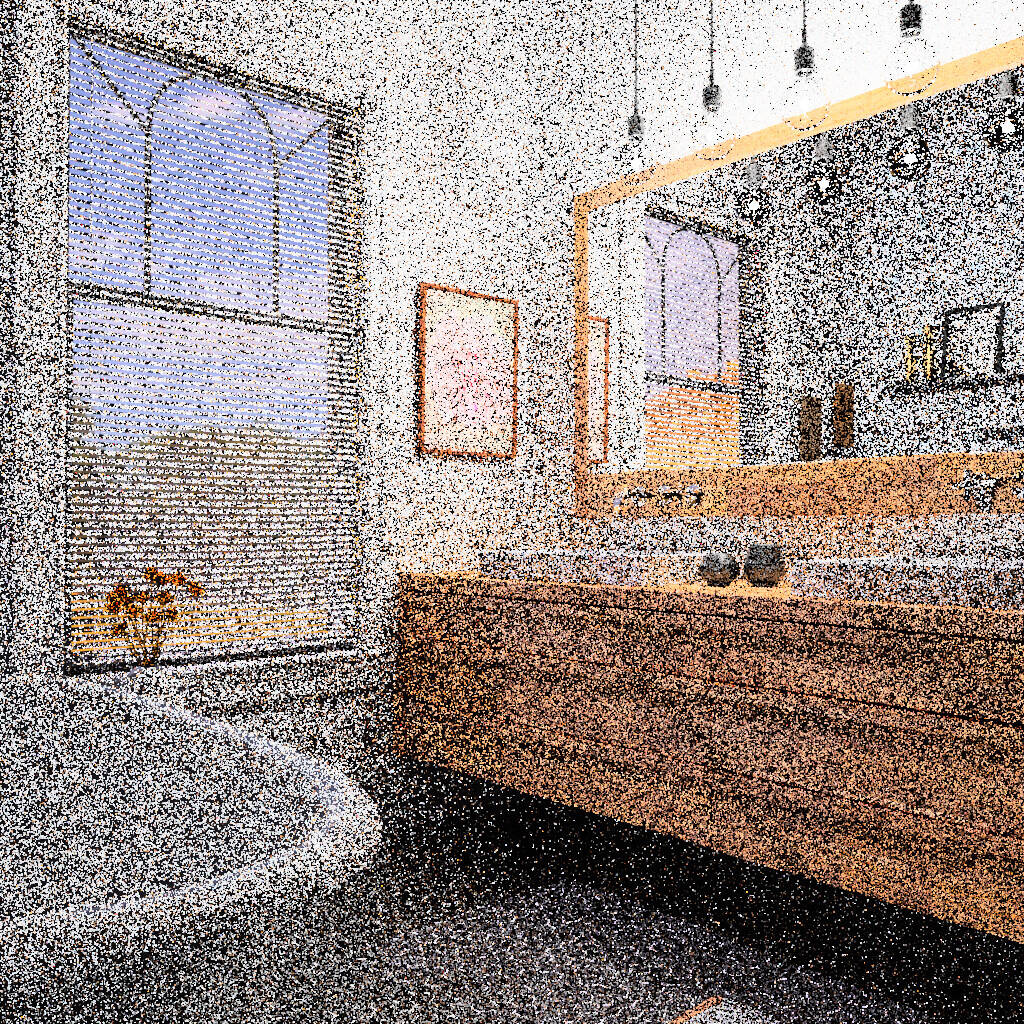}
    &
    \includegraphics[width=0.125\textwidth]{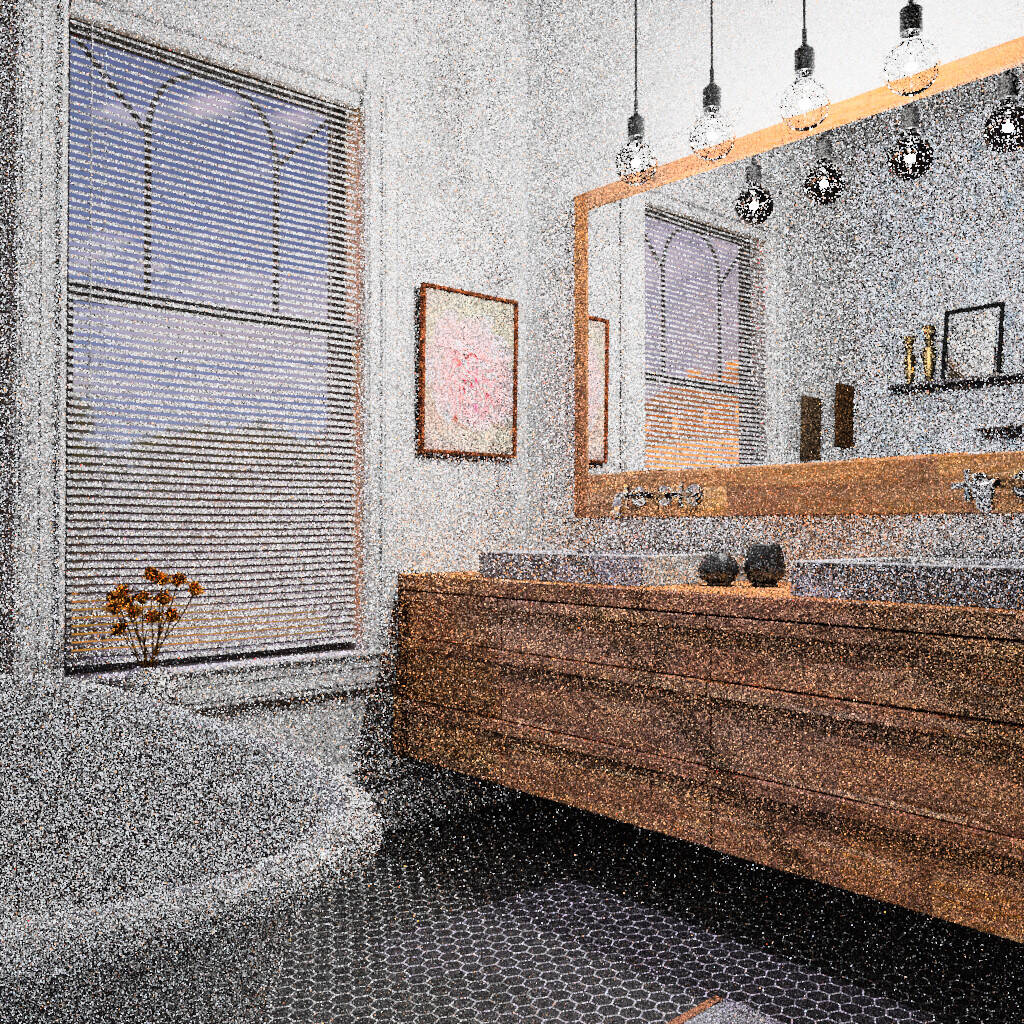}
    &
    \includegraphics[width=0.125\textwidth]{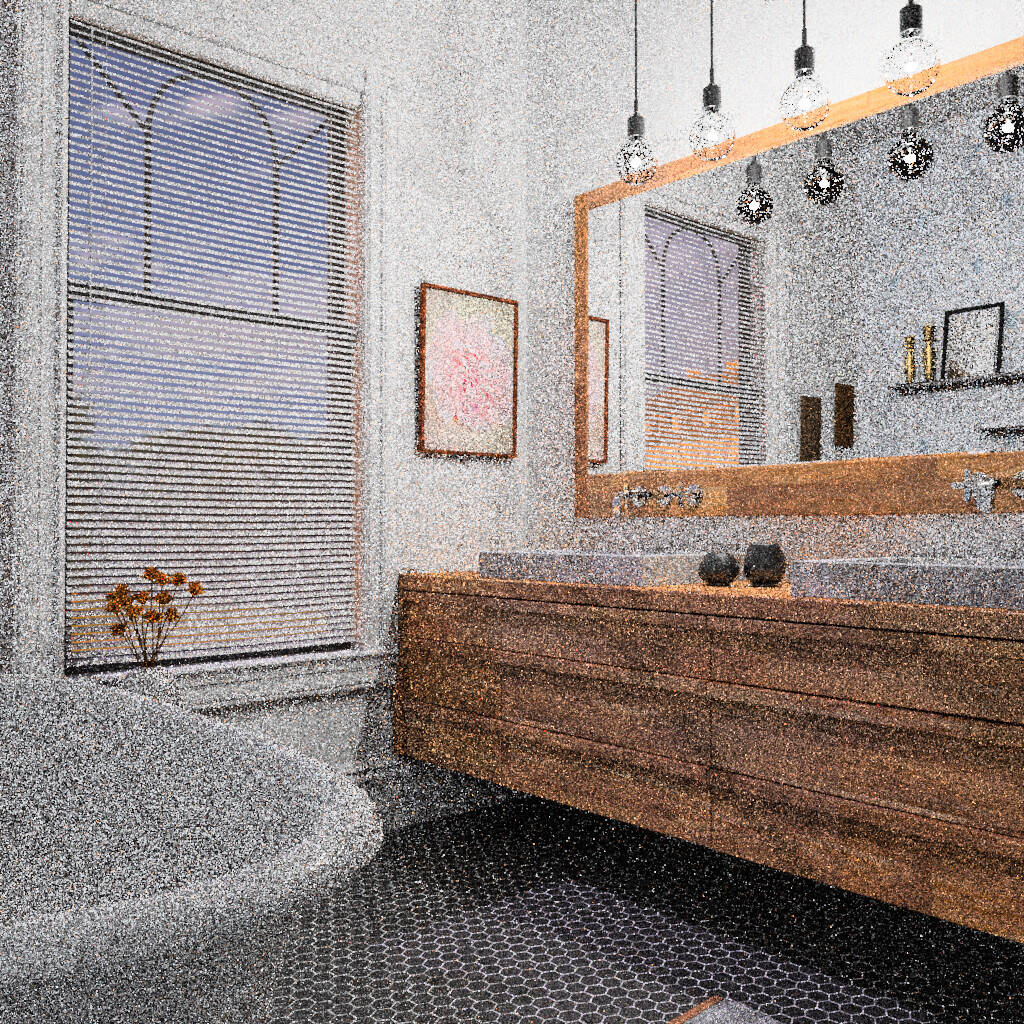}
    &
    \includegraphics[width=0.125\textwidth]{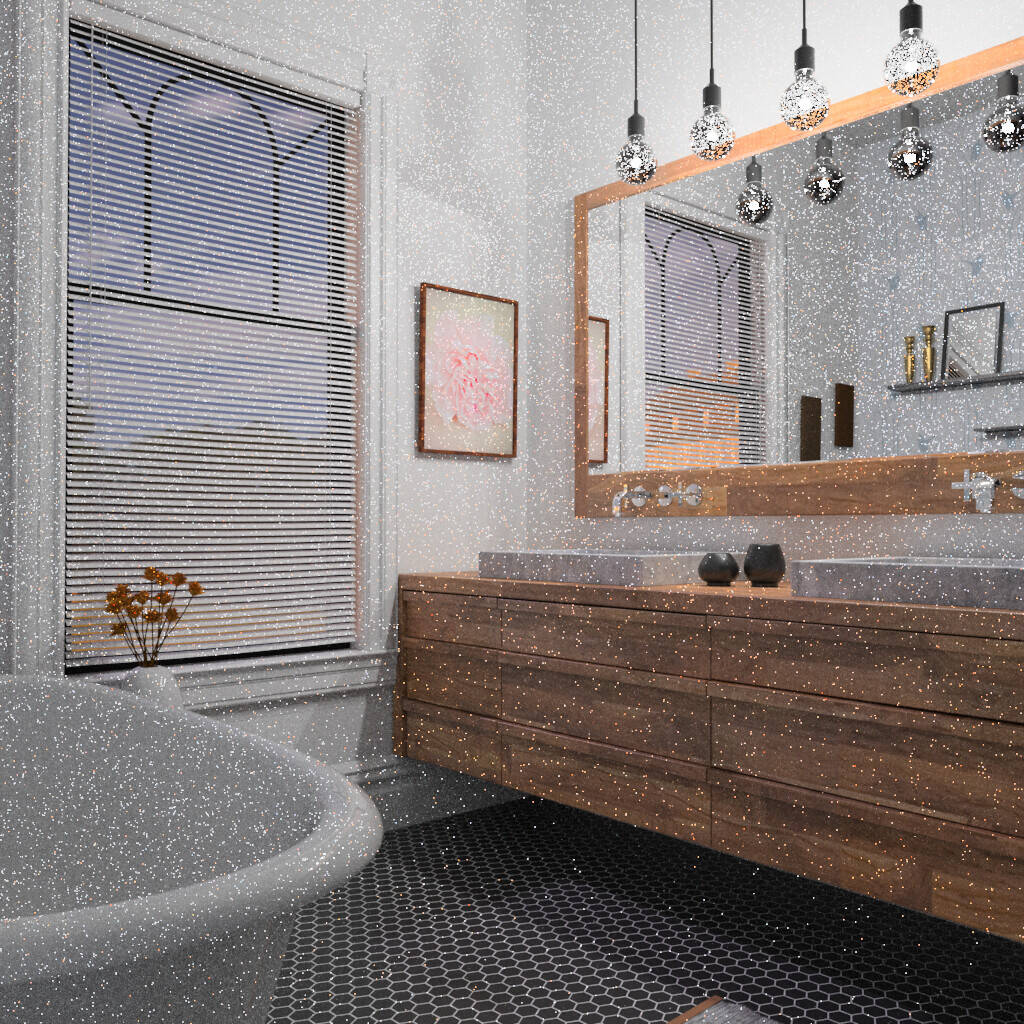}
    &
    \includegraphics[width=0.125\textwidth]{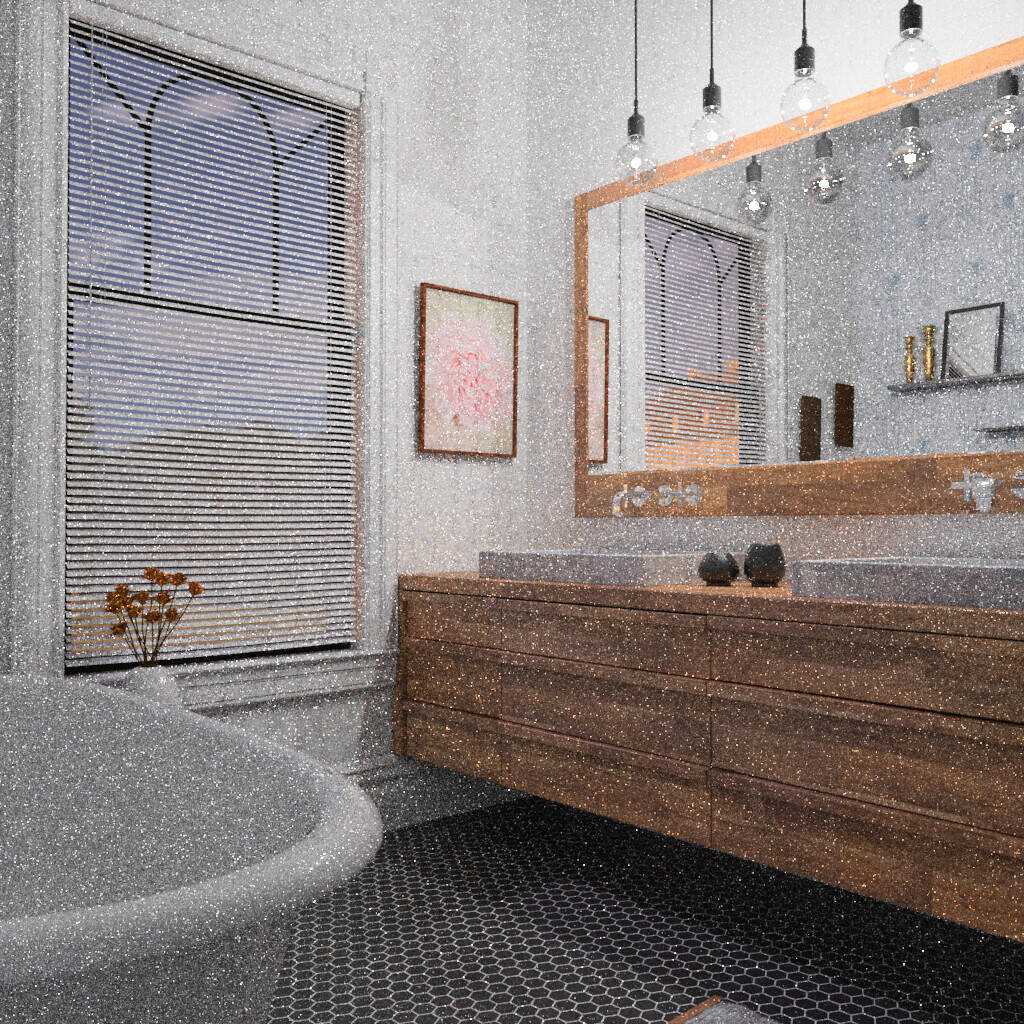}
    \\
    &
    \multirow{1}{*}[4em]{\rotatebox[origin=l]{90}{\textsc{Mitsuba}}}
    &
    %\multirow{1}{*}[3.4em]{FLIP Heat Map}
    \includegraphics[width=0.125\textwidth]{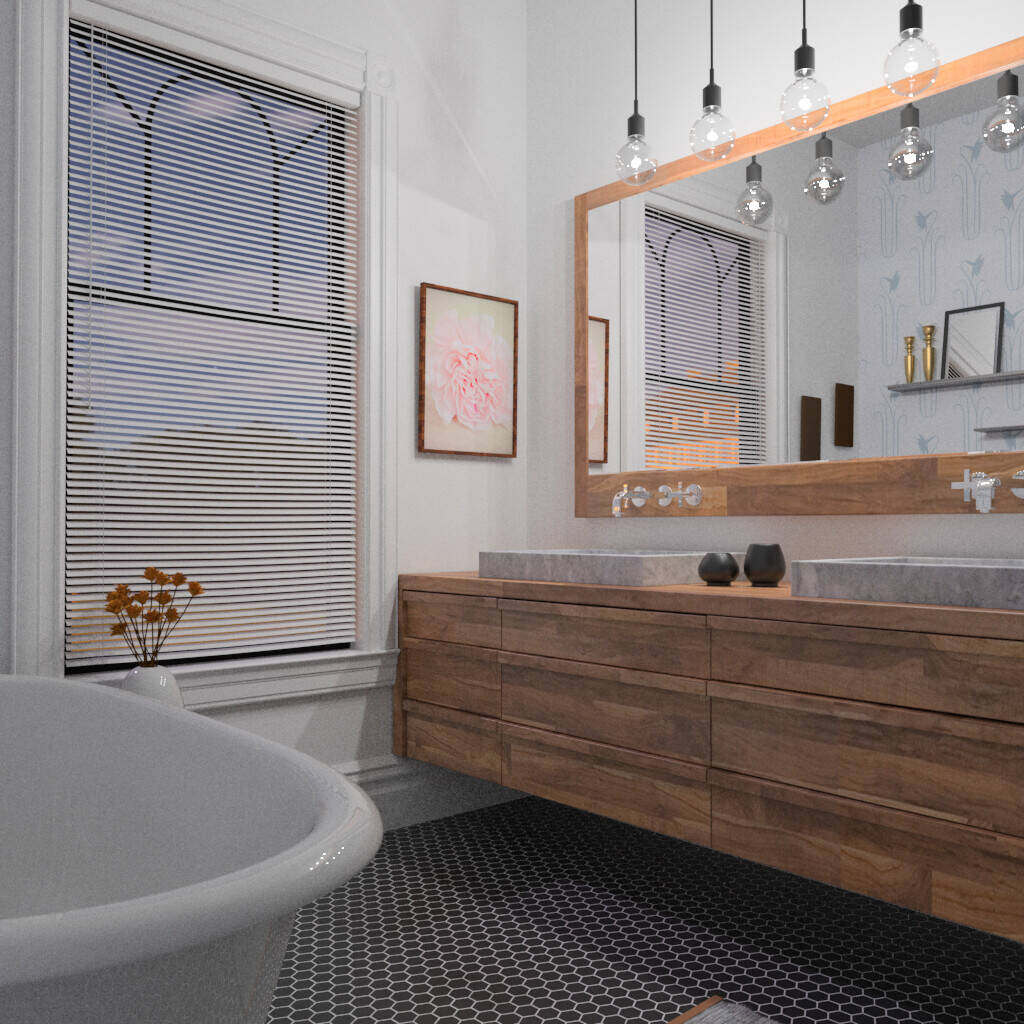} 
    &
    \includegraphics[width=0.125\textwidth]{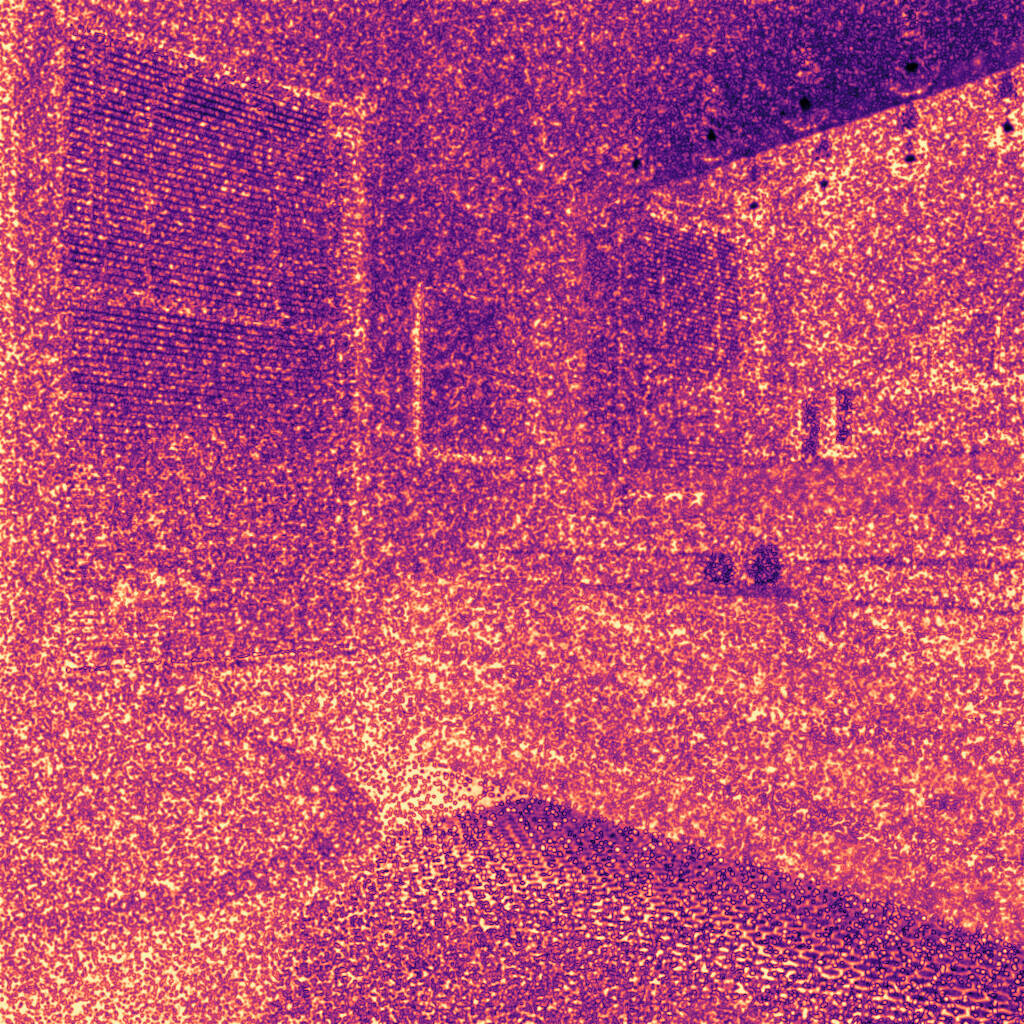} 
    &
    \includegraphics[width=0.125\textwidth]{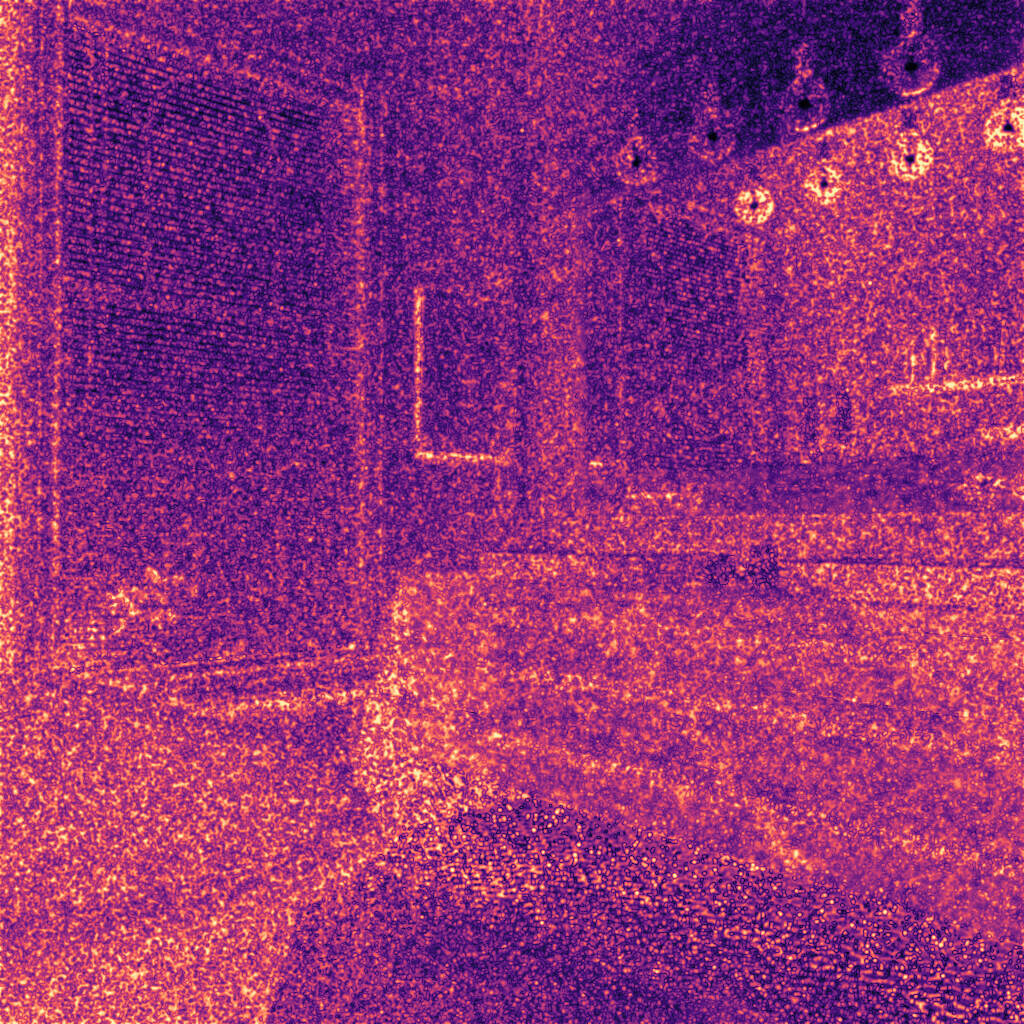}
    &
    \includegraphics[width=0.125\textwidth]{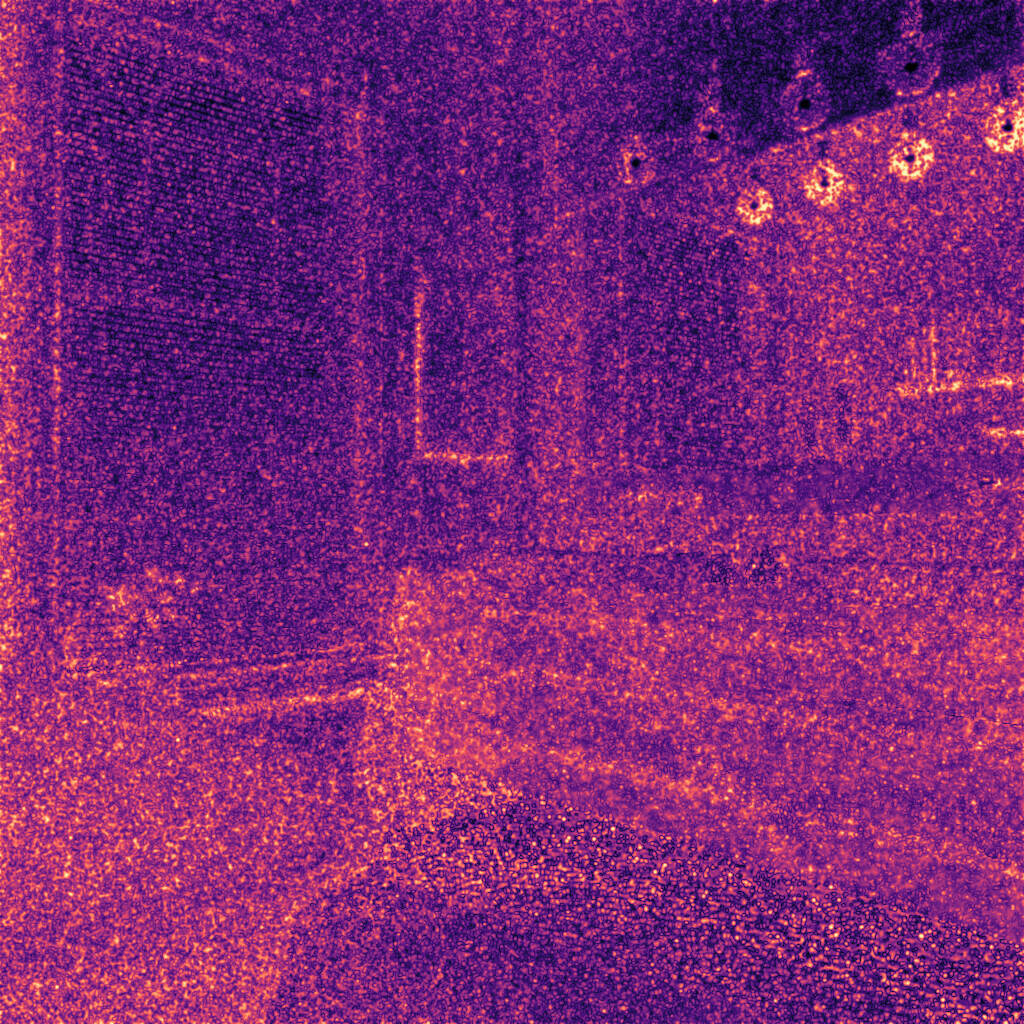}
    &
    \includegraphics[width=0.125\textwidth]{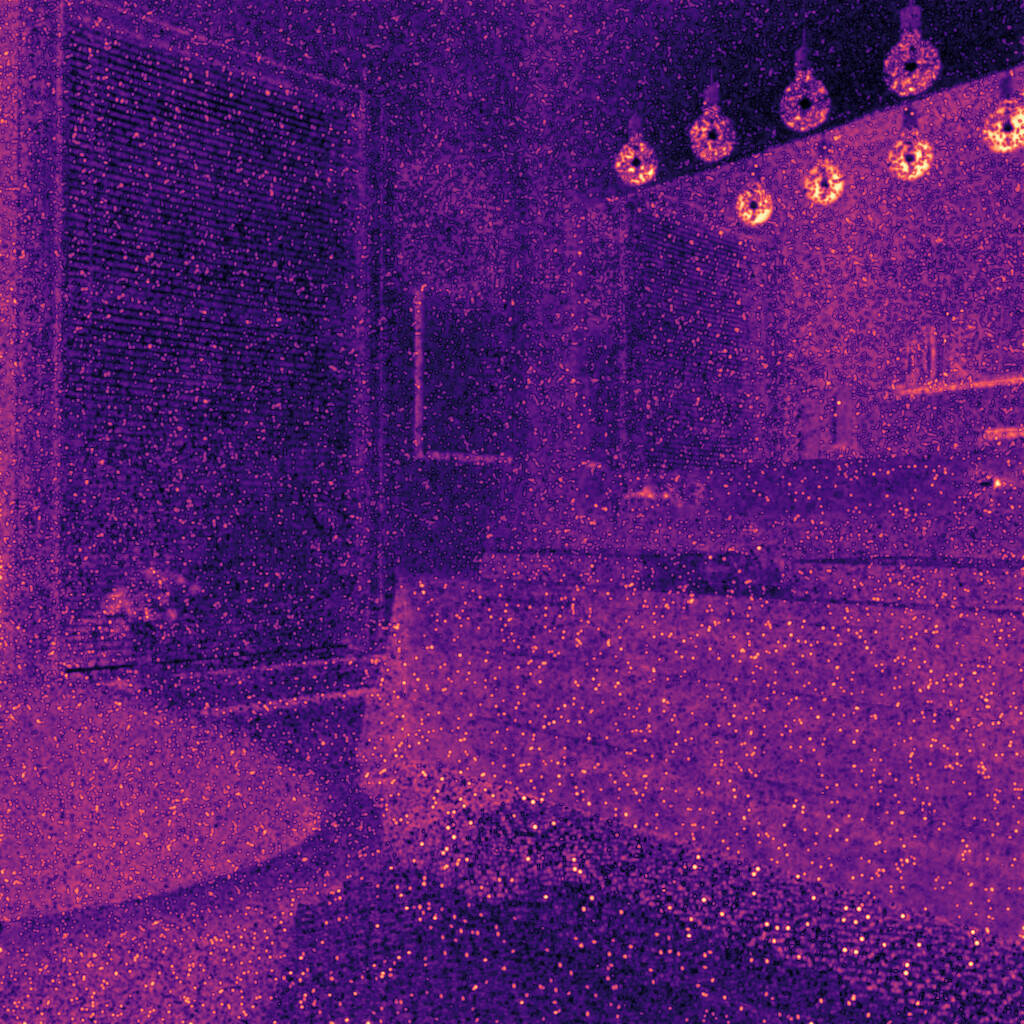}
    &
    \includegraphics[width=0.125\textwidth]{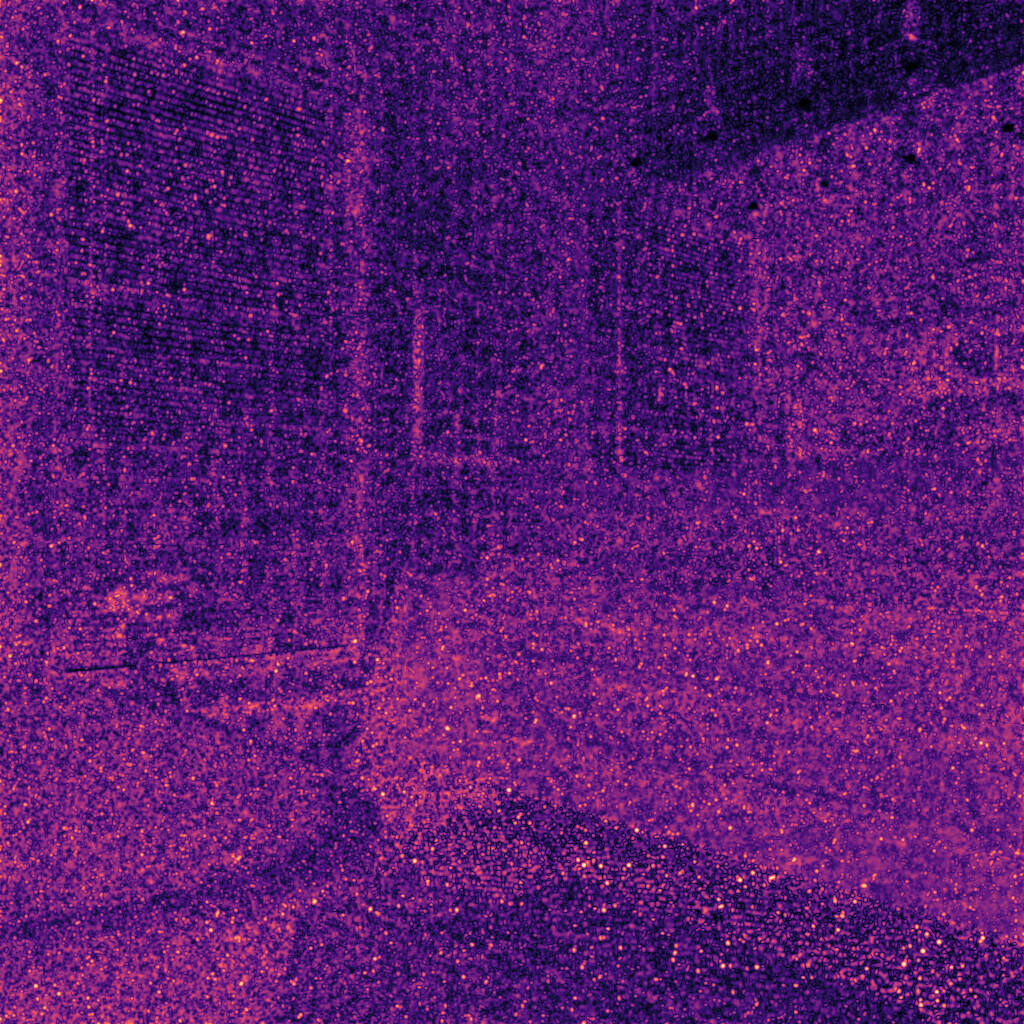}
    \\
    && FLIP Mean &
    0.533 &
    0.446 &
    0.363 &
    0.272 &
    \textbf{0.270}
    \\
    && MSE &
    0.125 &
    0.114 &
    \textbf{0.092} &
    0.113 &
    0.098
    \\
    \hline

    %=======================================%
    % Veach Door Scene
    %=======================================%
    & & & \multicolumn{3}{c|}{96spp} & \multicolumn{2}{c}{32t + 64spp}  \\
    \multirow{3}{*}[1em]{\rotatebox[origin=l]{90}{\textsc{\normalsize VeachDoor}}} 
    &
    \multirow{1}{*}[1.7em]{\rotatebox[origin=l]{90}{\textsc{Ours}}}
    &
    \includegraphics[width=0.125\textwidth]{figures/comparisons/VeachDoor/mray-door-ref}
    &
    \includegraphics[width=0.125\textwidth]{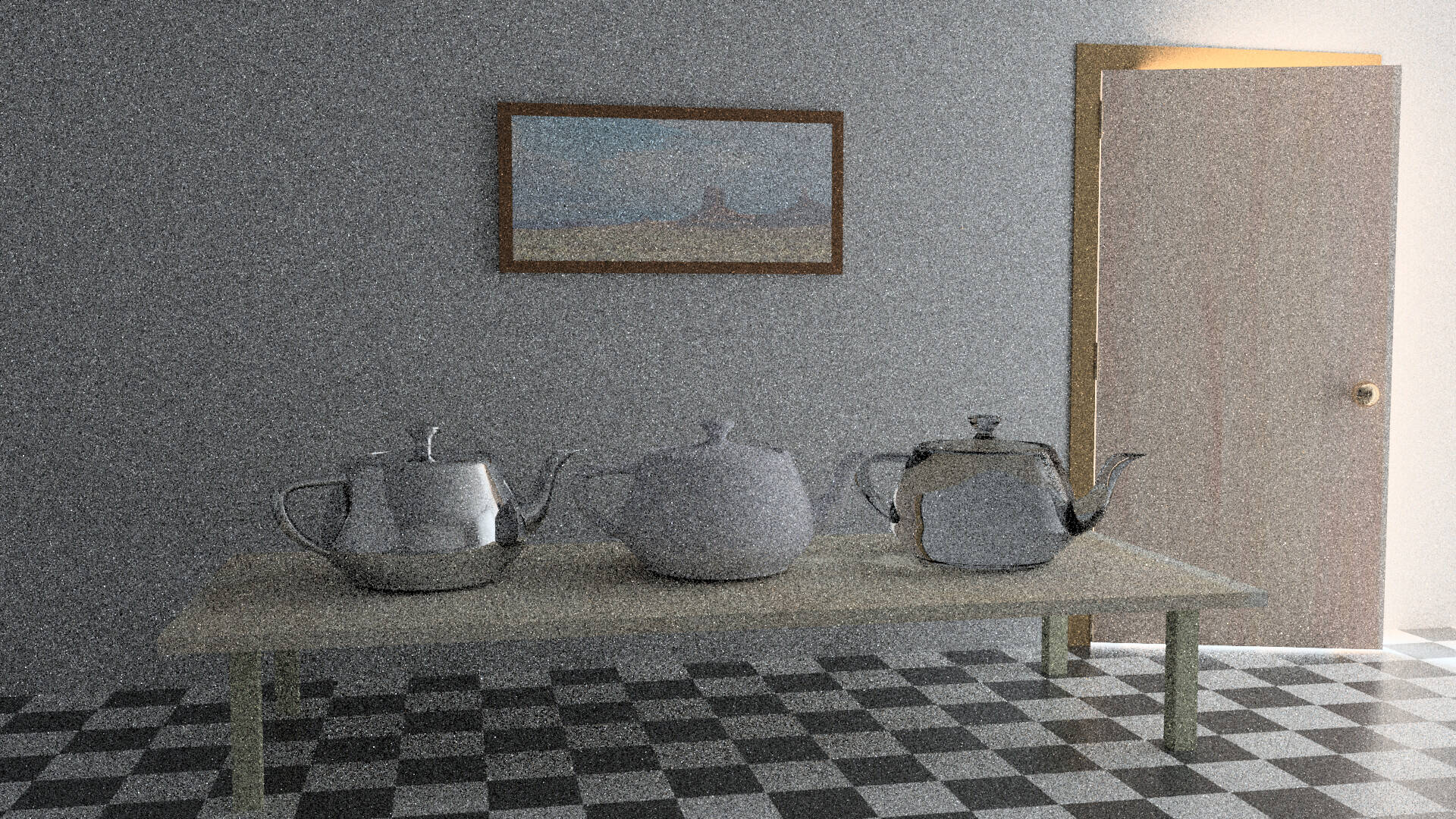}
    &
    \includegraphics[width=0.125\textwidth]{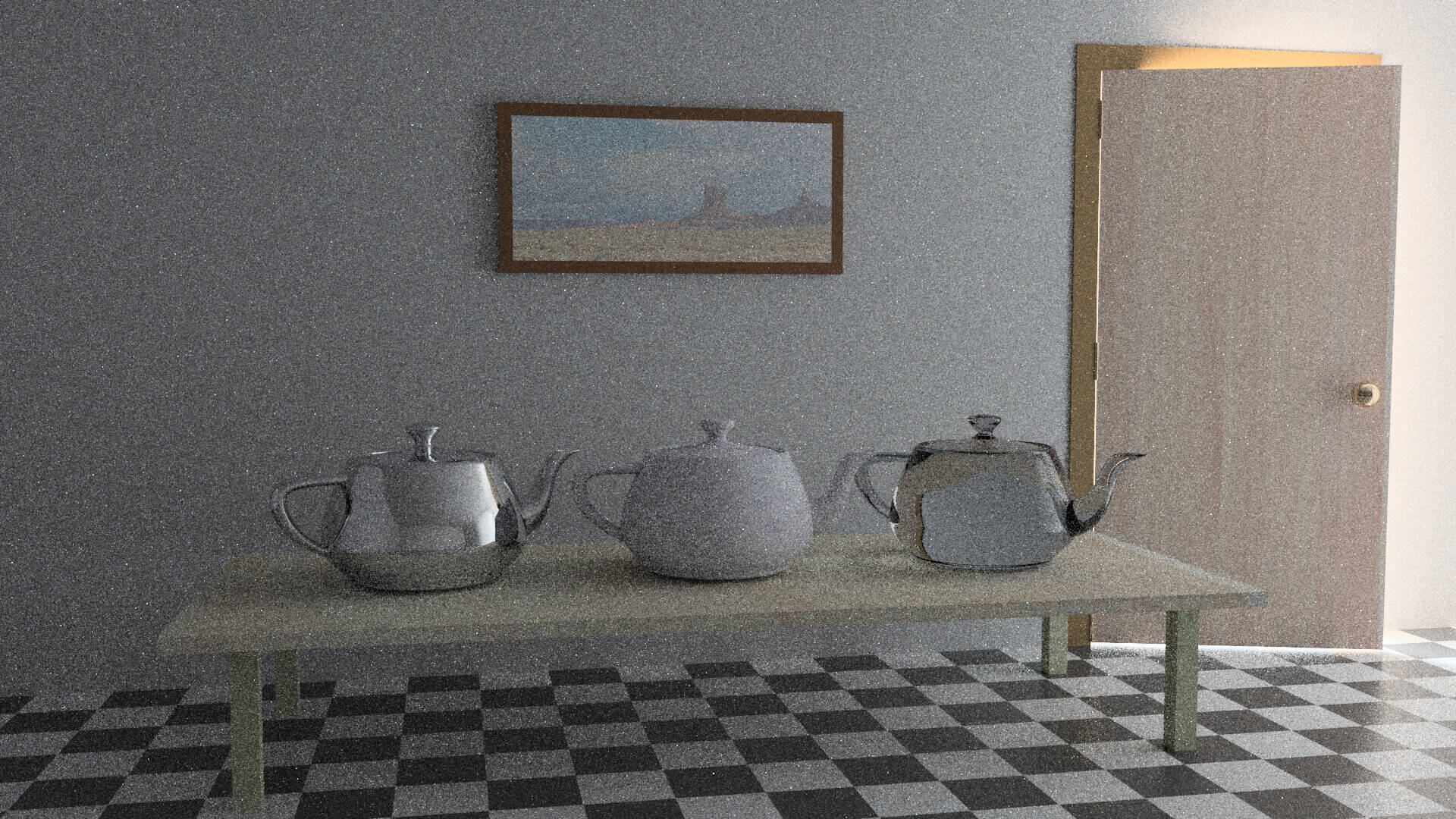}
    &    
    \includegraphics[width=0.125\textwidth]{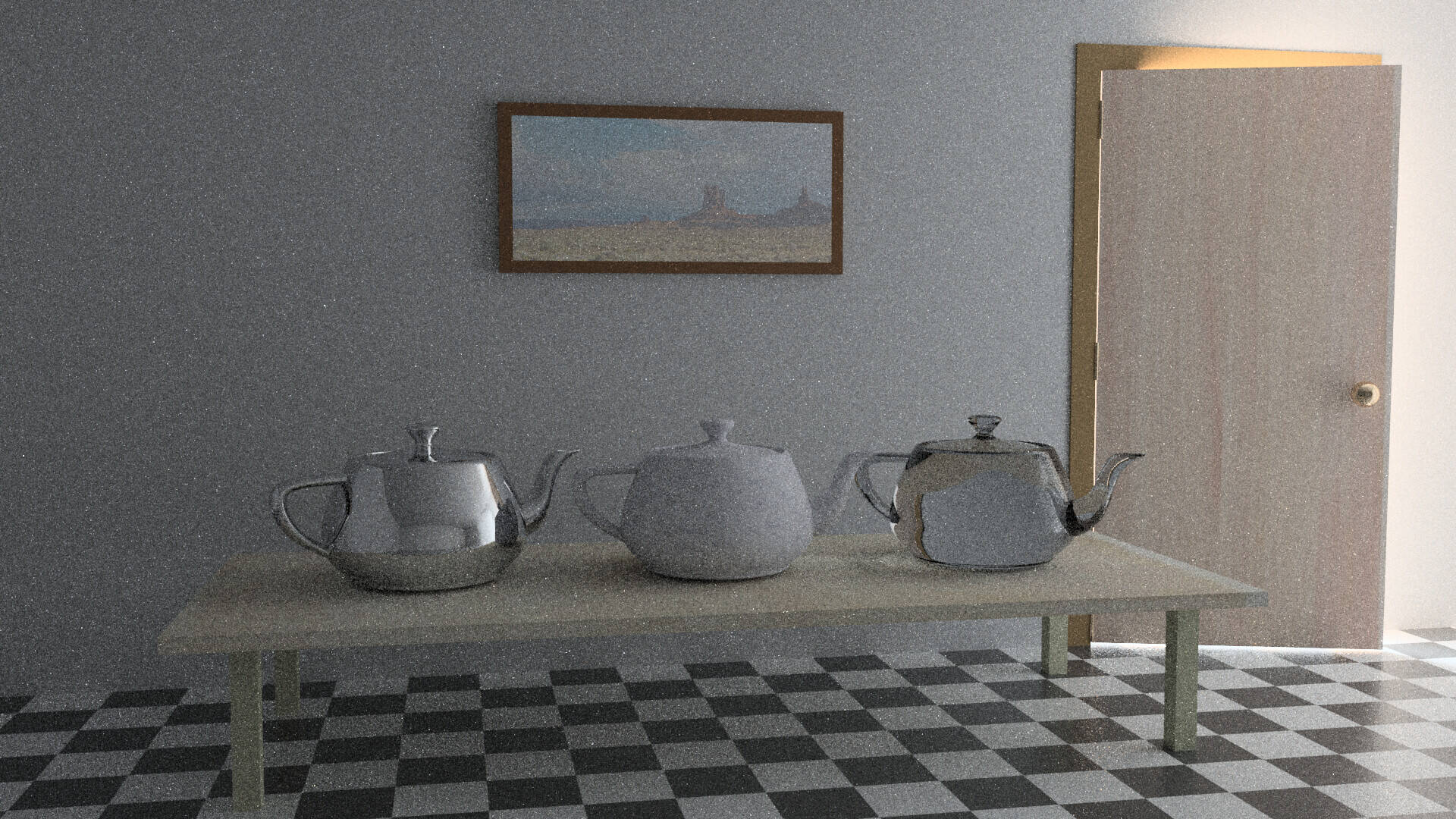}
    &
    \includegraphics[width=0.125\textwidth]{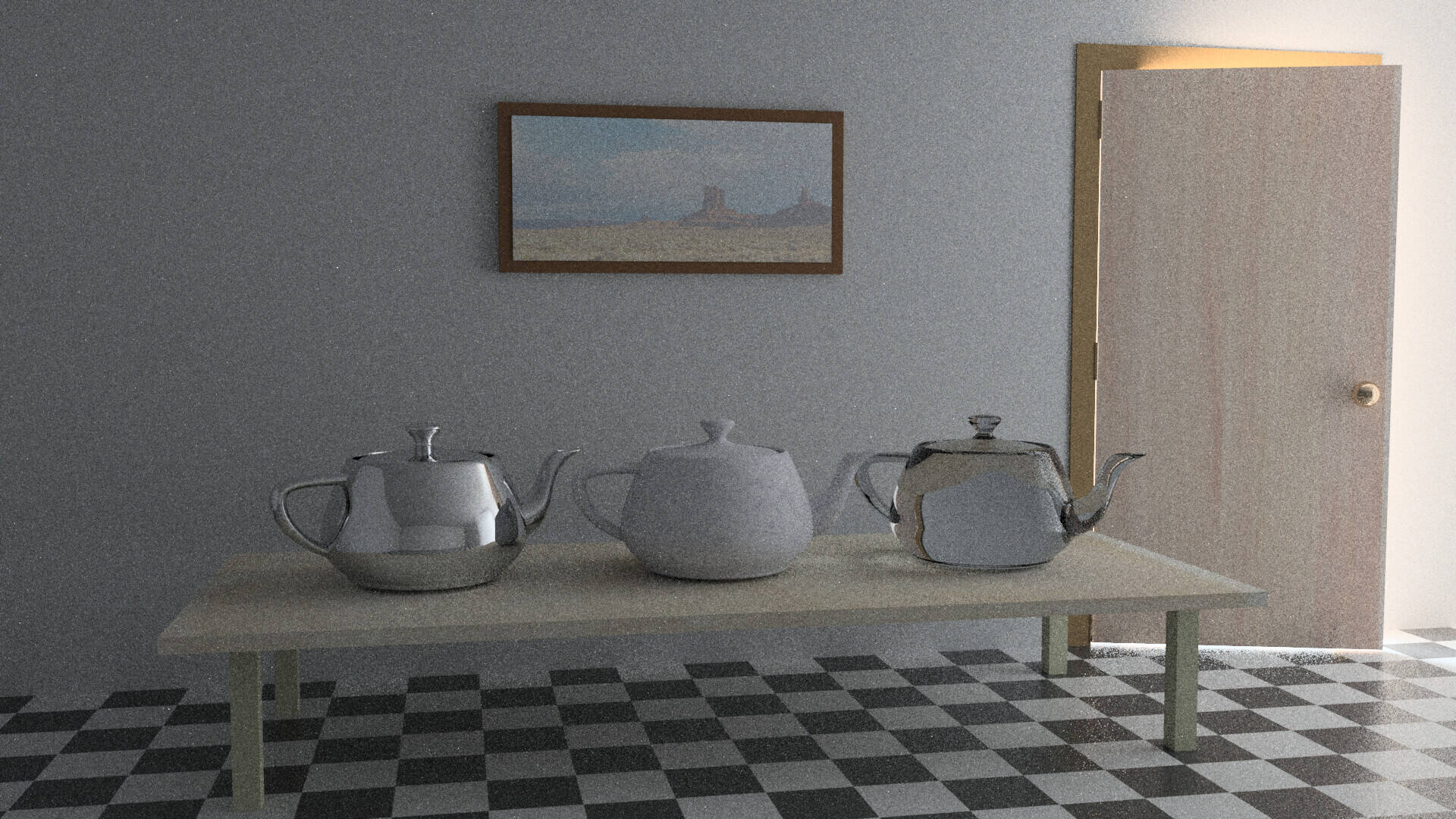}    
    &
    \includegraphics[width=0.125\textwidth]{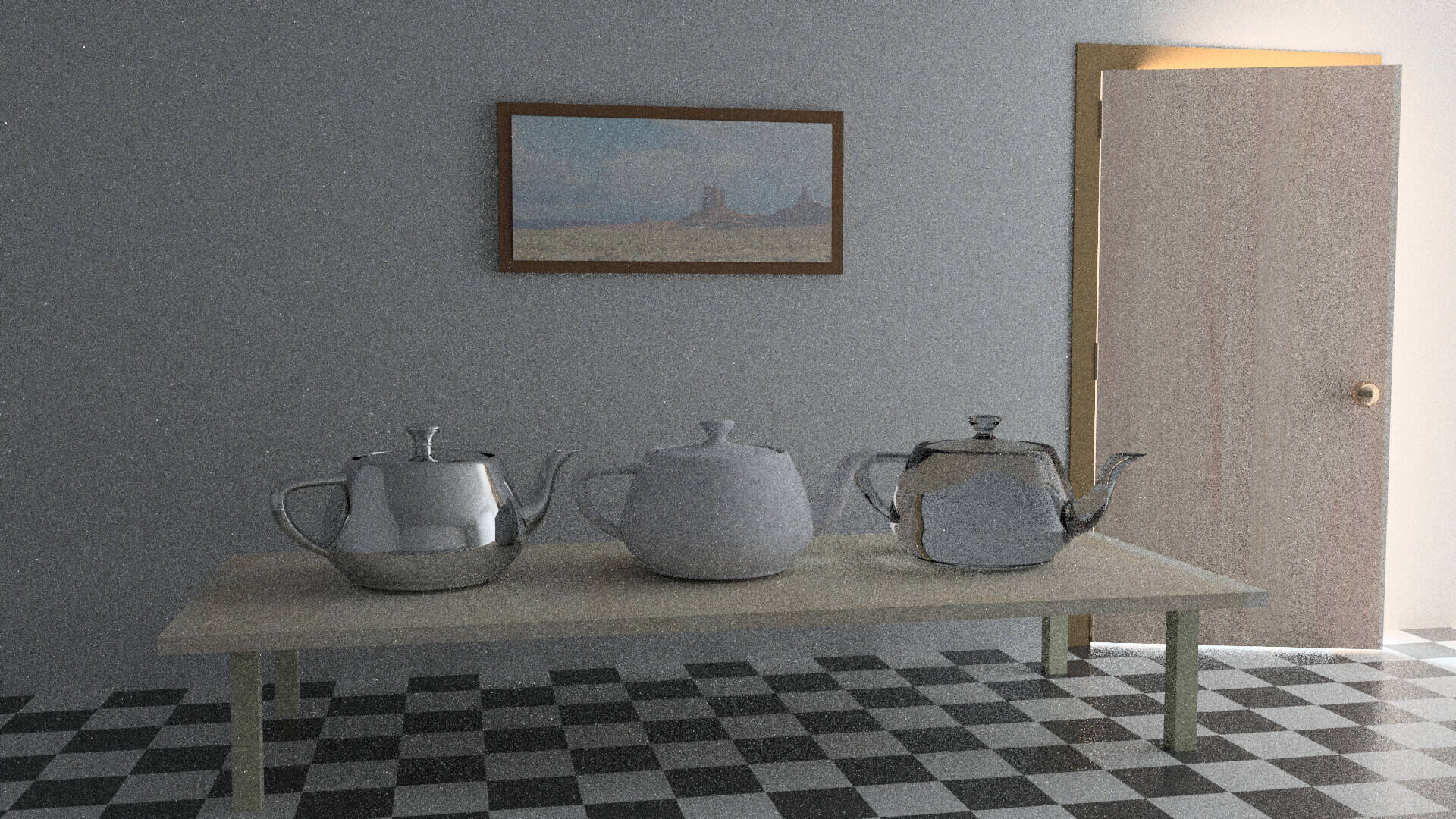}
    \\
    &
    \multirow{1}{*}[2.4em]{\rotatebox[origin=l]{90}{\textsc{Mitsuba}}}
    &
    %\multirow{1}{*}[1.7em]{FLIP Heat Map}
    \includegraphics[width=0.125\textwidth]{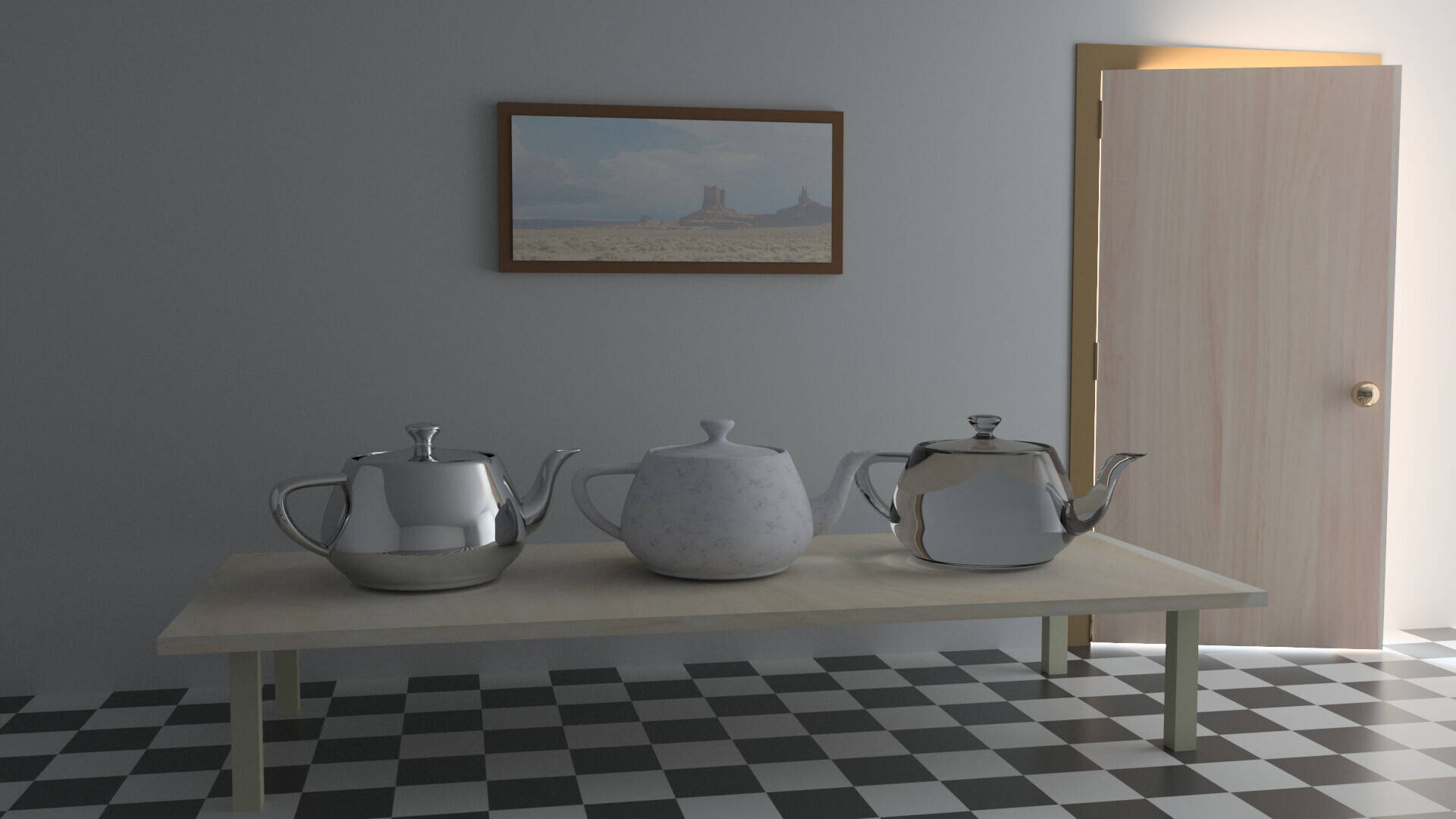}
    &
    \includegraphics[width=0.125\textwidth]{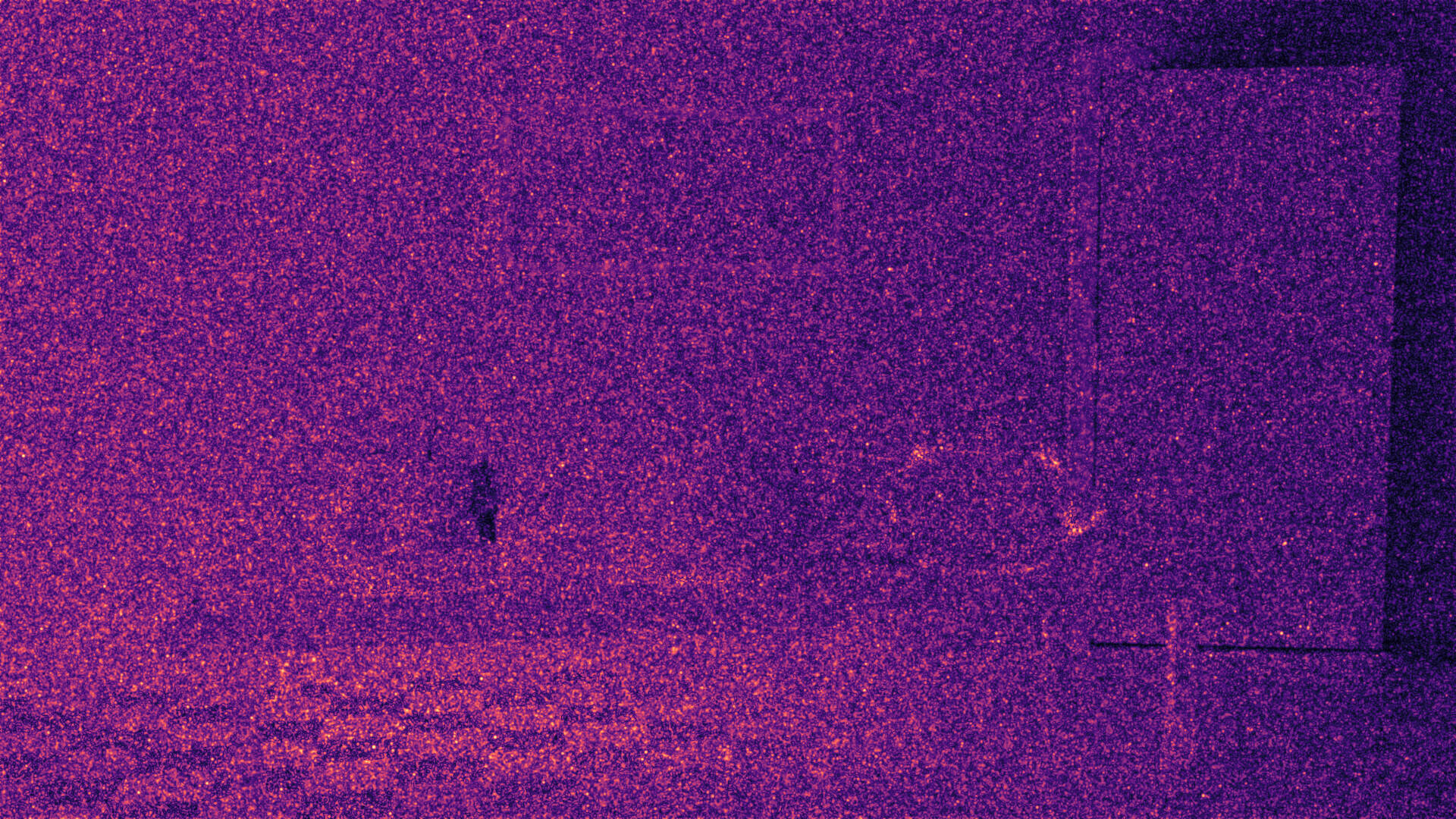}
    &
    \includegraphics[width=0.125\textwidth]{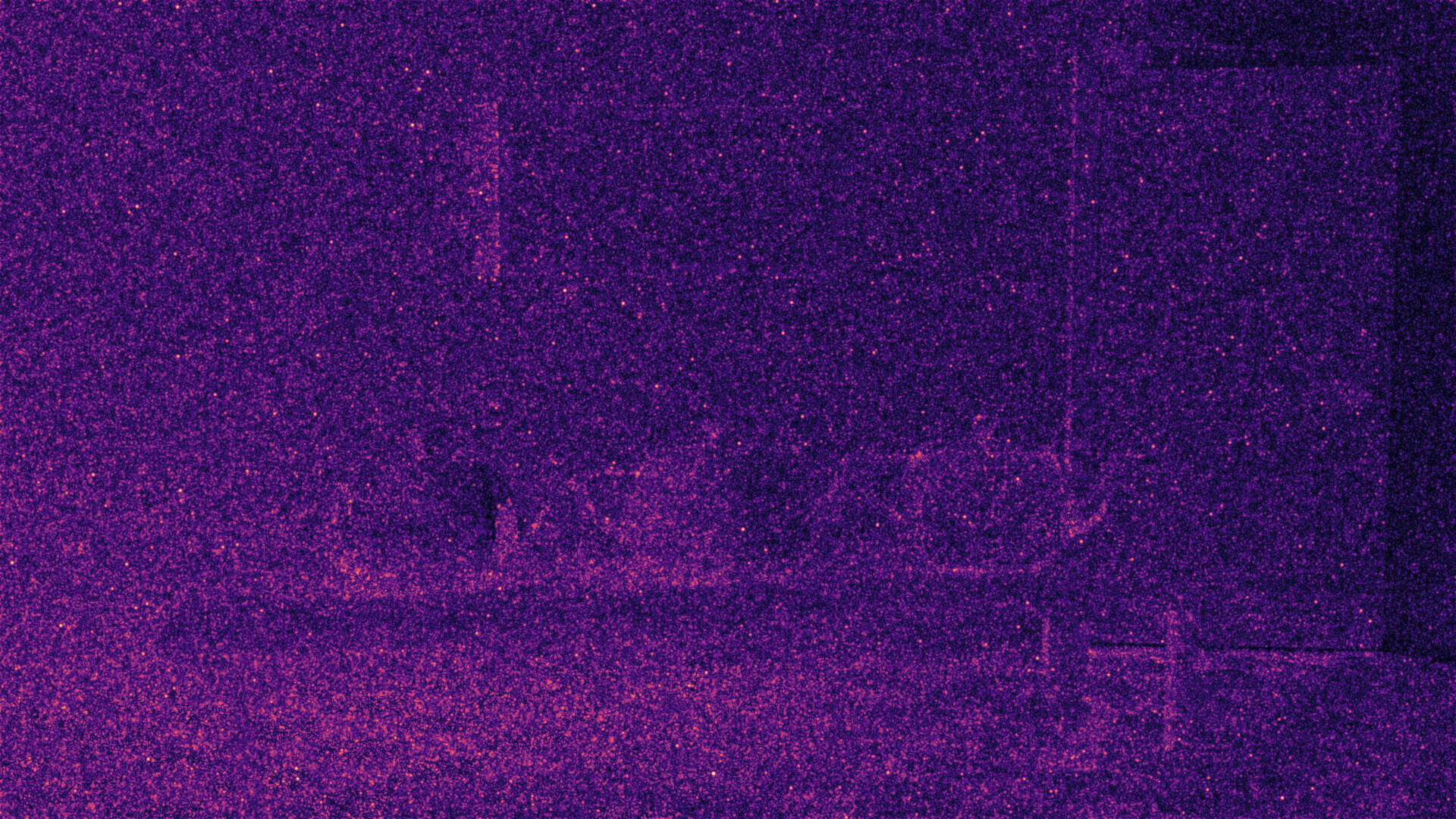}
    &
    \includegraphics[width=0.125\textwidth]{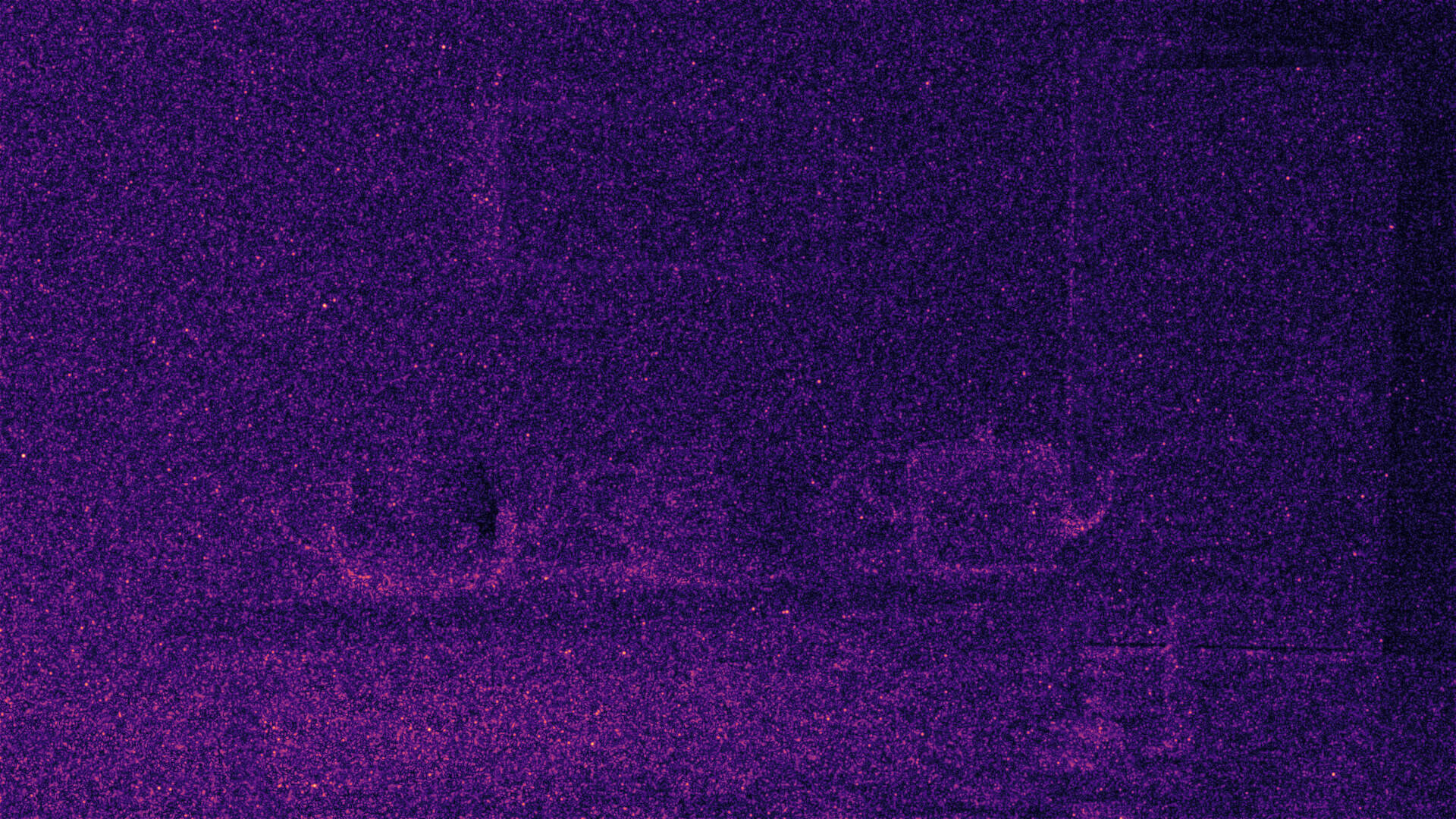}
    &
    \includegraphics[width=0.125\textwidth]{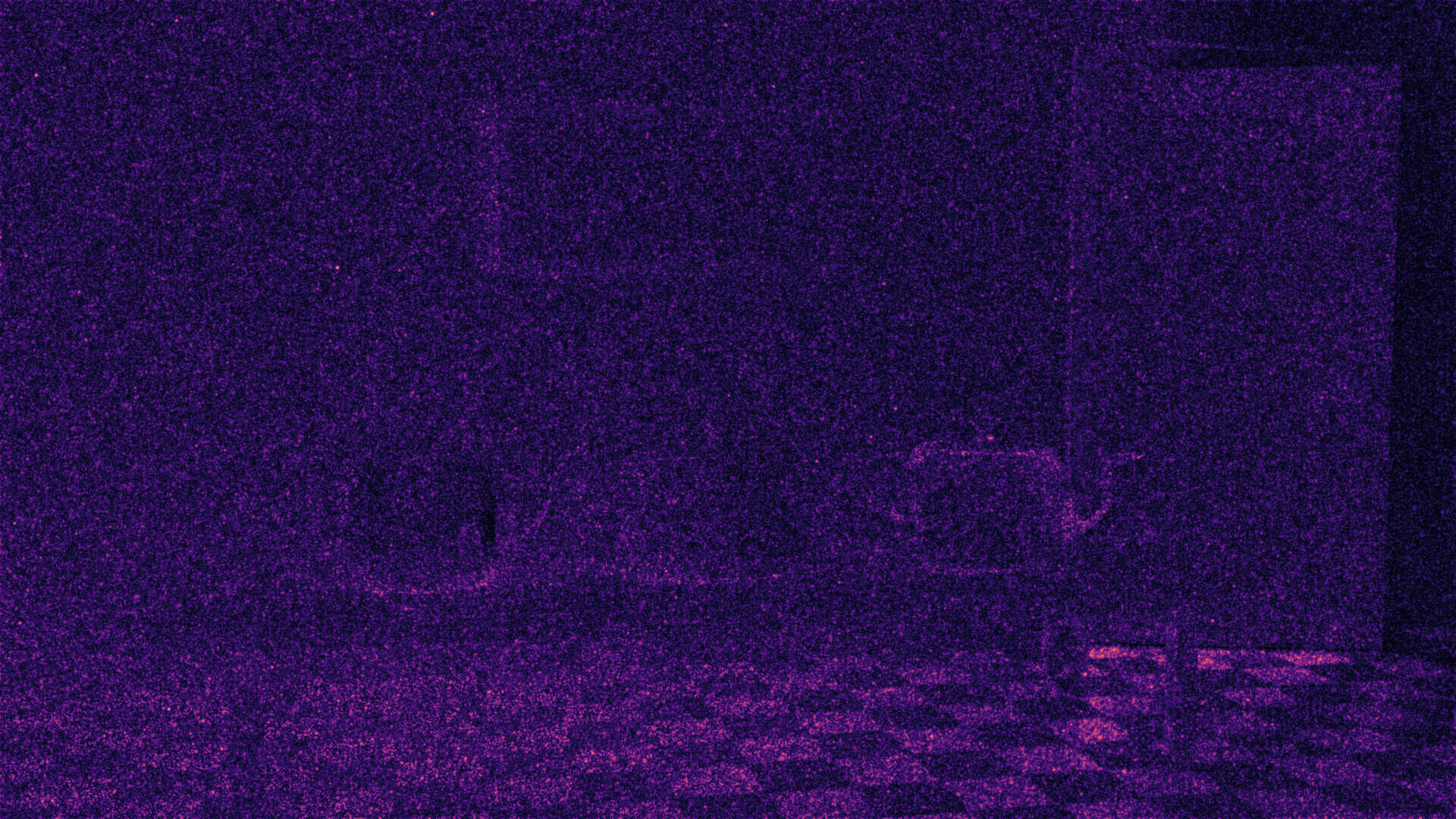}
    &
    \includegraphics[width=0.125\textwidth]{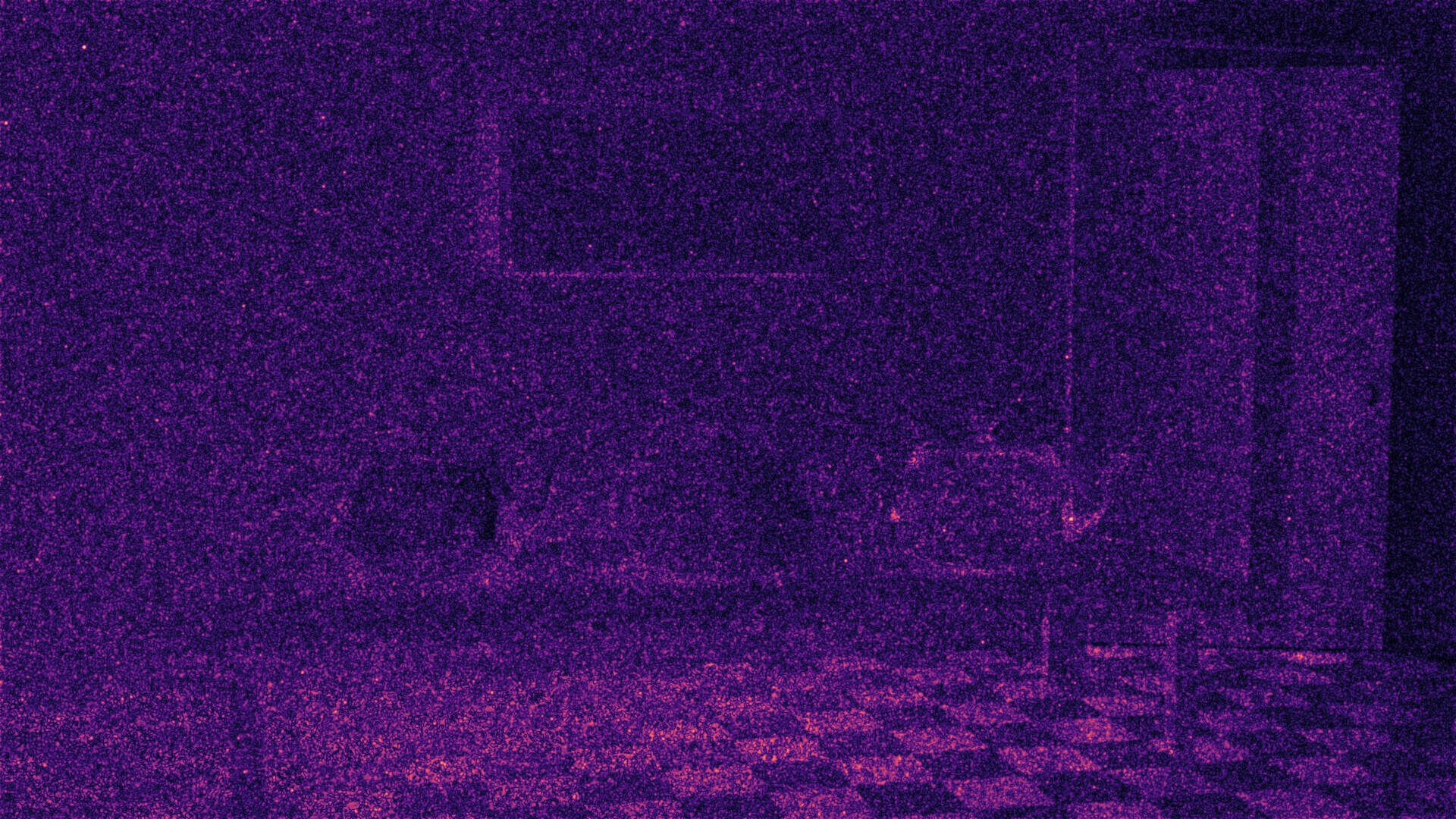}
    \\
    && FLIP Mean &
    0.295 &
    0.207 & 
    0.162 &
    \textbf{0.142} &
    0.177
    \\
    && MSE &
    0.089 &
    0.069 &
    0.036 &
    \textbf{0.015} &
    0.023
    \\
    \hline
    
    %=======================================%
    % Sponza Scene
    %=======================================%
    & & & \multicolumn{3}{c|}{48spp} & \multicolumn{2}{c}{16t + 32spp}  \\    
    \multirow{5}{*}[-3em]{\rotatebox[origin=l]{90}{\textsc{\normalsize CrySponza}}} 
    &
    \multirow{1}{*}[1.7em]{\rotatebox[origin=l]{90}{\textsc{Ours}}}
    &
    \begin{tikzpicture}
        \node[anchor=south west,inner sep=0] (sponzaRef) at (0,0) {\includegraphics[width=0.125\textwidth]{figures/comparisons/Sponza/sponza_ref}};
        \begin{scope}[x={(sponzaRef.south east)},y={(sponzaRef.north west)}]
            \draw[red,thick] (0.025,0.627) rectangle (0.15,0.752);
            \draw[green,thick] (0.465,0.52) rectangle (0.59,0.645); 
        \end{scope}
    \end{tikzpicture}
    &
    \includegraphics[width=0.125\textwidth]{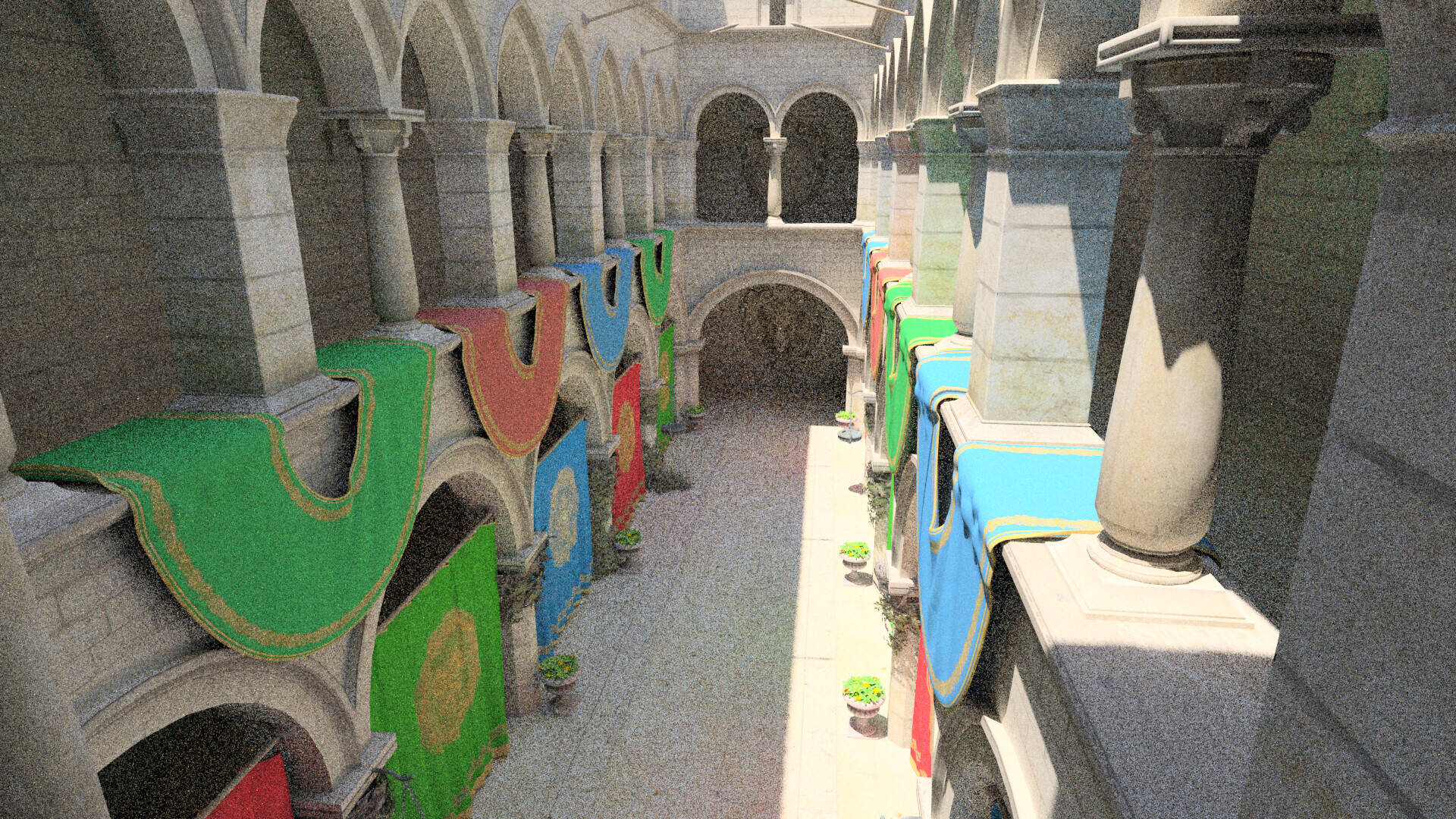}
    &
    \includegraphics[width=0.125\textwidth]{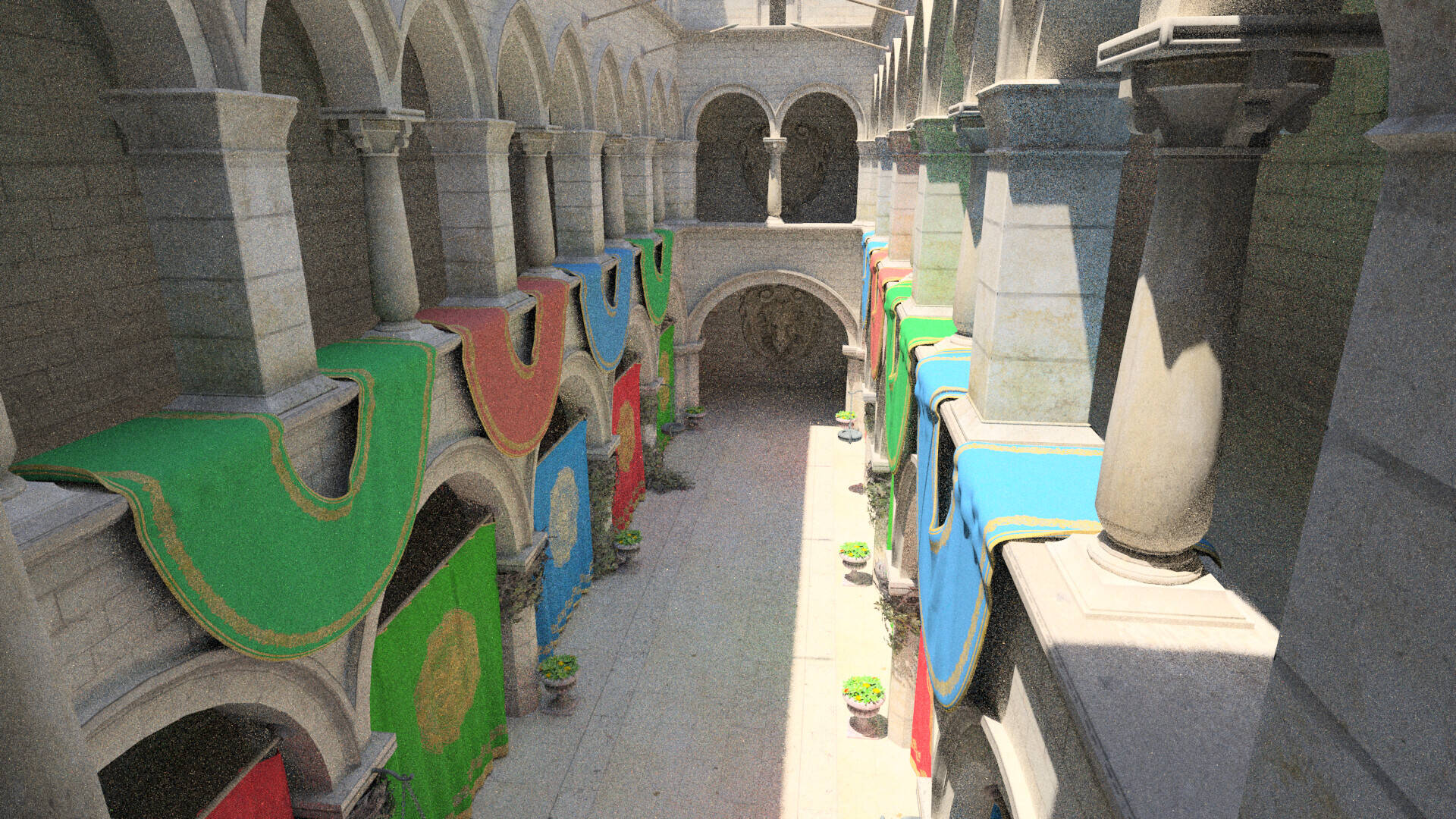}
    &
    \includegraphics[width=0.125\textwidth]{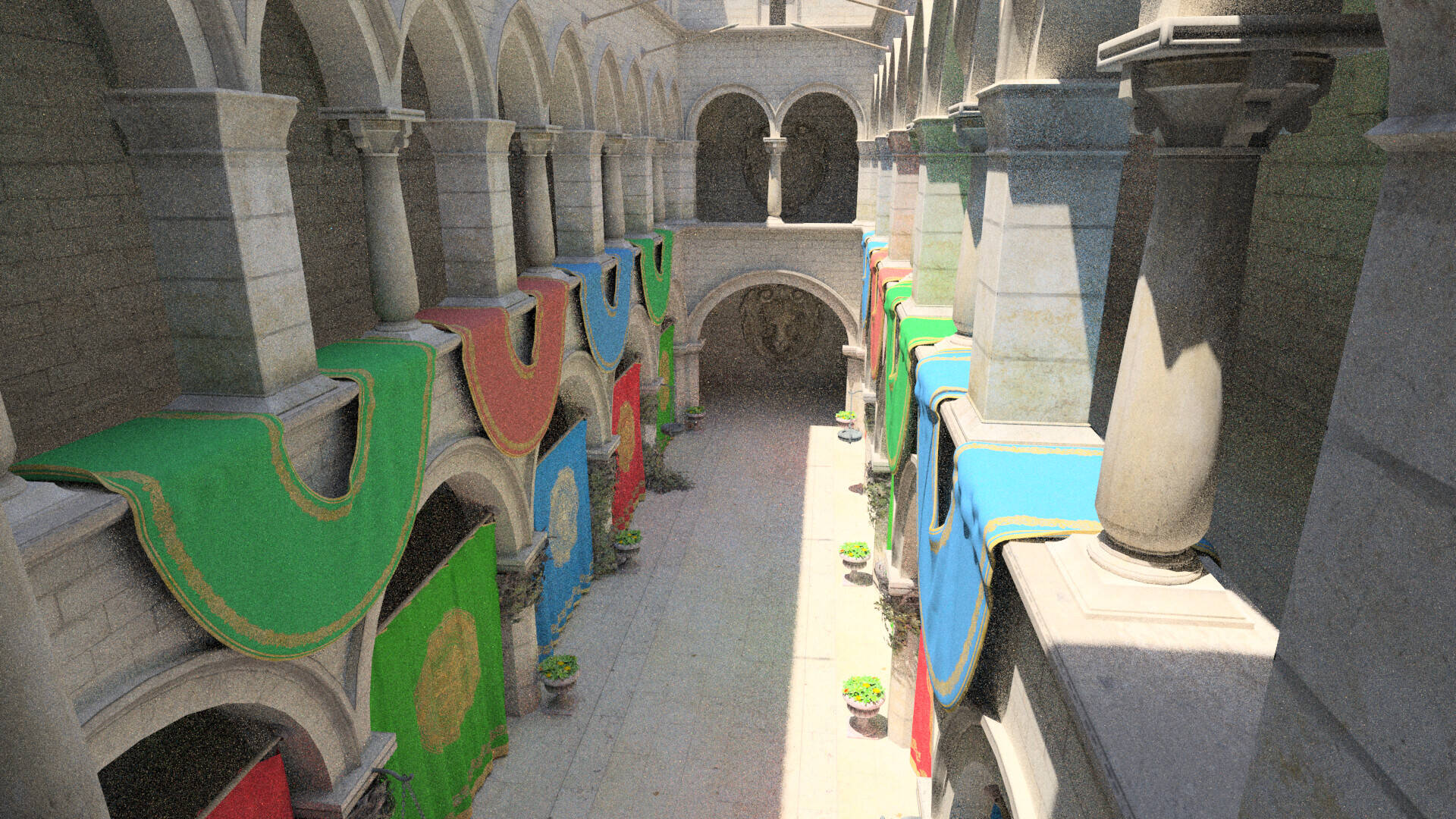}
    &
    \includegraphics[width=0.125\textwidth]{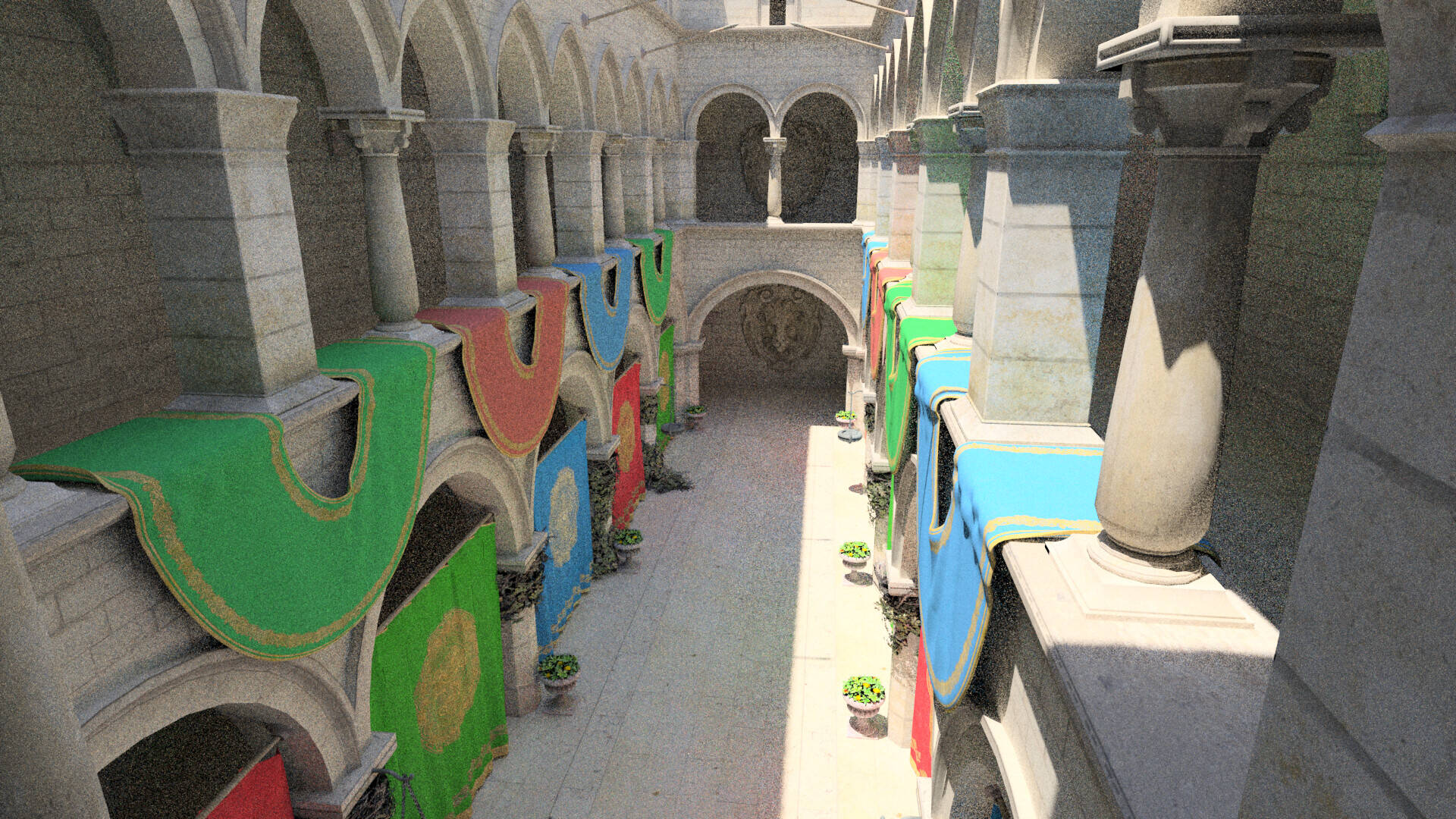}
    &
    \includegraphics[width=0.125\textwidth]{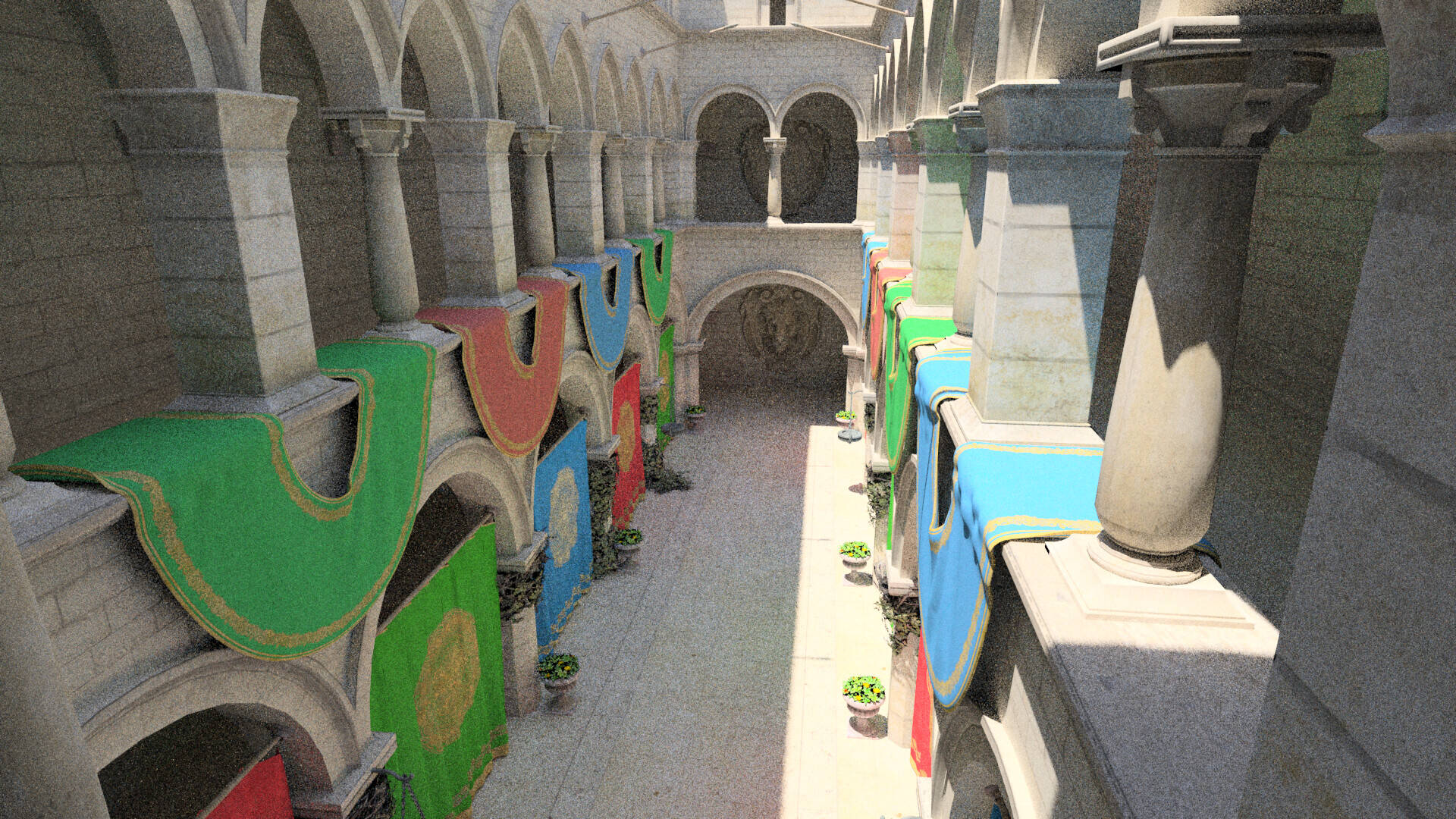}
    \\
    &
    \multirow{1}{*}[2.4em]{\rotatebox[origin=l]{90}{\textsc{Mitsuba}}}
    &
    %\multirow{1}{*}[1.7em]{FLIP Heat Map}
    \includegraphics[width=0.125\textwidth]{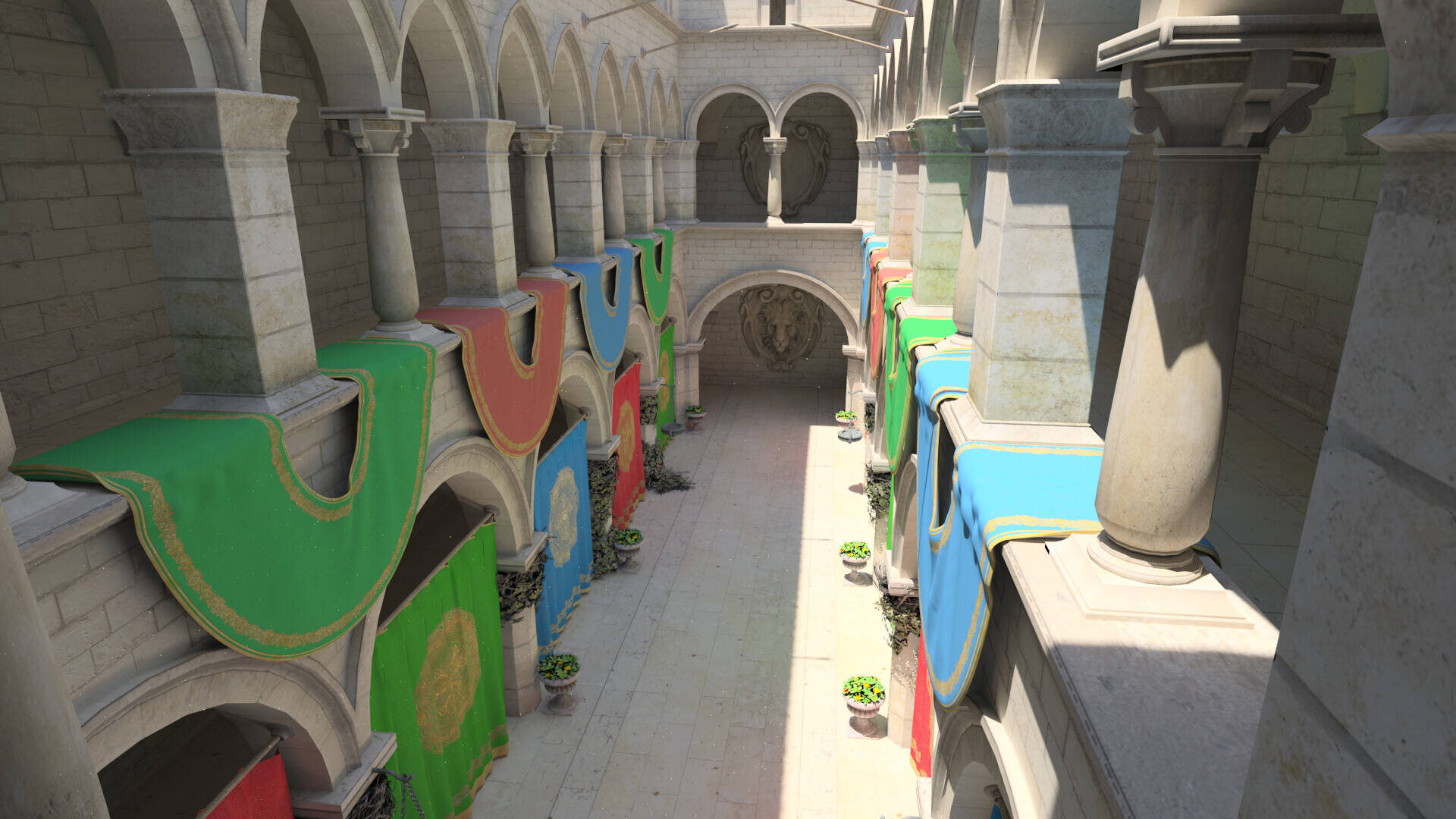} 
    &
    \includegraphics[width=0.125\textwidth]{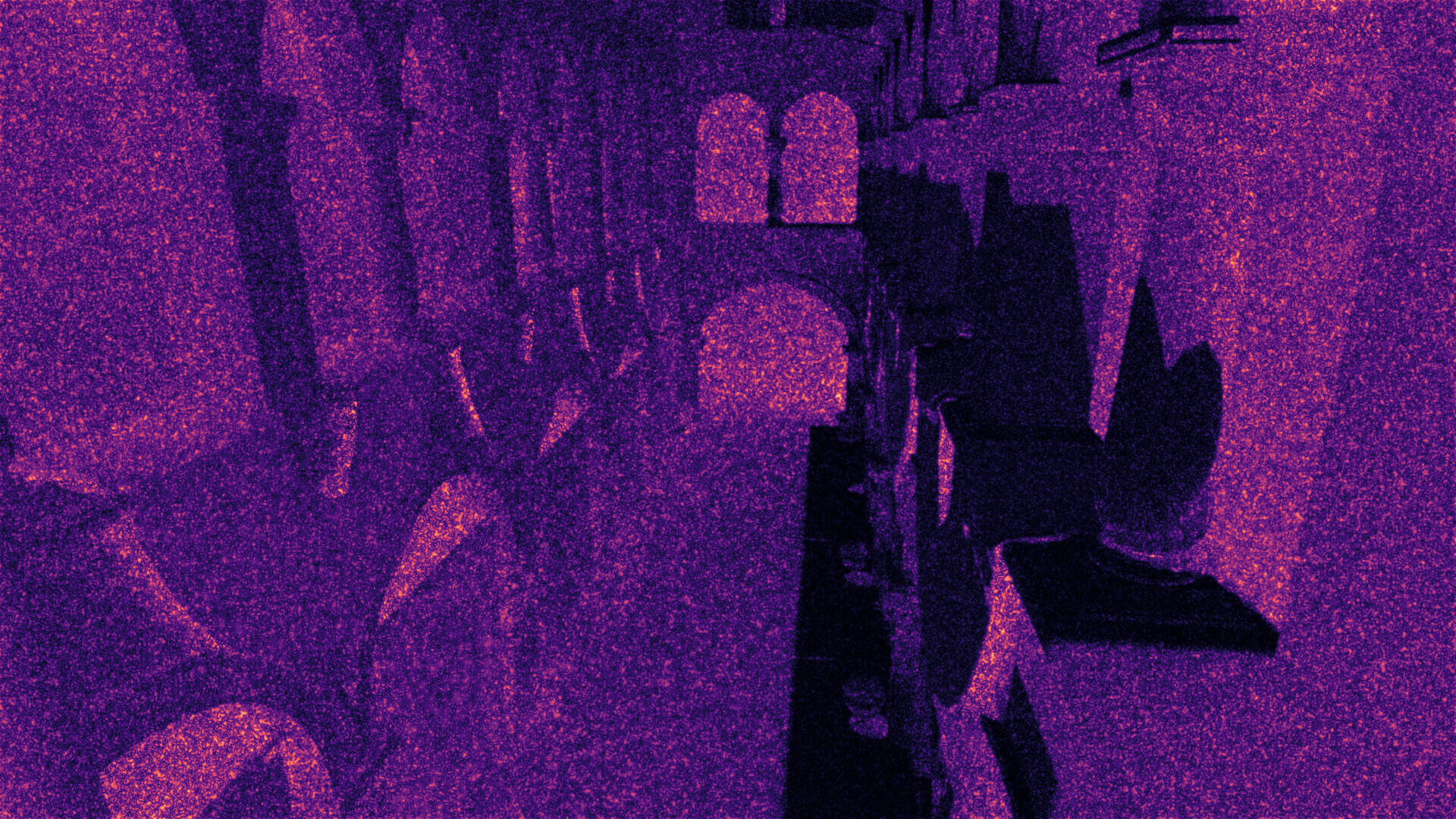} 
    &
    \includegraphics[width=0.125\textwidth]{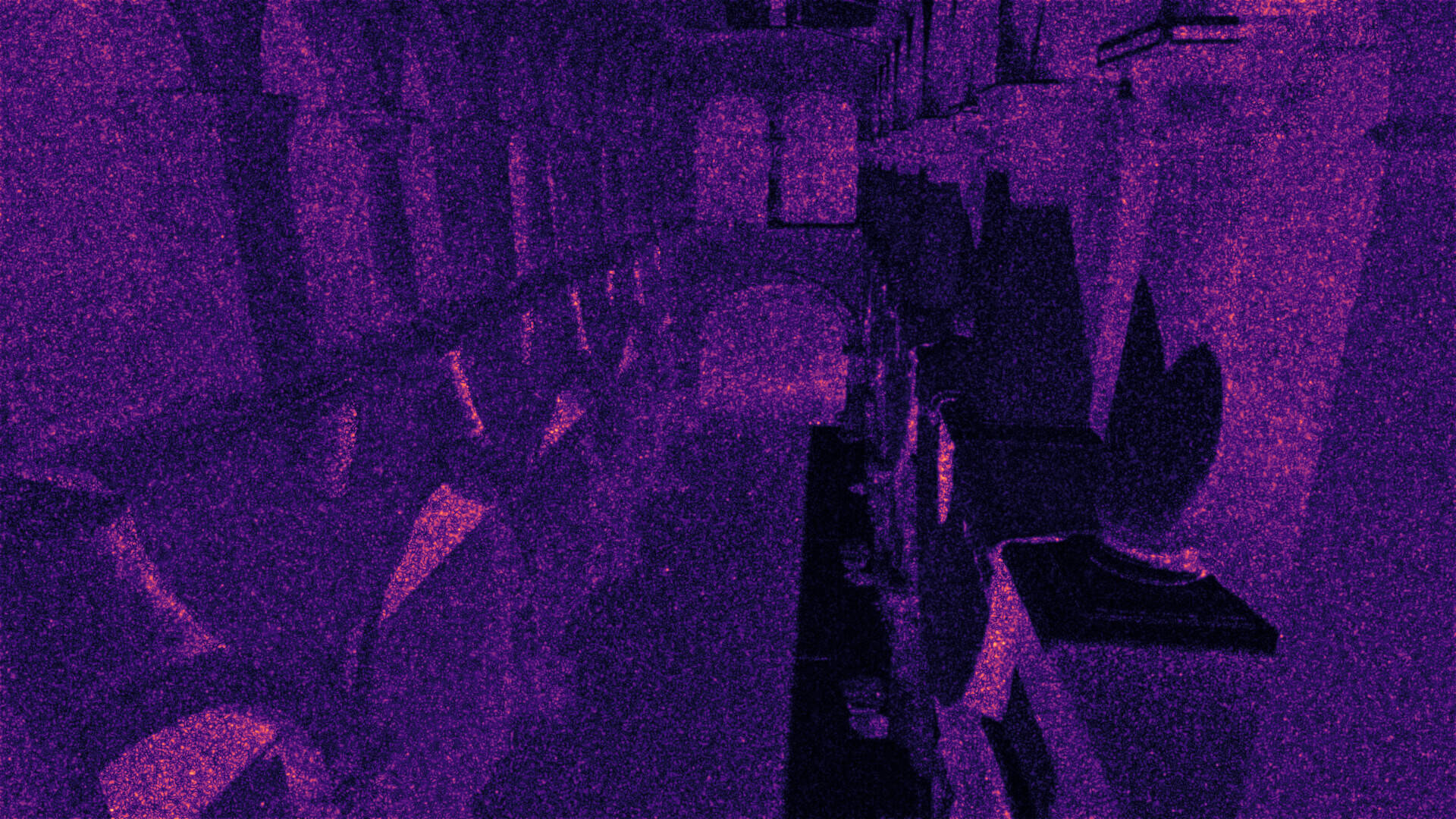}
    &
    \includegraphics[width=0.125\textwidth]{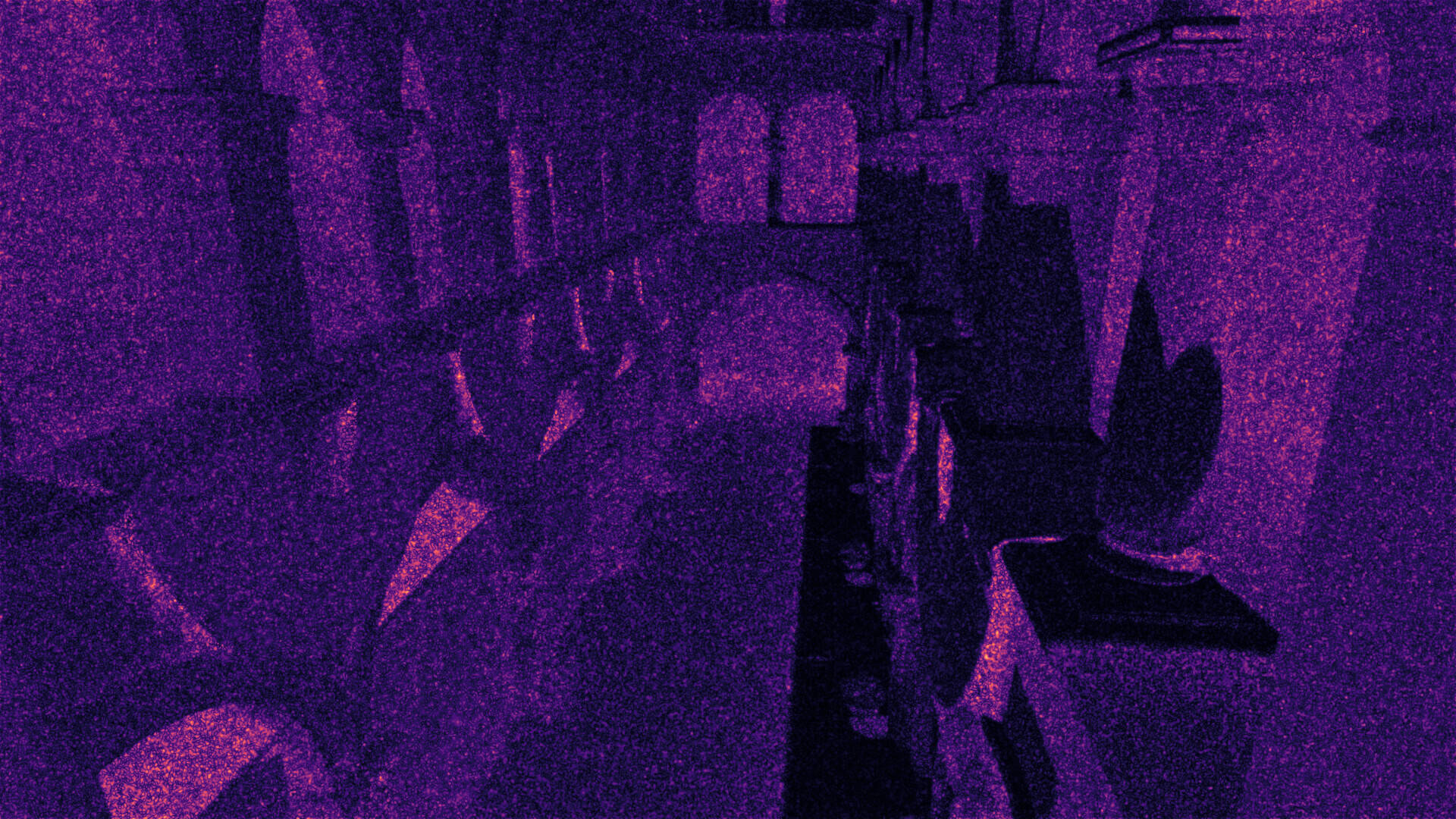}
    &
    \includegraphics[width=0.125\textwidth]{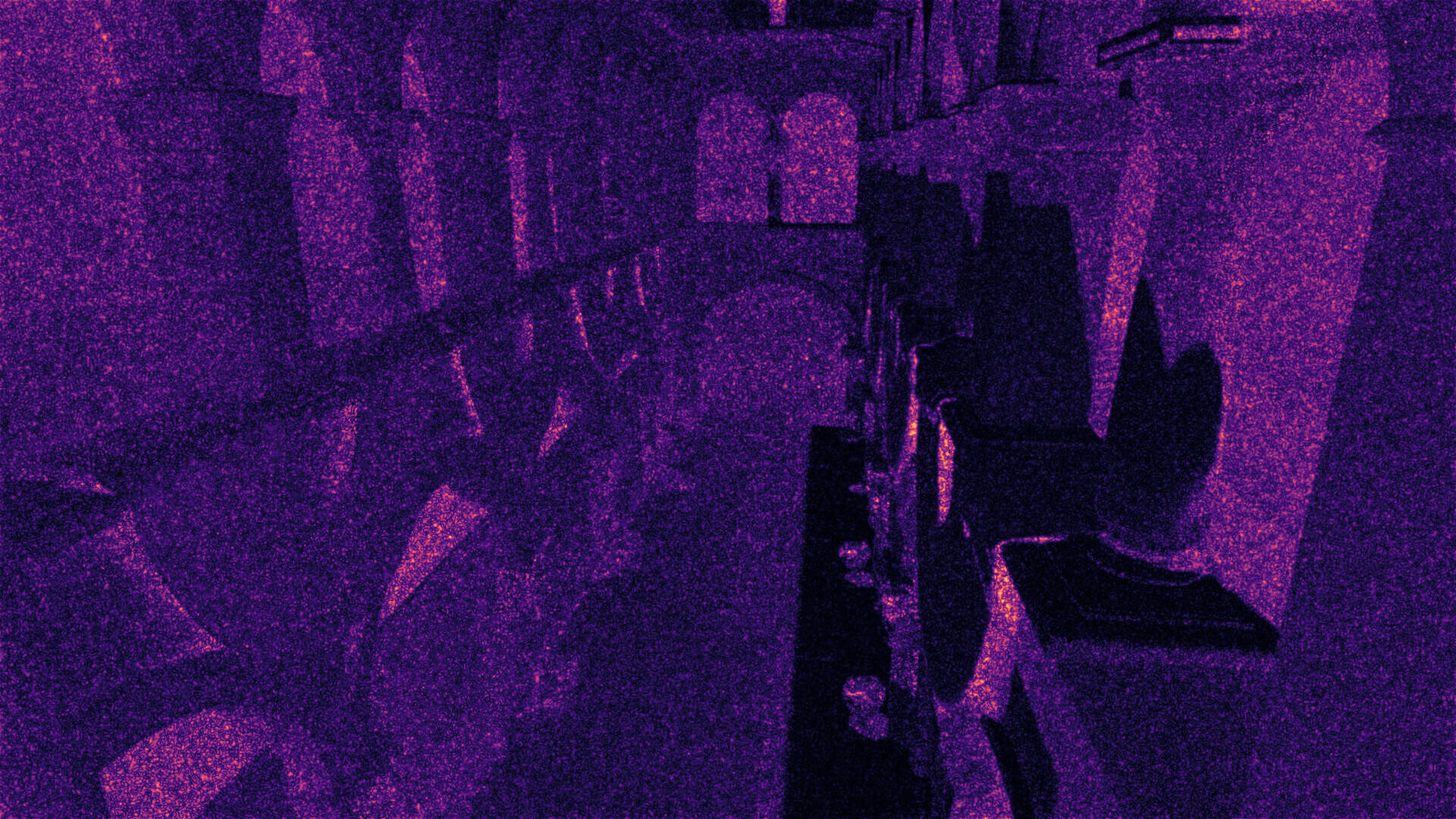}
    &
    \includegraphics[width=0.125\textwidth]{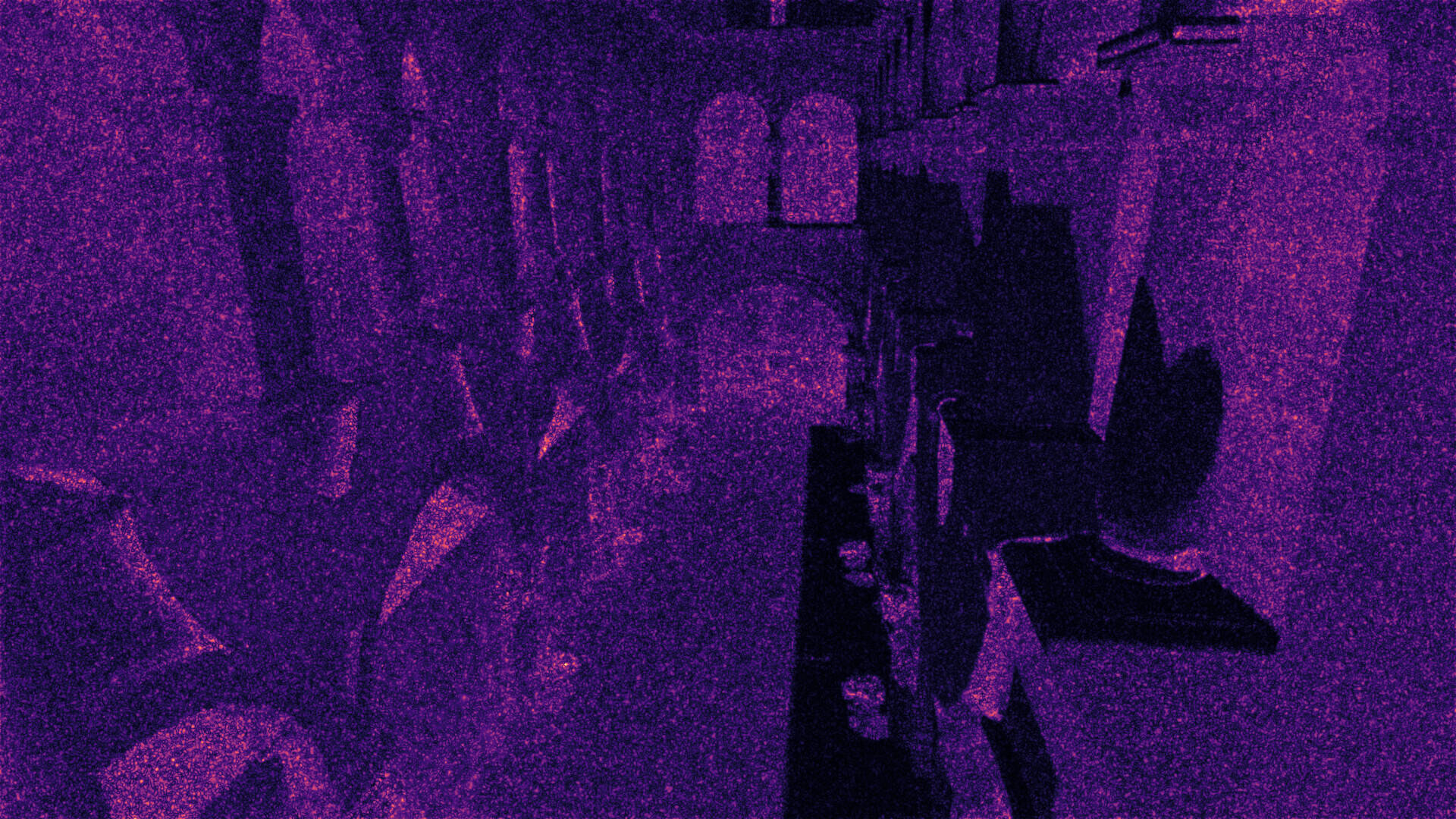}
    \\
    && FLIP Mean &
    0.226 &
    0.170 & 
    \textbf{0.165} &
    0.168 &
    0.176
    \\
    && MSE &
    0.001 &
    0.001 &
    0.001 &
    0.001 &
    0.001
    \\
    %============================%
    % RED REGION                 %
    %============================%
    &
    & 
    \colorBoxGraphics{spnzRefRed}{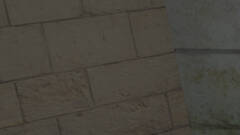}
    {red}{.125\textwidth}{ultra thick}
    &
    \colorBoxGraphics{spnzPtRed}{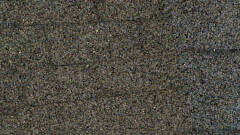}
    {red}{.125\textwidth}{ultra thick}
    &
    \colorBoxGraphics{spnzWFPGNPRed}{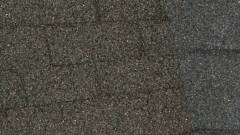}
    {red}{.125\textwidth}{ultra thick}
    &
    \colorBoxGraphics{spnzWFPGPRed}{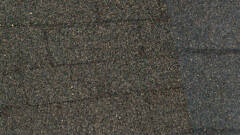}
    {red}{.125\textwidth}{ultra thick}
    &
    \colorBoxGraphics{spnzVMMRed}{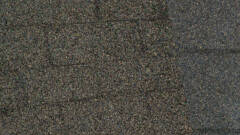}
    {red}{.125\textwidth}{ultra thick}
    &
    \colorBoxGraphics{spnzPPGRed}{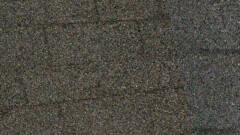}
    {red}{.125\textwidth}{ultra thick}
    \\[0mm]
    %============================%
    % GREEN REGION               %
    %============================%
    &
    & 
    \colorBoxGraphics{spnzRefGreen}{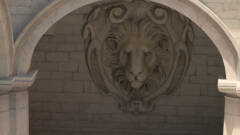}
    {green}{.125\textwidth}{ultra thick}
    &
    \colorBoxGraphics{spnzPtGreen}{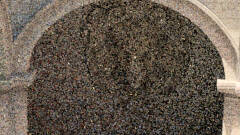}
    {green}{.125\textwidth}{ultra thick}
    
    &
    \colorBoxGraphics{spnzWFPGNPGreen}{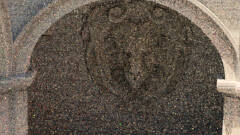}
    {green}{.125\textwidth}{ultra thick}
    &
    \colorBoxGraphics{spnzWFPGPGreen}{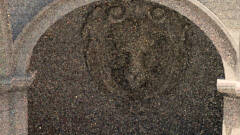}
    {green}{.125\textwidth}{ultra thick}
    &
    \colorBoxGraphics{spnzVMMGreen}{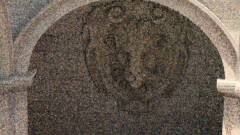}
    {green}{.125\textwidth}{ultra thick}
    &
    \colorBoxGraphics{spnzPPGGreen}{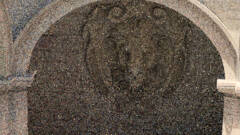}
    {green}{.125\textwidth}{ultra thick}
    \\
    \hline
    %=======================================%
    % Sponza Lion Camera
    %=======================================%
    % & & \multicolumn{3}{c|}{48spp} & \multicolumn{2}{c}{16t + 32spp}  \\
    % \multirow{3}{*}[1em]{\rotatebox[origin=l]{90}{\textsc{\large SponzaLion}}} 
    % &
    % \includegraphics[width=0.13\textwidth]{figures/comparisons/SponzaLion/sponza-lion-ref}
    % &
    % \includegraphics[width=0.13\textwidth]{figures/comparisons/SponzaLion/pt-48}
    % &
    % \includegraphics[width=0.13\textwidth]{figures/comparisons/SponzaLion/wfpg-np-48spp}
    % &    
    % \includegraphics[width=0.13\textwidth]{figures/comparisons/SponzaLion/wfpg-p-48spp}
    % &
    % \includegraphics[width=0.13\textwidth]{figures/comparisons/SponzaLion/vmm-16-32}    
    % &
    % \includegraphics[width=0.13\textwidth]{figures/comparisons/SponzaLion/ppg-16-32}
    % \\
    % &
    % \multirow{1}{*}[1.7em]{FLIP Heat Map}
    % &
    % \includegraphics[width=0.13\textwidth]{figures/comparisons/SponzaLion/flip-pt-48}
    % &
    % \includegraphics[width=0.13\textwidth]{figures/comparisons/SponzaLion/flip-wfpg-np-48}
    % &
    % \includegraphics[width=0.13\textwidth]{figures/comparisons/SponzaLion/flip-wfpg-p-48}
    % &
    % \includegraphics[width=0.13\textwidth]{figures/comparisons/SponzaLion/flip-vmm-16-32}
    % &
    % \includegraphics[width=0.13\textwidth]{figures/comparisons/SponzaLion/flip-ppg-16-32}
    % \\
    % & FLIP Mean &
    % 0.352 &
    % 0.194 & 
    % 0.219 &
    % \textbf{0.186} &
    % 0.212
    \end{tabular}
\end{table*}

Results can be seen in Table~\ref{tab:vmmComparison}. Comparisons provide HDR-FLIP heat maps and mean HDR-FLIP values~\cite{Andersson:2021:HDRFLIP}. On the left of the FLIP heat maps is the reference image of the Mitsuba Renderer. The images' mean square error (MSE) is also given as a separate row. Both compared method parameters are run with their default parameters. As both methods require training and rendering samples, we set the sample count of our process to the sum of these values.

We opted for a \(256^3\) resolution for the SVO, although Figure~\ref{fig:mem:ablation} suggests that \(128^3\) resolution is adequate for capturing the radiant exitance field. This is true for the ``VeachDoor'' scene, but the smaller resolution may not be sufficient on larger scenes such as the ``Sponza''. Because the memory overhead of increasing the resolution to $256^3$ is insignificant, we opted for this higher resolution for the comparisons.

For the ``Bathroom'' scene, Müller et al.'s and Ruppert et al.'s methods yield very similar error scores, which are both lower than our error score (lower is better in this case). This scene represents a worst-case scenario for our algorithm due to the ideal specular reflection of the mirror. As we represent the illumination using radiant exitance, we guide more rays toward the mirror, despite the fact that the light reflected off the mirror only illuminates the perfect reflection directions.

For the ``Veach Door'' scene, our error score for product path guiding lies in between the other two methods. In this scene, our product path-guiding version outperforms the regular path-guiding version due to the reason explained in the previous section. Finally, for the ``Sponza'' scene, WFPG product path guiding yields the lowest error score with a small margin.

Given that the compared methods use different renderers (an earlier version of the Mitsuba renderer) and architectures (GPU vs. CPU), the high degree of similarity between the algorithms suggests that our approach demonstrates competitive performance despite having a smaller memory impact.

\section{Limitations \& Future Work}
\label{sec:futureWork}

There are several limitations of the proposed method, some of which are shared by the other path-guiding methods as well. Here, we highlight the most important ones that can be addressed by future work.

\textbf{Densely generated radiance fields:} Generated radiance fields may not capture high-frequency features. These would require a higher resolution capture, which asymptotically requires \(\mathcal{O}(n^2)\) amount of work. As a future work, asymmetric cones could be used to query the incident location. A minimal data structure could orchestrate this approach. ``Compressed Directional Quadtree'' (CDQ), proposed by Dittebrandt et al., can be a candidate for this~\cite{Dittebrandt:2020:Quake}.

\textbf{Radiant exitance and highly specular objects:} To minimize memory footprint, we deemed it necessary to hold only the radiant exitance in the SVO data structure. However, in scenarios such as the ``Bathroom'' scene, this approach proves insufficient, as discussed earlier.  To address this issue, cones can be bounced from specular surfaces to continue to query the next hit location. Alternatively, the aforementioned CDQ can be used to segment the radiant exitance on regions with high specularity. Determination of these regions could be done at initialization time during SVO generation since it does not rely on light interaction. The refinement of the CDQ, however, would be performed during the runtime phase.

\textbf{Volumetric subdivision of the scene:} In the context of most path-guiding methods, spatial subdivision schemes are often volumetric. Similarly, in our case, the spatial binning scheme is also volumetric. This creates problems when an infinitely thin and two-sided surface occupies this volume. As discussed in this paper, product path guiding could mitigate this issue. Arguably, most scenes involve mostly reflective materials; a surface-based subdivision approach would be beneficial. 
    
\textbf{Selective path guiding:} Since our method generates radiance fields on the fly, it would reduce computational cost to avoid these calculations in regions that would minimally benefit from path guiding, such as those that receive strong direct lighting. This optimization would dramatically improve the computation time of scenes containing mixed regions dominated by direct and indirect illumination.

\section{Conclusion}
\label{sec:conc}

In this paper, we proposed the first GPU-oriented wavefront style path guiding method that does not rely on dynamic memory management -- an operation that does not suit the GPU architecture. The proposed method pre-generates an SVO data structure by voxelizing surfaces and refines the radiant exitance field during rendering. This structure is then utilized to generate PDFs to guide rays on the fly. This leads to a smaller memory requirement than the existing methods without hampering image quality. We also showed how to perform product sampling under this setting. By sharing our source code, we hope to stimulate future research for GPU path guiding, which could be vital for real-time path tracing.

\vfill\null
%\columnbreak

%\section*{Acknowledgments}
%Acknowledgments are withheld due to anonymity.
%The reference images reported in this paper were partially performed at TUBITAK ULAKBIM, High Performance and Grid Computing Center (TRUBA resources).

% Bibs...
\bibliographystyle{cag-num-names}
\bibliography{references, pathGuide, octree}

\appendix
\section{Validation}
\label{app:validation}

We share the results of an evaluation to compare the run-time performance of our ray tracing architecture with well-known architectures in the literature \cite{Bitterli:2016:Resources, PBRBOOK:2023, Jakob:2022:DrJit, NVIDIA:2023:PTSDK}. As seen in Table~\ref{tab:validation}, the run-time performance of our GPU-based WFPT implementation is similar to PBRT-v4's WFPT implementation. Both ours and PBRT-v4's WFPT results are somewhat slower than megakernel-based architectures due to the implementation overhead of manually managing path states.

\begin{table}[!ht]
    \renewcommand{\arraystretch}{1.3}
    \setlength{\tabcolsep}{1.5mm}
    \small
    \centering
    \caption{Comparison between different GPU-based path tracer implementations and ours using the ``Kitchen'' scene in $1920\times1080$ resolution. As can be seen from the per-sample timings, wavefront path tracers have an inherent implementation overhead compared to MegaKernel-based ones. However, our timings are on par with well-known ray tracer architectures. The image shown is generated by our renderer.}
    \label{tab:validation}

    \includegraphics[width=.65\columnwidth]{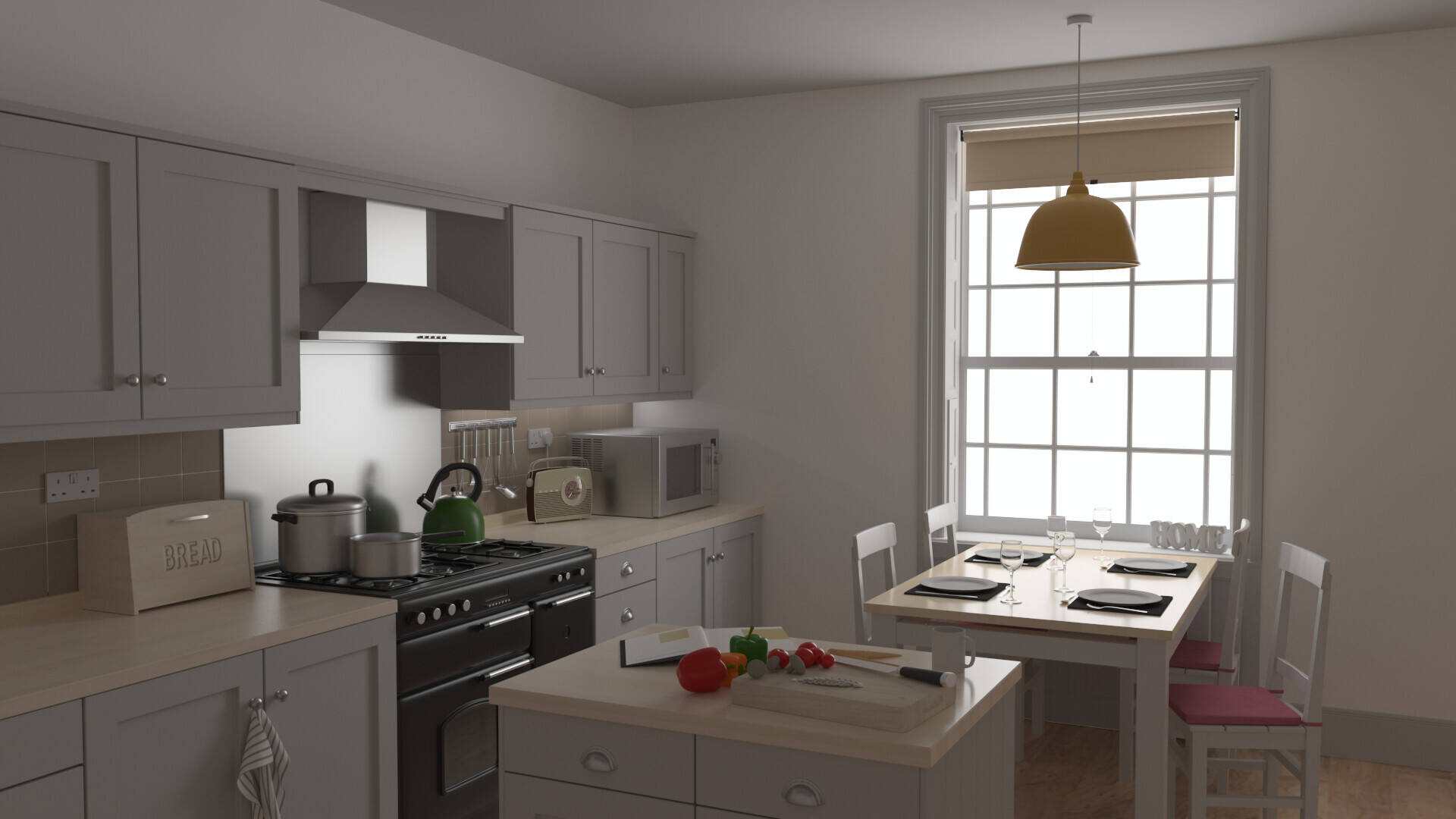}
    
    \begin{tabular}{c c c}
        \toprule
        \textbf{Renderer - Framework} & \textbf{Type} & \textbf{Time per sample (ms)} 
        \\ 
        \cellW NVIDIA-PT-SDK (DXR) & 
        \cellW MegaKernel & 
        \cellW 40.2 (NEE On)
        \\
        \cellW & \cellW & \cellW 26.0 (NEE Off)
        \\
        \cellG PBRT-v4 (CUDA) & 
        \cellG Wavefront & 
        \cellG 71.2 
        \\
        \cellW Mitsuba3 (CUDA) & 
        \cellW MegaKernel & 
        \cellW 48.8 
        \\
        \cellG Ours (CUDA) & 
        \cellG Wavefront & 
        \cellG 67.2 
        \\        
        \bottomrule
    \end{tabular}
\end{table}

\section{Occupancy Analysis}
\label{app:occAnalyze}

We have conducted GPU resource usage of our product sampling version of radiance field generation kernels. Results can be seen in Table~\ref{tab:gpuResource}. Shared memory utilization is at its limit for \(128 \times 128\) field generation kernel (68.3 KiB). Each multiprocessor has 128KiB of shared memory, of which 102KiB is available for the user on a 3070Ti mobile GPU. Doubling the radiance field resolution will quadruple the memory requirement, which will not fit into the shared memory. To achieve maximum utilization of a multiprocessor, we select a block size of 512, which results in 33\% occupancy. Achieving a higher occupancy would have required simplifying the kernels, but this is not feasible due to the complexity of the sampling routines.

\begin{table}[!ht]
    \renewcommand{\arraystretch}{1.3}
    \setlength{\tabcolsep}{1.5mm}
    \small
    \centering
    \caption{GPU resource usage statistics of the radiance field generation kernels. All kernels are launched via 512 threads per block.}
    \label{tab:gpuResource}

    \begin{tabular}{c c c c}
        \toprule
        \textbf{Field Resolution} & \textbf{Shared Mem. (KiB)} & \textbf{Registers} & \textbf{Occupancy}
        \\ 
        \midrule        
        \cellW \(128 \times 128\) & 
        \cellW 68.3 & 
        \cellW 128 & 
        \cellW 33\%
        \\
        \cellG \(64 \times 64\) & 
        \cellG 19.01 & 
        \cellG 128 & 
        \cellG 33\% 
        \\
        \cellW \(32 \times 32\) & 
        \cellW 6.72 & 
        \cellW 128 & 
        \cellW 33\% 
        \\
        \bottomrule
    \end{tabular}
\end{table}

\end{document}